\documentclass[12pt]{iopart}
\usepackage{tikz}
\usepackage{iopams}
\usepackage{bbm}
\usepackage{graphicx}
\usepackage{esvect}
\usepackage{color,cite}
\usepackage{bm}
\usepackage{amsmath}
\usepackage{url}
\usepackage{subcaption}
\usepackage{diagbox}
\usepackage[colorlinks=true,allcolors=blue]{hyperref}
\usepackage[normalem]{ulem}
\usepackage{booktabs}

\newcommand{\erf}[1]{\mathrm{erf}\!\left( {#1} \right)}
\newcommand{\erfc}[1]{\mathrm{erfc}\! \left( {#1} \right)}
\newcommand{\cumul}[1]{\langle\!\langle #1 \rangle\!\rangle}
\newcommand{\As}[1]{A\!\left(#1\right)}
\newcommand{\pr}[1]{\mathbb{P}\!\left(#1\right)}
\newcommand{\Qq}{\mathcal{Q}}

\newcounter{subsubsubsection}[subsubsection]

\usepackage{tikz}
\newcommand{\diagcorner}[2]{% #1 = row label (i,j),  #2 = column label (omega)
  \begin{tikzpicture}[baseline=(bb.center)]
    \node[minimum width=5em, minimum height=3.3em, inner sep=0pt] (bb) {};
    \draw (bb.north west) -- (bb.south east);
    \node[anchor=east,  inner xsep=3pt] at (bb.east)  {#2};  % omega -> right-middle
    \node[anchor=south, inner ysep=3pt] at (bb.south) {#1};  % (i,j) -> bottom-middle
  \end{tikzpicture}}

\begin{document}

\title[Macroscopic fluctuation theory for multi-time statistics]{Macroscopic fluctuation theory for the multi-time statistics of current in non-stationary diffusive systems}

\author{Sabyasachi Chowdhury, Kapil Sharma, Sandeep Jangid, Tridib Sadhu$^\star$}

\address{Department of Theoretical Physics, Tata Institute of Fundamental Research, Homi Bhabha Road, Mumbai 400005, India}

\ead{$^\star$tridib@theory.tifr.res.in}

\begin{abstract} The statistics of current fluctuations has long been a central object of study in non-equilibrium physics. Most existing work has focused on one-time statistics, while their multi-time generalisation remains comparatively less explored. We address this gap by extending the fluctuating hydrodynamics framework of Macroscopic Fluctuation Theory (MFT) to study multi-time statistics in the non-stationary state of a diffusive system on an infinite line. For the simplest cases of a non-interacting lattice gas and hard-core Brownian point particles, we present explicit solution of the MFT leading to multi-time large-deviation statistics. For generic systems, the MFT is solved perturbatively, yielding explicit results for two-time correlations. These reveal that the connection between current fluctuations and fractional Brownian motion, previously observed for flat initial conditions, does not persist for step initial conditions. We independently verify these hydrodynamic results by solving the corresponding microscopic dynamics for the non-interacting gas and for the symmetric simple exclusion process. Additional confirmation comes from numerical simulations.

\end{abstract}


\submitto{A special issue of JSTAT for StatPhys29}

\maketitle

%\tableo

\section{Introduction}

The study of current fluctuations has been an important enterprise in non-equilibrium statistical mechanics~\cite{Derrida2007, Derrida2025,Mallick2015,Prahofer2002,Bertini2005Current,Bertini2006Current,2015_Lazarescu_The,Dandekar2023,2020_Banerjee_Current,2023_Bello_Current,1995_Lee_Universal,2023_McCulloch_Full}. In this line of investigation, a major interest has been in estimating the long-time statistics of the current and its probability distribution in terms of large deviation function~\cite{TOUCHETTE,Derrida2007,Derrida2025,Mallick2015}. However, estimating the large-deviation statistics is generally a challenging task.
In specific cases, the full distribution of current fluctuations has been obtained, such as for boundary-driven diffusive systems\cite{Lecomte2010, Akkermans2013,Bodineau2004,Harris_2005,Bodineau2006}, diffusive systems on a ring\cite{Bodineau2005,Appert2008,Mallick2015}, and exclusion processes on infinite~\cite{Gerschenfeld2009Bethe,Gerschenfeld2009,Mallick2022PRL} and semi-infinite~\cite{semi_inf_2026, Kirone_semi_2024} geometries. 

Most of these studies have focused on single-time statistics, namely the probability distribution of the integrated current measured over a single time interval. In the stationary state of a finite system \cite{Bodineau2004,Harris_2005,Bodineau2006}, current fluctuations are short-range correlated in time, and  their multi-time statistics is therefore easily inferred from single-time statistics. In the non-stationary state on an infinite line, however, current fluctuations retain~\cite{Gerschenfeld2009} memory of the initial state even at long times, making it important to understand how currents measured over different time intervals are statistically related. Furthermore, currents in non-stationary systems can have probability distributions that are non-Gaussian~\cite{Gerschenfeld2009, Gerschenfeld2009Bethe,Mallick2022PRL}, which necessitates knowledge of higher-order time correlations. However, exact results on multi-time correlations have so far remained limited~\cite{Sadhu_2016,Krapivsky_2015,Sadhu_2015, Benichou_multi_2025}.

In this work, we investigate the multi-time statistics of current within the framework of Macroscopic Fluctuation Theory (MFT) for generic diffusive systems on a one-dimensional infinite line with a domain-wall initial state. MFT, developed by Bertini, De Sole, Gabrielli, Jona-Lasinio, and Landim~\cite{Bertini2014,Bertini2002,Bertini2001} has been a powerful hydrodynamic framework for characterizing large-deviation statistics for generic diffusive systems. This hydrodynamic theory has reproduced several results known earlier from microscopic computations~\cite{Derrida2007,Derrida2025,Mallick2015} and even led to exact results that are beyond the microscopic tractability~\cite{saha2026_slow,semi_inf_2026, Kirone_semi_2024,2026_Saha_Bottom}. MFT is a coarse-grained macroscopic description~\cite{Bertini2014,2026_Saha_Bottom,Spohn1991} of diffusive dynamics suitably defined in rescaled coordinates $(x,\tau)\equiv(\tfrac{X}{\ell},\tfrac{t}{\ell^2})$ of microscopic space-time with a length scale $\ell$ much larger than equilibrium correlation lengths. For systems with a single locally conserved quantity, the macroscopic dynamics is characterized by a coarse-grained density $\rho(x,\tau)$. In this description, all the microscopic details are encoded~\cite{Bertini2014,2026_Saha_Bottom,Derrida2007,Derrida_2011,Derrida2025} in two macroscopic transport coefficients: diffusivity $D(\rho)$ and mobility $\sigma(\rho)$.  

\begin{figure}[htbp]
     \centering
         \includegraphics[width=14cm, height = 3cm]{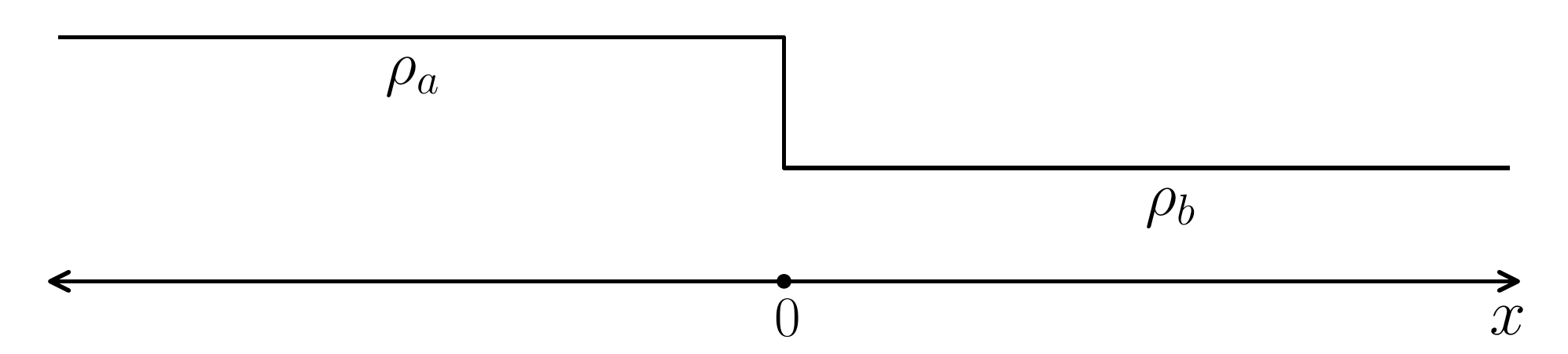}  
\caption{The domain-wall initial state, where the negative half-line ($x\le 0$) is in equilibrium with average density $r(x)=\rho_a$ while the positive half-line ($x>0$) has average density $r(x)=\rho_b$.}
\label{fig:initial_condition}    
\end{figure}

The MFT framework has been extensively used for studying single-time current statistics in different geometries. Within MFT, the long-time current statistics in terms of large-deviation theory is formulated as a variational problem of minimizing the associated dynamical action~\cite{Gerschenfeld2009,Bertini2014,Derrida_2011}. For the stationary state in a finite geometry, the solution to the variational problem for generic systems follows from the additivity principle~\cite{Bodineau2004,Bertini2005Current,Akkermans2013,BodineauDerrida2026}. For the non-stationary scenario, the problem remains challenging and only a limited number of systems are tractable. Particularly among them, on the infinite line, a recent remarkable connection \cite{Mallick2022PRL, KMP_exact_2022, KPZ_exact_2021} of the variational problem with classical integrability has led to explicit large-deviation results for single-time current statistics in a symmetric simple exclusion process (SSEP). The corresponding problem has also been solved recently in the semi-infinite geometry~\cite{semi_inf_2026, Kirone_semi_2024} and on an infinite line with defect bonds~\cite{saha2026_slow}. 

In this article, we explore how the framework of MFT can be extended to study the multi-time statistics of current in the non-stationary scenario of a diffusive system on an infinite line with an initial domain-wall state, as illustrated in the Figure~\ref{fig:initial_condition}. We shall closely rely on a similar earlier work~\cite{Krapivsky_2015} on the multi-time statistics of tracer particles. The time-integrated current $Q(t)$ is defined as the total flux of particles across the origin in a given time interval $t$. Its multi-time statistics is described in terms of a cumulant generating functional. It has been noted earlier that one-dimensional transport exhibits an unusual long-time memory of the initial state~\cite{2013_Leibovich_Everlasting, KMS_JSP, Sadhu_2015, Lizana2014,Gerschenfeld2009,Krapivsky2012PRE}. In the literature, this has been commonly demonstrated by averaging over initial fluctuations in ways that are analogous~\cite{Sadhu_2015} to annealed and quenched ensembles in disordered systems.
In annealed averaging, the initial configuration is sampled from a stationary measure, including rare fluctuations. On the other hand, quenched averaging considers only typical initial configurations. For the multi-time statistics, we define this precisely by the following cumulant generating functionals (cgf) for the time-integrated current:
\begin{subequations}
\label{eq:generating functional unscaled}
\begin{align}
\mu_{\mathcal{A}}(\lambda(t)) = \log\overline{\left\langle \exp\left( \int_0^T \mathrm{d}t \, \lambda(t) Q(t) \right) \right\rangle} \label{eq:annealed generating functional unscaled} \\ 
\mu_{\mathcal{Q}}(\lambda(t)) = \overline{\log\left\langle \exp\left( \int_0^T \mathrm{d}t \, \lambda(t) Q(t) \right) \right\rangle}\label{eq:quenched generating functional unscaled}
\end{align}
\end{subequations}
where $\lambda(t)$ is a fugacity parameter,  $\langle\cdots\rangle$ denotes the average over the stochastic evolution, and $\overline{\cdots}$ represents the average over initial fluctuations. The subscripts $\mathcal{A}$ and $\mathcal{Q}$ denote annealed and quenched ensembles, respectively. Here, $T$ is an arbitrarily large time, which in principle can be infinite. We set it to a finite value to keep the integral in \eqref{eq:generating functional unscaled} finite. The conventional~\cite{Gerschenfeld2009Bethe,Gerschenfeld2009,Mallick2022PRL} cgf for single-time statistics corresponds to the case $\lambda(t)=\lambda \, \delta(T-t)$.

The cgf in \eqref{eq:generating functional unscaled} encodes all multi-time correlations of $Q(t)$ and is obtained by functional derivatives
\begin{align}
    \cumul{ Q(t_1)\cdots Q(t_n)} =\left.\frac{\delta^n \mu(\lambda(t))}{\delta\lambda(t_1)\cdots\delta\lambda(t_n)}\right\vert_{\lambda=0}
\end{align}
where $\cumul{\ldots}$ denotes a cumulant.
Our interest in this article is in correlations for times $t_{i}$ that are comparable to the hydrodynamic time scale $\ell^2$. More precisely, we consider $T\sim \ell^2$ and $t_i\sim T$. We shall describe how multi-time statistics in this limit can be obtained by extending the MFT framework that was earlier developed~\cite{Gerschenfeld2009} for single-time statistics. Our main results obtained from this formalism are as follows.

For the simplest cases of a non-interacting diffusive lattice gas and for Brownian hard-core point particles the theory corresponds to ($D(\rho)=1$, $\sigma(\rho)=2\rho$) and it is tractable. For this we obtain explicit expressions for the correlations. In the annealed setting, \begin{subequations}\label{eq:QA multi corr}
\begin{align}
    \cumul{ Q(t_1) \cdots Q(t_n)}_{\mathcal{A}}\simeq \left(\rho_a+(-1)^n\rho_b\right)\int_{-\infty}^0 \mathrm{d}z_0\,\Psi_{z_0}\left(t_1,  \ldots , t_n\right)\label{eq:non_annealed_n_point}
\end{align}    
for large $t_i$ with $t_1<t_2<\cdots<t_n$ and 
\begin{align}
    \Psi_{z_0}(t_1,  \ldots , t_n) = \int_0^{\infty}\prod_{k=1}^{n}dz_{k}\,\frac{e^{
    -\frac{(z_{k}-z_{k-1})^{2}}{4(t_k-t_{k-1})}}}{\sqrt{4\pi (t_k-t_{k-1})}}\label{eq:Psi n point}
\end{align}\label{eq:non_annealed_n_point_full}\end{subequations}
with $t_0=0$ by convention. Similar results for the multi-time statistics of a tracer were obtained earlier for hard-core Brownian particles~\cite{Sadhu_2015} and for the SSEP in the high-density limit~\cite{Benichou_multi_2025} using microscopic techniques.

In the quenched setting, \begin{subequations}\label{eq:QQ multi corr}
\begin{align}
    \cumul{ Q(t_1)\cdots Q(t_n)}_{\mathcal{Q}} \simeq \left(\rho_a + (-1)^n\rho_b\right)  \int_{-\infty}^{0}\mathrm{d}z_0 \,\Phi_{z_0}\left(t_1, \ldots ,t_n\right)\label{eq:non_quenched_n_point}
\end{align}
with
\begin{align}\label{eq:phiz0 intro}
\Phi_{z_0}(t_1,\ldots,t_n)
=
\sum_{\omega\in\Omega_n}
(-1)^{|\omega|-1} (|\omega|-1)!
\prod_{B_\omega\in\omega}
\Psi_{z_0}\left(t_{b_1},\ldots,t_{b_{|B_\omega|}}\right)
\end{align}\label{eq:non_quenched_n_point_full}\end{subequations}
where $\Omega_n$ denotes the set of all partitions of
the index set $\{1,\ldots,n\}$. For each partition
$\omega\in\Omega_n$,  $|\omega|$ denotes the number of blocks $B_\omega$ with ordered elements. For a block $B_\omega\equiv\{b_1,b_2,\cdots,b_{\lvert B_\omega\rvert}\}$ of size $\lvert B_\omega\rvert$, with its elements in increasing order, the function $\Psi_{z_0}$ is evaluated at the corresponding times $(t_{b_1},t_{b_2},\ldots,t_{b_{\lvert B_\omega\rvert}})$.

As an example, for $n=3$, considering the five partitions of $\{1,2,3\}$, we get
\begin{align}
\Phi_{z_0}(t_1,t_2,t_3)
=&
\Psi_{z_0}(t_1,t_2,t_3)
-\Psi_{z_0}(t_1)\Psi_{z_0}(t_2,t_3)
-\Psi_{z_0}(t_2)\Psi_{z_0}(t_1,t_3)
-\Psi_{z_0}(t_3)\Psi_{z_0}(t_1,t_2)
\cr
&\qquad
+2\,\Psi_{z_0}(t_1)\Psi_{z_0}(t_2)\Psi_{z_0}(t_3).\nonumber
\end{align}

Explicit expressions for the correlations~(\ref{eq:QA multi corr},\ref{eq:non_quenched_n_point_full}) can be obtained by evaluating the Gaussian integrals. For the annealed ensemble, this yields
\begin{align}\label{eq:explicit_annealed}
    \cumul{ Q(t_1)....Q(t_n)}_{\mathcal{A}}\simeq \frac{\left(\rho_a +(-1)^n\rho_b\right)}{\sqrt{\pi}}\sum_{0\leq i < j\leq n}^n \epsilon_{ij} P^{(n)}_{ij}\sqrt{t_j-t_i}
\end{align}
with the signs $\epsilon_{0j}=+1$ and $\epsilon_{ij}=-1$ for $i\ge 1$, while the coefficients $P^{(n)}_{i,j}$ are intensive (scale-invariant) functions of time involving only time ratios. For the quenched ensemble, there is a similar expression 
\begin{align}\label{eq:explicit_quenched}
    \cumul{ Q(t_1)\cdots Q(t_n)}_{\mathcal{Q}}\simeq\;
\frac{\left(\rho_a+(-1)^n\rho_b\right)}{\sqrt{\pi}}
\left\{\sum_{0\le i<j\le n} D^{(n)}_{ij}\,\sqrt{t_j-t_i}
\;+\sum_{1\le i<j\le n} S^{(n)}_{ij}\,\sqrt{t_i+t_j}\right\}
\end{align}
with the time-intensive coefficients $D^{(n)}_{ij}$ and $S^{(n)}_{ij}$. Explicit expressions for these coefficients are presented in \sref{sec: mft infinite non-interacting}. 

These results for the non-interacting limit are obtained by an explicit solution of the corresponding MFT and independently by a solution of the microscopic dynamics. For generic diffusive systems, characterized by diffusivity $D(\rho)$ and mobility $\sigma(\rho)$, the corresponding MFT is challenging and there are only limited results available even for single-time statistics~\cite{Krapivsky2014, Benichou_2024, Berlioz2025PRL}. A standard approach~\cite{Krapivsky2012PRE,Saha_2023,Berlioz_2024,Sadhu2023,Benichou_2024,KMS_interface,Dandekar2023,Krapivsky_2015} is a perturbative solution in the fugacity $\lambda$, which systematically gives correlations order by order. A tractable scenario is the case of constant diffusivity $D(\rho) = 1$ (set to unity by an appropriate rescaling of time) and arbitrary $\sigma(\rho)$. In this case, we show that the two-time correlation in the annealed setting has an explicit expression for $t_2>t_1$,
\begin{align}
    \cumul{ Q(t_1) Q(t_2)}_{\mathcal{A}} \simeq & \frac{\sigma\left(\bar{\rho}\right)}{2 \sqrt{ \pi}}\left(\sqrt{t_1} + \sqrt{t_2} - \sqrt{t_2 - t_1}\right)
        -\frac{(\Delta \rho)^2 \, 
        \sigma^{\prime\prime}\left( \bar{\rho} \right)}{8 \pi^{3/2} } 
    \Bigg\{ 
        \frac{\pi}{2}\left(\sqrt{t_1} + \sqrt{t_2} \right) \nonumber \\
    & - 2 \sqrt{t_1 + t_2} \arctan\left( \sqrt{ \frac{t_1 + t_2}{t_2 - t_1} } \right) - 2 \sqrt{t_1} \arctan\left( \sqrt{ \frac{t_2}{t_1} - 1 } \right) 
    \Bigg\}
\label{eq:annealed_two_time_correlation_general_sigma_constant_D_intro}
\end{align}
up to quadratic order in $\Delta \rho = \rho_a - \rho_b$ around an average density $\bar{\rho} = \tfrac{\rho_a + \rho_b}{2}$. Here, the sign $\simeq$ denotes that the above expression holds in the large-$t_1$ and large-$t_2$ limit.
\begin{figure}[t]
    \centering
    \includegraphics[width=0.8\linewidth]{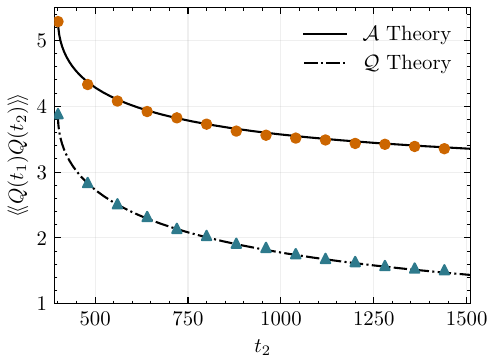}
    \caption{\textbf{SSEP two-time correlation:} Autocorrelations of the integrated current $Q(t)$ in an SSEP for an initial domain-wall state with $\rho_a = 0.5$ and $\rho_b = 0.3$. The plot shows the variation of $\cumul{ Q(t_1)Q(t_2)}$ with $t_2> t_1$ for a fixed $t_1 = 400$ for both ensembles. The circles and triangles represent the data points obtained by Monte Carlo simulations and averaged over $10^8$ samples. The lines correspond to the theoretical results \eqref{eq:annealed_two_time_correlation_general_sigma_constant_D_intro} and \eqref{eq:quenched_two_time_correlation_general_sigma_constant_D_intro} for $D=1$ and $\sigma=2\rho(1-\rho)$. The slow decay reflects the long-range nature of the autocorrelations.}
    \label{fig:SSEP_two_time}
\end{figure}

A few remarks are in order. First, the higher-order contributions in $\Delta\rho$ contain higher-order derivatives of the mobility $\sigma(\rho)$. For a quadratic mobility, the expression \eqref{eq:annealed_two_time_correlation_general_sigma_constant_D_intro} is exact. This includes well-known lattice transport models, such as the symmetric simple exclusion process (SSEP)~\cite{Derrida2007}  $(D(\rho) = 1$ and $\sigma(\rho) = 2 \rho (1 - \rho))$,  the Kipnis-Marchioro-Presutti (KMP) model of heat conduction~\cite{1982_Kipnis_Heat}  $(D(\rho) = 1$ and $\sigma(\rho) = 2 \rho^2$), and the symmetric simple inclusion process~\cite{2007_Giardina_Duality} $(D(\rho) = 1$ and $\sigma(\rho) = 2 \rho (1 + \rho))$. See \cite{2026_saha_density,2026_Saha_Bottom} for a list of such models.
Second, for a stationary initial condition $(\Delta \rho = 0)$, the correlation \eqref{eq:annealed_two_time_correlation_general_sigma_constant_D_intro} follows the covariance of fractional Brownian motion ~\cite{1968_Mandelbrot_Fractional,Sadhu2018ArcsineFBM,Sadhu2021FunctionalsFBM} with Hurst exponent $H=\tfrac{1}{4}$. This has been noted earlier in several model systems~\cite{2013_Leibovich_Everlasting,Dandekar2023,Krapivsky_2015,Sadhu_2016}. However, the expression \eqref{eq:annealed_two_time_correlation_general_sigma_constant_D_intro} shows that for a non-stationary state, the process differs from fractional Brownian motion. Interestingly, the leading non-trivial correction appears at quadratic order in $\Delta \rho$.

In the quenched setting, the two-time correlation of the current has a similar expression in powers of $\Delta \rho$. 
\begin{align}
      \cumul{ Q(t_1) Q(t_2) }_{\mathcal{Q}} = 
          \frac{\sigma\left(\bar{\rho}\right)}{2 \sqrt{ \pi}} 
     & \left(
          \sqrt{t_1 + t_2} 
          - \sqrt{t_2 - t_1}    \right) -\frac{(\Delta \rho)^2\sigma^{\prime\prime}\left( \bar{\rho} \right)}{8 \pi^{3/2} } 
    \Bigg\{ 
          \frac{\pi}{2}\left(\sqrt{t_1} + \sqrt{t_2} - \sqrt{t_2 + t_1} \right) \nonumber \\
       +\frac{\pi}{2}\left(\sqrt{t_1} + \sqrt{t_2}\right) -  2 & \sqrt{t_1 + t_2} \arctan\left( \sqrt{ \frac{t_1 + t_2}{t_2 - t_1} } \right) - 2 \sqrt{t_1} \arctan\left( \sqrt{ \frac{t_2}{t_1} - 1 } \right) 
      \Bigg\}
\label{eq:quenched_two_time_correlation_general_sigma_constant_D_intro}
\end{align}
Note the similarity of the second line in \eqref{eq:quenched_two_time_correlation_general_sigma_constant_D_intro} to the expression in \eqref{eq:annealed_two_time_correlation_general_sigma_constant_D_intro}. Both expressions have been validated through numerical simulations for the SSEP and shown in Figure~\ref{fig:SSEP_two_time}. For both ensembles, the autocorrelation depends explicitly on $t_1$ and $t_2$, rather than solely on their time difference, reminiscent of aging in disordered systems \cite{2008_Henkel_Local,2011_Berthier_Theoretical}.

MFT, being a hydrodynamic theory, rests on the construction of a large-scale description that often relies on reasonable assumptions (see, for example, \cite{2026_Saha_Bottom}). For this reason, it is common practice to independently verify generic results from MFT in specific tractable models using exact microscopic solution. We do the same for the current correlations in the SSEP, which is integrable via Bethe ansatz \cite{Gerschenfeld2009Bethe}. 

Besides verifying (\ref{eq:annealed_two_time_correlation_general_sigma_constant_D_intro},\ref{eq:quenched_two_time_correlation_general_sigma_constant_D_intro}), we compute the equal-time correlations of the current with the background density. The single-file constraint in the SSEP induces a non-trivial density profile in order to produce an atypical current $Q(t)$. Quantifying this change in the surrounding density has drawn recent interest~\cite{Grabsch2022,Benichou2021PRL}. A measure of this density variation is the correlation $\cumul{n_i(t)Q(t)}$ between the occupation variable $n_i$ of a site $i$ in SSEP and the current $Q(t)$ across the origin. For the domain-wall initial state in the annealed setting, we show using microscopic solution that the correlation follows a scaling property in the long-time limit, 
\begin{subequations}\label{eq:tau_Q_annealed_explicit_SSEP}
\begin{equation}
    \cumul{ n_i(t)Q(t) }_{\mathcal{A}}\simeq h_{\mathcal{A}}\left(\frac{i}{2\sqrt{t}} \right) \qquad \textrm{for } i> 0,
\end{equation}
where the scaling function, for $x>0$, has a simple expression
\begin{equation}
    h_{\mathcal{A}}(x) = \frac{\overline{\rho}\,(1-\overline{\rho})}{2}\,\erfc{x}
+ \frac{(\Delta\rho)^2}{8}\left(\erfc{x} - \erfc{\frac{x}{\sqrt{2}}}^{2}\right)
\end{equation}
\end{subequations}
For $i<0$, the correlation is obtained by odd parity, $\cumul{ n_i(t)Q(t) }_{\mathcal{A}}\simeq - h_{\mathcal{A}}\left(-\frac{i}{2\sqrt{t}} \right)$.

For the quenched setting, the correlation also follows a similar scaling dependence, $\cumul{n_i(t)Q(t)}_{\mathcal{Q}}\simeq h_{\mathcal{Q}}\left(\frac{i}{2\sqrt{t}} \right)$, with the scaling function for $x>0$,
\begin{align}
	h_{\mathcal{Q}}\left(x \right)=h_{\mathcal{A}}\left(x \right)- \frac{\rho_a(1 - \rho_a)}{4}\phi\left(x\right)  + \frac{\rho_b(1 - \rho_b)}{4}\phi\left(-x\right)
	\label{eq:annealed_quenched_difference_tauQ}
\end{align}
where $\phi(x) = \Erfc{x} - \frac{1}{2}\Erfc{\frac{x}{\sqrt{2}}}^2$. 

\begin{figure}
     \centering
         \includegraphics[width=0.8\linewidth]{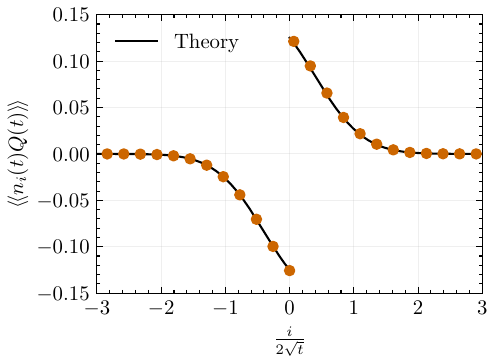}
     \caption{\textbf{Density-current correlation in SSEP:} The solid line indicates the theoretical result \eqref{eq:tau_Q_annealed_explicit_SSEP} while the data points indicate numerical simulation results averaged over $10^7$ samples for $t=1500$ for a step-initial profile with densities $\rho_a = 1$ and $\rho_b = 0$. }    
     \label{fig:density_current_annealed}
\end{figure}

For equilibrium ($\Delta\rho=0$), the expression simplifies to, $h_{\mathcal{Q}}(x) = \frac{\overline{\rho}(1 - \overline{\rho})}{2}\Erfc{\frac{x}{\sqrt{2}}}$, compared with $h_{\mathcal{A}}(x) = \frac{\overline{\rho}(1 - \overline{\rho})}{2}\Erfc{x}$. The equilibrium result for the annealed setting recovers the expression obtained in~\cite{Grabsch2022,Benichou2021PRL} following a closure scheme for the density profile in a biased ensemble. We have verified the non-equilibrium result~(\ref{eq:tau_Q_annealed_explicit_SSEP},\ref{eq:annealed_quenched_difference_tauQ}) independently by Monte Carlo simulation in Figure~\ref{fig:density_current_annealed} for $(\rho_a=1,\rho_b=0)$, where the results for both ensembles coincide. These exact results (\ref{eq:tau_Q_annealed_explicit_SSEP},\ref{eq:annealed_quenched_difference_tauQ}) provide an independent verification of the optimal profile within MFT, which we present in~\sref{sec:micro_SSEP}.

In the rest of this article, we discuss how these results are obtained in the following order. In \sref{sec: mft infinite}, we introduce the MFT formalism for computing multi-time statistics of the current $Q(t)$ for diffusive systems evolving on an infinite one-dimensional line. In~\sref{sec: mft infinite non-interacting}, we solve the MFT for non-interacting particles, leading to the explicit results (\ref{eq:QA multi corr},\ref{eq:QQ multi corr}), and in \sref{sec:mft_infinite_interacting}, we discuss a series solution for systems with constant diffusivity $D(\rho)$ and generic mobility $\sigma(\rho)$, leading to the two-time correlation functions (\ref{eq:annealed_two_time_correlation_general_sigma_constant_D_intro},\ref{eq:quenched_two_time_correlation_general_sigma_constant_D_intro}). In \sref{sec:micro_non_interacting}, we reproduce the results for non-interacting particles using a microscopic analysis, while in \sref{sec:micro_SSEP}, we do the same for the two-time correlation function for the SSEP. Additional details are given in the appendices.

\section{Macroscopic fluctuation theory for multi-time statistics \label{sec: mft infinite}}

We build on an earlier discussion~\cite{Krapivsky_2015} of the MFT framework for analyzing multi-time statistics of time-integrated observables. The MFT provides~\cite{Bertini2014,Bertini2002,Bertini2001,Derrida_2011,Derrida2025} a variational framework for estimating the large deviations associated with a fluctuating observable in generic diffusive systems. Essentially, it takes a fluctuating-hydrodynamics description of a many-body system  and casts it~\cite{Derrida2007,Derrida_2011,Tailleur2007,Tailleur2008} into a classical field theory for the computation of large-deviation statistics. The basic idea of fluctuating hydrodynamics~\cite{Spohn1991,2026_Saha_Bottom} is that on large spatial and temporal scales, defined in terms of rescaled coordinates $(x,\tau)\equiv(\tfrac{X}{\ell}, \tfrac{t}{\ell^2})$ with large $\ell$, a diffusive system is characterized by coarse-grained density fields whose dynamics are governed by stochastic differential equations with multiplicative noise. For diffusive systems with a single locally conserved field $\rho(x,\tau)$ at the macroscopic scale, fluctuating hydrodynamics is described by a conservation equation~\cite{Derrida2007, Derrida2025, 2026_Saha_Bottom}
\begin{align}\label{eq: fluctuatig_hyfro_equation}
    \partial_{\tau} \rho(x,\tau)=-\partial_x j(x,\tau)\quad\textrm{with }  j(x,\tau)=-D(\rho(x,\tau))\partial_x\rho(x,\tau)+\eta(x,\tau)
\end{align}
where $\eta(x,\tau)$ is a multiplicative Gaussian white noise with zero mean and covariance 
\begin{equation}\label{eq:noise covariance}
\langle \eta(x,\tau)\eta(x^{\prime},\tau^{\prime})\rangle=\frac{\sigma(\rho(x,\tau))}{\ell}\delta(x-x^{\prime})\delta(\tau-\tau^{\prime})
\end{equation}
with the angular brackets denoting the ensemble average over histories of evolution. At this macroscopic scale, all microscopic details are captured in two transport coefficients: the diffusivity $D(\rho)$ and the mobility $\sigma(\rho)$, which are related by the fluctuation relation $2D(\rho)=\sigma(\rho)f''(\rho)$, where $f(\rho)$ is the canonical free-energy density~\cite{Derrida2007, Derrida2025}. 

The hydrodynamic scale $\ell$ is considered larger than the equilibrium correlation length, such that local fluctuations are governed by an effective equilibrium with a local chemical potential. For details on the construction of~\eqref{eq: fluctuatig_hyfro_equation} for both lattice and continuum models, we refer to our recent article~\cite{2026_Saha_Bottom}.

In this article, we focus on characterizing the current $Q(t)$ measured across the origin over a time $t$ on an infinite line (see \fref{fig:initial_condition}) for systems governed by the fluctuating hydrodynamics equation \eqref{eq: fluctuatig_hyfro_equation}. Due to conservation, $Q(t)$ is the net change in the `charge' on the positive half-line, which can be expressed in terms of the coarse-grained density $\rho(x,\tau)$ as
\begin{align}
	Q(t)=\ell \int_{0}^{\infty} \mathrm{d}x\,\left(\rho(x,\tau)-\rho(x,0)\right)\label{eq:QT definition}
\end{align}
where $\tau=\tfrac{t}{\ell^2}$ is the rescaled time.

The observable $Q(t)$ is a fluctuating quantity with its value depending on the evolving density profile. Considering the stochastic evolution \eqref{eq: fluctuatig_hyfro_equation}, the generating functional characterizing the multi-time statistics of $Q(t)$ can be written~\cite{Derrida2007,Derrida_2011} as a path integral
\begin{align}\label{eq:path integral one}
    \left\langle \mathrm{e}^{\int_0^T \mathrm{d}t \lambda(t)Q(t)}\right\rangle \simeq &\int_{\rho(x,0)}\!\!\!\!\!\!\!\!\!\!\!\mathcal{D}\left[\rho,j\right]  \exp{\left\{\ell^3 \int_0^{\frac{T}{\ell^2}} \mathrm{d}\tau\, \lambda(\ell^2\tau)\int_{0}^{\infty}\mathrm{d}x\left(\rho(x,\tau)-\rho(x,0)\right)\right\}}  \cr
  & \exp{\left\{-\ell\int_{-\infty}^{\infty}\mathrm{d}x\int_0^{\frac{T}{\ell^2}}\mathrm{d}\tau\frac{\left(j+D(\rho)\partial_x\rho\right)^2}{2\sigma(\rho)}\right\}}  \prod_{x,\tau}\delta\left(\partial_{\tau} \rho+\partial_x j\right)
\end{align}
where $T$ is a large time, $\simeq$ indicates the asymptotic behavior in the large-$\ell$ limit, and $\langle \ldots \rangle$ indicates an average over evolutions starting with a density profile $\rho(x,0)$, also indicated by the lower limit of the path integral. The exponential term in the second line of \eqref{eq:path integral one} is the Gaussian probability weight of current fluctuations $j$ due to~\eqref{eq:noise covariance}, while the delta function is due to the continuity equation \eqref{eq: fluctuatig_hyfro_equation}.

Using an integral representation of the delta function $\delta(u)=\frac{1}{2\pi \mathrm{i}}\int d\hat{\rho}\,\exp(-\hat{\rho}u)$ at every space-time point and subsequently performing the Gaussian integration over $j(x,\tau)$ after integrating by parts, the generating function
\begin{align}\label{eq:path integral two}
 \bigg\langle \!\! \mathrm{e}^{\int_0^T \mathrm{d}t \,\lambda(t)Q(t)}\!\!\bigg\rangle & \simeq \int_{\rho(x,0)}\!\!\!\!\!\!\!\!\!\!\!\mathcal{D}\left[\rho,\hat{\rho}\right]\exp\left\{\ell^3 \int_0^{\frac{T}{\ell^2}} \mathrm{d}\tau\, \lambda(\ell^2\tau)\int_{0}^{\infty}\mathrm{d}x\left(\rho(x,\tau)-\rho(x,0)\right)\right\}\cr
   \exp&\left\{-\ell \int_{0}^{\frac{T}{\ell^2}}\mathrm{d}\tau
    \int_{-\infty}^{\infty}\mathrm{d}x\,\left\{ \hat{\rho}\partial_{\tau}\rho -
    \frac{\sigma(\rho)}{2}\left( \partial_{x}\hat{\rho}
    \right)^{2}+D(\rho)\left( \partial_{x}\rho \right)\left(
    \partial_{x}\hat{\rho}
    \right) \right\}\right\}
\end{align}
This is the Martin–Siggia–Rose-Janssen-DeDominicis (MSRJDD)~\cite{Martin1973, DeDominicis1975, Janssen1976} formalism for the stochastic differential equation \eqref{eq: fluctuatig_hyfro_equation}, where $\hat{\rho}$ is the response field.

For the spatial integrals in \eqref{eq:path integral two} to be well defined, we consider spatial boundary conditions
\begin{equation}\label{eq:spatial bc}
    \rho(x,\tau)=r(x)\quad \textrm{and } \quad \partial_x\hat{\rho}(x,\tau)=0\qquad \textrm{at } x\to \pm \infty \textrm{ for all $\tau$}
\end{equation}
where $r(x)$ is the initial average density profile in \fref{fig:initial_condition}. This amounts to assuming no fluctuations far from the origin. This condition does not affect the current statistics at the origin at a finite time.

\subsection{Annealed Ensemble \label{sec:anneal formal}}

For the domain-wall initial state (see \fref{fig:initial_condition}) with average density profile 
\begin{align}
    r(x) = \rho_{a}\Theta(-x) + \rho_{b}\Theta(x),
    \label{eq: initial density profile}
\end{align}
the probability of an initial density fluctuation $\rho(x,0)$, for large $\ell$, has the large-deviation asymptotic~\cite{Derrida2007,Derrida_2011}
\begin{equation}\label{eq:prob initial}
    \pr{\rho(x,0)}\simeq \mathrm{e}^{ -\ell \,\mathcal{F}(\rho(x,0))}
\end{equation}
where the free energy functional
\begin{align}
    \mathcal{F}\left( \rho(x, 0) \right)=\int_{-\infty}^{\infty}\mathrm{d}x
    \int_{r(x)}^{\rho(x, 0)}\mathrm{d}z\,\left( \rho(x, 0) - z
    \right)\frac{2D(z)}{\sigma(z)}
    \label{eq:F}
\end{align}

The dependence on $\ell$ in \eqref{eq:path integral two} and \eqref{eq:prob initial} suggests defining $\lambda(\ell^2\tau)=\tfrac{\Lambda(\tau)}{\ell^2}$ and taking $\ell=\sqrt{T}$, such that the generating functional averaged over the initial density fluctuations 
\begin{align}\label{eq:path integral three}
    \overline{\left\langle \mathrm{e}^ {\int_0^T \mathrm{d}t \lambda(t)Q(t)}\right\rangle}&\simeq \int\mathcal{D}\left[\rho,\hat{\rho}\right]\mathrm{e}^{-\sqrt{T} \, S_{\Lambda}(\rho,\hat{\rho})}
\end{align}
with an action functional
\begin{align}
    S_{\Lambda}(\rho,\hat{\rho})&=-\int_0^1 \mathrm{d}\tau\,\Lambda(\tau) \int_{0}^{\infty}\mathrm{d}x\,(\rho(x,\tau) - \rho(x,0))+\mathcal{F}(\rho(x,0))\nonumber \\
    &\qquad +\int_{0}^{1}\mathrm{d}\tau
    \int_{-\infty}^{\infty}\mathrm{d}x\,\left\{ \hat{\rho}\partial_{\tau}\rho -
    \frac{\sigma(\rho)}{2}\left( \partial_{x}\hat{\rho}
    \right)^{2}+D(\rho)\left( \partial_{x}\rho \right)\left(
    \partial_{x}\hat{\rho}
    \right) \right\}
    \label{eq:action}
\end{align}
The choice of the scale $\ell$ and the scaling of the fugacity in~\eqref{eq:path integral three} capture multi-time statistics of $Q(t)$ for times $t$ that are much larger than the equilibrium relaxation time scales.

The term in the exponential in \eqref{eq:path integral three} scales extensively with $\sqrt{T}$, implying that, in the large-$T$ limit, the path integral is dominated by the path of least action, with $S_{\rm min}(\Lambda(\tau))\equiv\min_{(\rho,\hat{\rho})}S_{\Lambda}(\rho,\hat{\rho})$,
which leads to

\begin{align}\label{eq:saddle point one}
    \overline{\left\langle \mathrm{e}^{\int_0^T \mathrm{d}t \lambda(t)Q(t)}\right\rangle}&\sim \mathrm{e}^{-\sqrt{T} \, S_{\rm min}(\Lambda(\tau))}
\end{align}
This implies that, for large $T$, the CGF~\eqref{eq:annealed generating functional unscaled} in the annealed setting has the scaling property
\begin{equation}\label{eq:anneal scaling scgf}
    \mu_{\mathcal{A}}\left(\lambda(t)\right)\simeq \sqrt{T}\,\chi_{\mathcal{A}}\left(\Lambda\left(\tau\right)\right)\quad\textrm{with }
    \chi_{\mathcal{A}}\left(\Lambda(\tau)\right)=-S_{\rm min}(\Lambda(\tau))
\end{equation}
The scaling in \eqref{eq:anneal scaling scgf} implies the following scaling property of correlations
\begin{equation}\label{eq: anneal scaled correlations}
    \cumul{ Q(t_1)\ldots Q(t_n)}_{\mathcal{A}} \simeq \sqrt{T}\,C_{\mathcal{A}}\left(\frac{t_1}{T},\ldots,\frac{t_n}{T}\right)
\end{equation}
with
\begin{equation}\label{eq:CA}
    C_{\mathcal{A}}(\tau_1,\ldots,\tau_n)=\left.\frac{\delta^n \chi_{\mathcal{A}}(\Lambda(\tau))}{\delta\Lambda(\tau_1)\cdots\delta\Lambda(\tau_n)}\right\vert_{\Lambda=0}
\end{equation} 

\subsubsection*{Least action path:}
The least-action path $(\rho, \hat{\rho})\equiv(q,p)$ is a solution of the Euler-Lagrange equations
\begin{subequations}\label{eq:optimal}
\begin{align}
    \partial_{\tau}q-\partial_{x}\left(D(q)\partial_{x}q\right)&=-\partial_{x}\left(\sigma(q) \partial_{x}p
    \right)\label{eq:optimal q2}\\
    \partial_{\tau}p +D(q)\partial_{x}^2p&=-\frac{1}{2}\,\sigma^{\prime}(q)\left( \partial_{x}p
    \right)^{2}-\Lambda(\tau)\Theta(x)\label{eq:optimal p2}
\end{align}
\end{subequations}
subject to the boundary conditions
\begin{equation}
   p(x,1)=0\quad \textrm{and}\quad p(x,0)=\Theta(x)\int_0^1 \mathrm{d}\tau\,\Lambda(\tau) + 2\int_{r(x)}^{q(x,0)}\mathrm{d}z\,\frac{D(z)}{\sigma(z)} \label{eq:boundary anneal}
\end{equation}
where $\Theta(x)$ is the Heaviside step function. These Euler-Lagrange equations and boundary conditions are obtained by variational calculus using the action \eqref{eq:action}. See \cite{Gerschenfeld2009} for the corresponding theory for single-time statistics, which had similar Euler-Lagrange equations \eqref{eq:optimal}, but without the $\Lambda(\tau)\Theta(x)$ term.

\subsubsection*{Minimal action:} Using the Euler-Lagrange equations \eqref{eq:optimal}, the expression for the action in \eqref{eq:action} simplifies, leading to the scaled cumulant generating functional (scgf) \eqref{eq:anneal scaling scgf}, with
\begin{align}
    \chi_{\mathcal{A}}(\Lambda(\tau))=\int_0^1 \mathrm{d}\tau \Lambda(\tau) & \int_{0}^{\infty}\mathrm{d}x\,(q(x,\tau) - q(x,0)) -\mathcal{F}(q(x,0))\cr &-\frac{1}{2}\int_{0}^{1}\mathrm{d}\tau\int_{-\infty}^{\infty}\mathrm{d}x\,
    \sigma(q(x,\tau))\left( \partial_{x}p(x,\tau) \right)^{2}
    \label{eq:chi annealed}
\end{align}
This result \eqref{eq:chi annealed}, along with the Euler-Lagrange equations \eqref{eq:optimal} and the boundary conditions \eqref{eq:boundary anneal}, gives a formal solution for the multi-time correlations of the current $Q(t)$ for generic diffusive systems. The solution to the Euler-Lagrange equations~\eqref{eq:optimal} characterizes the optimal trajectory leading to an atypical current fluctuation. 

\subsection{Quenched ensemble}
For the cgf in the quenched setting \eqref{eq:quenched generating functional unscaled}, we consider the same scaling of parameters  $\lambda(\ell^2 \tau)=\tfrac{\Lambda(\tau)}{\ell^2}$ and $\ell=\sqrt{T}$ in the evolution average \eqref{eq:path integral two} for a fixed initial density $\rho(x,0)$. Taking the logarithm of the generating function \eqref{eq:path integral two} makes it a slowly varying function of $\rho(x,0)$ compared with the distribution \eqref{eq:prob initial}, which is sharply peaked around the average density $r(x)$. This means that in the large-$\ell$ limit (and hence large $T$ for our choice of parameters), the cgf \eqref{eq:quenched generating functional unscaled} is dominated by contributions from evolutions starting with the average profile $r(x)$,
\begin{equation}
\overline{\log\left\langle \mathrm{e}^ {\int_0^T \mathrm{d}t \, \lambda(t) Q(t) }\right\rangle}\simeq \log\left\langle \mathrm{e}^{\int_0^T \mathrm{d}t \, \lambda(t) Q(t) }\right\rangle_{r(x)} \label{eq:quenched generating functional peaked}
\end{equation}
for $\lambda(t)=\frac{1}{T}\Lambda\left(\frac{t}{T}\right)$, with the subscript $r(x)$ denoting the fixed initial profile $\rho(x,0)=r(x)$. 

Fixing the initial profile $\rho(x,0)$ to the average value $r(x)$ gives a result analogous to \eqref{eq:path integral three} and \eqref{eq:action} with $\mathcal{F}(\rho(x,0))=0$. The rest of the analysis proceeds similarly to that described for the annealed case in \sref{sec:anneal formal}. We obtain the formal result 
\begin{equation}\label{eq:quench scaling scgf}
    \mu_{\mathcal{Q}}\left(\lambda(t)\right)\simeq \sqrt{T}\,\chi_{\mathcal{Q}}\left(\Lambda\left(\tau\right)\right)
\end{equation}
with the corresponding scgf 
\begin{align}
    \chi_{\mathcal{Q}}(\Lambda(\tau))=\int_0^1 \mathrm{d}\tau \Lambda(\tau)  \int_{0}^{\infty}\mathrm{d}x\,(q(x,\tau) - q(x,0))-\frac{1}{2}\int_{0}^{1}\mathrm{d}\tau\int_{-\infty}^{\infty}\mathrm{d}x\,
    \sigma(q(x,\tau))\left( \partial_{x}p(x,\tau) \right)^{2}
    \label{eq:chi quenched}
\end{align}
where $q(x,\tau)$ and $p(x,\tau)$ are solutions of the Euler-Lagrange equations \eqref{eq:optimal} with boundary conditions
\begin{align}
    q(x,0)=r(x) \qquad\textrm{and } p(x,1)=0\label{eq:q0 pT quench}
\end{align}

Similar to the annealed ensemble, correlations for the quenched case are given by
\begin{equation}\label{eq: quench scaled correlations}
    \cumul{ Q(t_1)\ldots Q(t_n) }_{\mathcal{Q}} \simeq \sqrt{T}\,C_{\mathcal{Q}}\left(\frac{t_1}{T},\ldots,\frac{t_n}{T}\right)
\end{equation}
with
\begin{equation}\label{eq:CQ}
    C_{\mathcal{Q}}(\tau_1,\ldots,\tau_n)=\left.\frac{\delta^n \chi_{\mathcal{Q}}(\Lambda(\tau))}{\delta\Lambda(\tau_1)\cdots\delta\Lambda(\tau_n)}\right\vert_{\Lambda=0}
\end{equation}

With the formal solutions \eqref{eq:chi annealed} and \eqref{eq:chi quenched}, the problem of characterizing multi-time statistics now boils down to solving the Euler-Lagrange equations~\eqref{eq:optimal} with the corresponding boundary conditions. However, this comes at the cost that the equations~\eqref{eq:optimal} are generally intractable for arbitrary $D(\rho)$ and $\sigma(\rho)$. Even for single-time statistics, obtaining exact solutions of the corresponding optimal equations for generic diffusive systems remains an open problem. Only recently have exact solutions been obtained for a handful of integrable models, such as the SSEP~\cite{Mallick2022PRL} and the Kipnis--Marchioro--Presutti model~\cite{KMP_exact_2022}, using highly nontrivial techniques of classical integrability. Another non-trivial class is the system of Brownian hard rods, where the corresponding MFT has been exactly solved~\cite{Jangid2026SingleFileMFT} using a canonical transformation inspired by an interacting to non-interacting mapping in ballistic systems~\cite{2026_saha_universal,2026_Kethepalli_Ballistic}. Extending those solutions to~\eqref{eq:optimal} remains a challenging open problem.

In the following section, we consider the simplest case of a non-interacting gas, where the problem is tractable. Despite its simplicity, the non-interacting limit captures most of the essential physics of one-dimensional transport~\cite{1965_Harris_Diffusion,1974_Percus_Anomalous,1978_Alexander_Diffusion,2003_Kollmann_Single}. 

\section{Non-interacting gas\label{sec: mft infinite non-interacting}}
We consider the MFT for systems with unit diffusivity $D(\rho)=1$ and linear mobility $\sigma(\rho)=2\rho$. This large-scale description corresponds~\cite{Gerschenfeld2009} to non-interacting particles on a lattice hopping to nearest-neighbor sites with unit rates, which is an effective description for the SSEP in the low density limit, where inter-particle interactions have sub-leading effects.  
Another system with the same hydrodynamics is a gas of Brownian point particles with either no interactions or with hard-core repulsion~\cite{Krapivsky2014,KMS_JSP,Gerschenfeld2009}. This point-particle limit also serves as the basis for characterizing Brownian hard rods with finite size~\cite{2026_Saha_Bottom} through a one-to-one mapping of trajectories~\cite{2026_saha_universal,2026_Kethepalli_Ballistic}.

\subsubsection*{Minimal action:} Before presenting the solutions of the corresponding Euler-Lagrange equations, we show how the action itself becomes remarkably simple for this case, even before obtaining the full solution. Explicitly writing the transport coefficients in \eqref{eq:chi annealed} and \eqref{eq:chi quenched} we get the scgf for both ensembles
\begin{align}
\chi\left(\Lambda(\tau)\right)\simeq\int_{0}^{1}\mathrm{d}\tau\, \Lambda(\tau)\int_{0}^{\infty}\mathrm{d}x\,&(q(x,\tau) - q(x,0)) -\mathcal{F}(q(x,0)) \, \mathbb{I}_{\mathcal{E}} \nonumber \\ & -
    \int_{-\infty}^{\infty}\mathrm{d}x\int_{0}^{1}\mathrm{d}\tau\,q(x, \tau)\left(\partial_{x}p\right)^{2}
    \label{eq:chi ni first}
\end{align}
with the indicator function $\mathbb{I}_{\mathcal{E}}=1$ for the annealed ensemble and zero for the quenched ensemble, while from \eqref{eq:F},
\begin{align}
\mathcal{F}(q(x,0))=\int_{-\infty}^{\infty}\mathrm{d}x\, q(x,0)\ln\left(\frac{q(x,0)}{r(x)}\right)-\int_{-\infty}^{\infty}\mathrm{d}x\,\left( q(x,0)-r(x) \right)
\end{align}

Similarly, the corresponding Euler-Lagrange equations~\eqref{eq:optimal} become
\begin{subequations}
\label{eq:optimalnoninteracting}
\begin{align}	
    \partial_{\tau}q-\partial_{x}^2q=&-\partial_{x}\left(2q \partial_{x}p \right)\\
    \partial_{\tau}p +\partial_{x}^2p=&-\left( \partial_{x}p
    \right)^{2}-\Lambda(\tau)\Theta(x)
\end{align}
\end{subequations}
with a shared final-time boundary conditions
\begin{align}
    p(x,1)=0 \label{eq:non_boundary_p1}
\end{align}
for both ensembles, and different initial conditions
\begin{align}
    \fl q(x,0)=\begin{cases} r(x)\, e^{p(x,0)-\Theta(x)\int_{0}^{1}\mathrm{d}\tau\,\Lambda(\tau)} & \text{for annealed ensemble}  ,\\
    r(x) & \text{for quenched ensemble},\end{cases}\label{eq:non_boundary_q0}
\end{align}
where $\Theta(x)$ is the Heaviside step function and $r(x)$ is the initial density profile~\eqref{eq: initial density profile}. The additional spatial boundary conditions come from \eqref{eq:spatial bc}. 

The expression \eqref{eq:chi ni first} for scgf further simplifies significantly using an identity
\begin{align}
    q(\partial_x p)^2=\partial_{\tau}(pq)-\partial_x\left[ p\partial_xq-q\partial_xp-2qp\partial_xp\right]+\Lambda(\tau)\Theta(x)q
\end{align}
derived from the optimal equations \eqref{eq:optimalnoninteracting}. Substituting the identity into the expression for the generating functional~\eqref{eq:chi ni first}, and using the boundary conditions (\ref{eq:spatial bc},\ref{eq:non_boundary_p1},\ref{eq:non_boundary_q0}), we get
\begin{align}
    \chi\left(\Lambda(\tau)\right)=\begin{cases}\int_{-\infty}^{\infty}\mathrm{d}x\,\left(e^{p(x,0)-\Theta(x)\int_{0}^{1}\mathrm{d}\tau\,\Lambda(\tau)}-1 \right)r(x) & \text{for annealed}, \\
    \int_{-\infty}^{\infty}\!\!\mathrm{d}x\,\left( 
    p(x,0)-\Theta(x)\int_{0}^{1}\!\mathrm{d}\tau\,\Lambda(\tau) \right) r(x) & \text{for quenched}. \end{cases}
    \label{eq:mu_simplified}
\end{align}
A similar simplification was introduced for single-time statistics in~\cite{Gerschenfeld2009}. In this simplified expression, the scgf depends only on the optimal field $p(x,0)$ at the initial time. Determining the initial field requires explicitly solving the Euler-Lagrange equations \eqref{eq:optimalnoninteracting}.

\subsubsection*{Optimal path:} We use a Cole-Hopf-type transformation, introduced earlier in~\cite{Gerschenfeld2009}, $p=\ln P_{\Lambda}$ and $q=R_{\Lambda}P_{\Lambda}$, with the subscript `$\Lambda$' denoting the dependence on the source term $\Lambda(\tau)$. This transformation decouples the Euler-Lagrange equations \eqref{eq:optimalnoninteracting} into an anti-diffusion and a diffusion equation
\begin{subequations}
\begin{align}
& \partial_{\tau} P_{\Lambda} + \partial_{x}^2 P_{\Lambda} = - \Lambda(\tau)\Theta(x)P_{\Lambda}\label{eq:P_lambda} \\ &
    \partial_{\tau}R_{\Lambda}-\partial_{x}^2R_{\Lambda}= \Lambda(\tau)\Theta(x)R_{\Lambda}\label{eq:R_lambda}
\end{align}
\end{subequations}
respectively, with the source terms. 

In terms of the new fields, the boundary conditions~\eqref{eq:non_boundary_p1} and~\eqref{eq:non_boundary_q0} read
\begin{equation}
P_{\Lambda}(x, 1) = 1\qquad \textrm{and }
R_{\Lambda}(x, 0) = \begin{cases}
r(x) e^{-\Theta(x) \int_0^1 \mathrm{d}\tau\,\Lambda(\tau)} & \text{for annealed}, \\
\frac{r(x)}{P_{\Lambda}(x, 0)} & \text{for quenched}.
\end{cases} \label{eq:boundary_PR}
\end{equation}

The general solution for arbitrary $\Lambda(\tau)$ can be expressed as an integral equation
\begin{align}
    {P}_{\Lambda}(x,\tau)=1+\int_\tau^1 \mathrm{d}s \Lambda(s) \int_{0}^\infty \mathrm{d}y \, g(y, s \vert x,\tau)P_{\Lambda}(y,s)  
\end{align}
An iterative expansion of this solution gives  
\begin{align}\label{eq:P_Lambda expanded}
    {P}_{\Lambda}(x,\tau)=1+\sum_{n=1}^{\infty}\int_{\tau}^{1}\mathrm{d}\tau_1\int_{\tau_1}^{1}\mathrm{d}\tau_2\cdots\int_{\tau_{n-1}}^{1}\mathrm{d}\tau_{n}\,\Lambda(\tau_{1})\cdots
    \Lambda(\tau_{n}) K_{n}(x,\tau,\tau_{1},\cdots,\tau_{n}),
\end{align}
where we define
\begin{align}
    K_{n}(x,\tau,\tau_{1},\cdots,\tau_{n})=\int_{0}^\infty dz_{1}\cdots\int_{0}^\infty dz_{n}\,g(z_{n},\tau_{n}
    \mid z_{n-1},\tau_{n-1})\cdots g(z_{1},\tau_{1} \mid x,\tau). \label{eq:K_n}
\end{align}
with the diffusion propagator
\begin{align}\label{eq:brownian propagator}
     g(x_2, \tau_2 \vert x_1,\tau_1)=\frac{e^{ 
     -\frac{(x_2-x_1)^{2}}{4(\tau_2-\tau_1)}} }{\sqrt{4\pi (\tau_2-\tau_1)}}\qquad \textrm{for all \(\tau_1 < \tau_2\).}
\end{align}

\subsubsection*{Scgf:}

\begin{subequations}\label{eq:mu_nonint_combined}
In terms of the new fields, the scgf~\eqref{eq:mu_simplified} gives
\begin{equation}
    \chi_{\mathcal{A}}\left(\Lambda(\tau)\right)=\int_{-\infty}^{0}\mathrm{d}x\Big(\rho_a\left( P_{\Lambda}(x,0)-1 \right)+\rho_b\left( P_{-\Lambda}(x,0)-1 \right)\Big)\label{eq:mu_annealed_non_interacting_MFT}
\end{equation}
for the annealed and
\begin{equation}
    \chi_{\mathcal{Q}}\left(\Lambda(\tau)\right)=\int_{-\infty}^{0}\mathrm{d}x\,\Big(\rho_a\ln P_{\Lambda}(x,0) +\rho_b\ln P_{-\Lambda}(x,0)\Big) \label{eq:mu_quenched_non_interacting_MFT}
\end{equation}\end{subequations}
for the quenched, where we used the initial average profile \eqref{eq: initial density profile} and the corresponding symmetry
\begin{align}
    P_{\Lambda}(-x,\tau)=P_{-\Lambda}(x,\tau)e^{\int_{\tau}^{1}\mathrm{d}\tau^{\prime}\Lambda(\tau^{\prime})}
\label{eq:P_symmetry}
\end{align}
which is evident from \eqref{eq:P_lambda}. Using the series solution~\eqref{eq:P_Lambda expanded} in the integral expressions for the scgf~(\ref{eq:mu_annealed_non_interacting_MFT},\ref{eq:mu_quenched_non_interacting_MFT}), the multi-time correlations~(\ref{eq:QA multi corr},\ref{eq:QQ multi corr}) can be extracted from the coefficients of powers of $\Lambda$ using the definitions (\ref{eq:CA},\ref{eq:CQ}). Note that for the uniform initial condition $(\rho_a = \rho_b)$, the scgfs are even functionals of $\Lambda(\tau)$, as expected from time-reversal symmetry. 

\subsubsection*{Remark:} The known result~\cite{Gerschenfeld2009} for the single-time scgf is recovered by setting $\Lambda(\tau)=\lambda \delta(1-\tau)$ in the series solution~\eqref{eq:P_Lambda expanded}, which gives $P_{\Lambda}(x,\tau) = 1+\sum_{n\ge 1}\tfrac{\lambda^n}{n!} K_1(x,\tau,1)$.  Evaluating the integral, we get $P_{\Lambda}(x,0) = 1 + \tfrac{1}{2}(e^\lambda-1)(1+\erf{\tfrac{x}{2}})$, which, together with, (\ref{eq:mu_annealed_non_interacting_MFT},\ref{eq:mu_quenched_non_interacting_MFT}) reproduces the results reported in~\cite{Gerschenfeld2009}. 

\subsection{Explicit expressions for correlations}
The integral expressions for the correlations~(\ref{eq:QA multi corr},\ref{eq:QQ multi corr}) can be cast in the forms (\ref{eq:explicit_annealed},\ref{eq:explicit_quenched}), which illustrate their time dependence. See~\ref{sec:derivation} and~\ref{sec:derivation_q} for their derivation. Here, we present explicit expressions for the correlations up to a few orders using~(\ref{eq:explicit_annealed},\ref{eq:explicit_quenched})

\subsubsection{Annealed.}
In the expression~\eqref{eq:explicit_annealed}, the coefficients $P^{(n)}_{ij}$ have the following interpretation as conditional probabilities for a single Brownian motion. Consider a standard Brownian motion $X(t)$ starting at the origin, whose position is measured stroboscopically at subsequent times $t_1< \cdots< t_n$, denoted by $X_1, \cdots, X_n$, respectively, where $X_i \equiv X(t_i)$. Then $P^{(n)}_{ij}$ is the probability that, conditioned on $X_i=X_j\ge 0$, all remaining positions satisfy $X_k \geq X_i$, for all $k\neq i, j$. 

Formally,
\begin{align}\label{eq:annealcoef}
    P^{(n)}_{ij} = \pr{ X_k\geq X_i\geq 0 \, \forall k\in\{1, \cdots, n\} \backslash  \{i,j\}\vert X_i =X_j}
\end{align}
They can be explicitly written for $i<j$,
\begin{equation}\label{eq:P_ij formula}
P^{(n)}_{ij}=
\begin{cases}
\displaystyle
\frac{\displaystyle\int_{0}^{\infty}dm\int_{\mathbf{X}_\perp\ge m}
G_{ij}(m,\mathbf{X}_\perp)\,d\mathbf{X}_\perp}
{\displaystyle\int_{-\infty}^{\infty}dm\int
G_{ij}(m,\mathbf{X}_\perp)\,d\mathbf{X}_\perp}
&\textrm{for } 1\le i<j\le n,\\[3.5ex]
\displaystyle
\frac{\displaystyle\int_{\mathbf{X}_\perp\ge 0}
G_{0j}(0,\mathbf{X}_\perp)\,d\mathbf{X}_\perp}
{\displaystyle\int G_{0j}(0,\mathbf{X}_\perp)\,d\mathbf{X}_\perp}
& \textrm{for } i=0,\ 1\le j\le n,
\end{cases}
\end{equation}
where $\mathbf{X}_\perp\equiv\left\{X_k\right\}_{k\ne i,j}$ and  $G_{ij}(m,\mathbf{X}_\perp)$ is the probability for $\{X_1,\ldots,X_n\}$ with $X_i=X_j=m$, written in terms of the Brownian propagator $g(x, t \mid x', t')$ in \eqref{eq:brownian propagator} as
\begin{equation}
G_{ij}(m,\mathbf{X}_\perp)\equiv
\left[\prod_{k=0}^{n-1}g(X_{k+1},t_{k+1}\mid X_k,t_k)
\right]_{X_i=X_j=m}
\end{equation}
with $X_0=0$ and $t_0=0$. Note that the denominator in \eqref{eq:P_ij formula} makes $P_{ij}^{(n)}$ a conditional probability. The piecewise form \eqref{eq:P_ij formula} is due to the initial condition that the Brownian motion starts at the origin.

The expression for $P^{(n)}_{ij}$ is related to the orthant probability~\cite{Bacon1963} of a multivariate Gaussian and can be evaluated explicitly. For $n=2$, $P^{(2)}_{ij} = \tfrac{1}{2}$, leading to the well-known result~\cite{Krapivsky_2015, Sadhu_2016}
\begin{align}
	\cumul{ Q(t_1)Q(t_2) }_{\mathcal{A}} = \frac{(\rho_a + \rho_b)}{2\sqrt{\pi}} \left(\sqrt{t_1} + \sqrt{t_2} - \sqrt{t_2 - t_1}\right)\label{eq:non_annealed_two_time_correlator}
\end{align}

For $n=3$, $P^{(3)}_{ij}=\tfrac14+\tfrac{1}{2\pi}\arcsin\kappa_{ij}$ with $\kappa_{01}=\sqrt{\tfrac{u_{21}-1}{u_{31}-1}}$, $\kappa_{03}=\sqrt{\tfrac{u_{32}-1}{u_{31}-1}}$, $\kappa_{23}=-\sqrt{1-u_{12}}$, and $\kappa_{ij}=0$ for the remaining cases, where we denote $u_{ij}=\tfrac{t_i}{t_j}$.
Using these results in~\eqref{eq:explicit_annealed} leads to an explicit expression for the three-time correlation
\begin{align}
\cumul{Q(t_1)Q(t_2)Q(t_3)}_{\mathcal{A}}\simeq
\frac{\rho_a-\rho_b}{4\sqrt{\pi}}&\Bigg\{
\left[1+\As{\tfrac{u_{21}-1}{u_{31}-1}}\right]\sqrt{t_1}
+\sqrt{t_2}
\nonumber\\
+\left[1+\As{\tfrac{u_{32}-1}{u_{31}-1}}\right]\sqrt{t_3}-&\sqrt{t_2-t_1}-\sqrt{t_3-t_1}
-\left[1-\As{1-u_{12}}\right]\sqrt{t_3-t_2}\Bigg\}
\label{eq:non_annealed_three_time_correlator}
\end{align}
where $A(x)=\frac{2}{\pi}\arcsin\sqrt{x}$.

For $n=4$, an explicit expression for $P^{(4)}_{ij}$ is given in Table~\ref{tab:pij anneal four} in \ref{appendix:derivation_four_time_correlator}, leading to an explicit expression for the four-time correlation from~\eqref{eq:explicit_annealed},
\begin{align}
&\cumul{ Q(t_1)Q(t_2)Q(t_3)Q(t_4) }_{\mathcal{A}} \simeq
\frac{\rho_a+\rho_b}{8\sqrt{\pi}}
\Bigg\{\sqrt{t_1}\bigg[1+
A\!\left(\frac{u_{21}-1}{u_{31}-1}\right)
+A\!\left(\frac{u_{31}-1}{u_{41}-1}\right)\cr 
&\qquad +A\!\left(\frac{u_{21}-1}{u_{41}-1}\right)
\bigg]
+\sqrt{t_2}\,
\left[1+A\!\left(\frac{u_{32}-1}{u_{42}-1}\right)\right]
+\sqrt{t_3}\,
\left[1+A\!\left(\frac{u_{32}-1}{u_{31}-1}\right)\right]
\cr
&\qquad
+\sqrt{t_4}\left[1+
A\!\left(\frac{u_{43}-1}{u_{41}-1}\right)
+A\!\left(\frac{u_{43}-1}{u_{42}-1}\right)
+A\!\left(\frac{u_{42}-1}{u_{41}-1}\right)
\right]\cr
&\qquad
-\sqrt{t_2-t_1}\,
\left[1+A\!\left(\frac{u_{32}-1}{u_{42}-1}\right)\right]-\sqrt{t_3-t_1}-\sqrt{t_4-t_1}\,
\left[1+A\!\left(
\frac{(u_{21}-1)(u_{34} - 1)}
{(u_{31}-1)(u_{24} - 1)}
\right)\right]\cr&\qquad
-\left(\sqrt{t_3-t_2}+\sqrt{t_4-t_2}\right)
\left[1-A\!\left(1 - u_{12}\right)\right]
\cr &\qquad -\sqrt{t_4-t_3}\left[
1-A\!\left(1 - u_{23}\right)
+
A\!\left(\frac{u_{23} - 1}{u_{13} - 1}\right)
-
A\!\left(1 - u_{13}\right)
\right]
\Bigg\}\label{eq:non_annealed_four_time_correlator}
\end{align}
To emphasize the relevance of these explicit expressions~(\ref{eq:non_annealed_two_time_correlator},\ref{eq:non_annealed_three_time_correlator},\ref{eq:non_annealed_four_time_correlator}),
we note that the same results, up to numerical prefactors, were found~\cite{Benichou_multi_2025}, using a microscopic approach, for multi-time correlations of the tracer position in the high-density limit of the SSEP for the uniform initial condition $\rho_a=\rho_b$. When comparing, note the difference in the definition of the function $A(x)$ in~\cite{Benichou_multi_2025}

\subsubsection{Quenched.}
In a similar fashion, the quenched correlations admit the form \eqref{eq:explicit_quenched}, where the time-intensive coefficients are defined as follows: for $i<j$,
\begin{equation}
D^{(n)}_{0j}=\sum_{\omega\in\Omega_n}\lambda_\omega\,P^{(n)}_{0j}[\omega],
~
D^{(n)}_{ij}=-\!\!\sum_{\omega:\,i\sim j}\!\!\lambda_\omega\,P^{(n)}_{ij}[\omega]
\ \ (\textrm{for } i\ge1), ~~ S^{(n)}_{ij}=-\!\!\sum_{\omega:\,i\not\sim j}\!\!\lambda_\omega\,P^{(n)}_{ij}[\omega]
\label{eq:quenchedcoef}
\end{equation}
Here, $\Omega_n$ is the set of all partitions of the index set $\{1,\dots,n\}$ associated with the times $t_i$ in the $n$-time correlation.  Each partition $\omega$ is assigned a weight $\lambda_\omega=(-1)^{|\omega|-1}(|\omega|-1)!$, where $|\omega|$ is the number of blocks in the partition $\omega$. We use
$i\sim j$ to denote that $i$ and $j$ lie in the same block, and $i \not\sim j$ when they lie in different blocks. For example, consider the index set $\{1,2,3\}$ for computing the three-time correlation $\cumul{Q(t_1)Q(t_2)Q(t_3)}$, where one of its partitions $\omega\equiv \left\{\{1\}\{2,3\}\right\}$ has two blocks, $\{1\}$ and $\{2,3\}$, giving $\vert \omega \vert =2$ and $\lambda_\omega=-1$. Here, $2\sim 3$ and $1 \not\sim 3$.

The coefficient $P^{(n)}_{ij}[\omega]$ is associated with each partition $\omega$ and is defined as a generalization of \eqref{eq:annealcoef}, as follows: for each block $B_\omega$ of the partition $\omega$, we assign an independent Brownian motion $X_{B_\omega}(t)$ starting from the origin. Its position is measured at times $t_k$ whose index $k$ belongs to that block $B_\omega$, i.e.,  $k\in B_\omega$. This gives a list of $n$ random variables
\begin{equation}\label{eq:walker_assign}
X_k\equiv\,X_{(B_\omega)}(t_k),\qquad \textrm{with } k\in B_\omega
\end{equation}
The coefficient $P_{ij}[\omega]$ is defined in terms of these random variables
$(X_1,\dots,X_n)$ as in~\eqref{eq:annealcoef}.

For the illustrative example of $\Omega_3$ and its partition $\omega$ discussed above, there are two Brownian motions, $X_{1}(t)$ and $X_{2}(t)$, associated with the blocks $\{1\}$ and $\{2,3\}$, respectively. The random variables are $X_1\equiv X_1(t_1)$, $X_2\equiv X_2(t_2)$, and $X_3 \equiv X_2(t_3)$, and the coefficient $P_{ij}^{(3)}[\omega]$ is defined following~\eqref{eq:annealcoef}. 

To write this generally as in~\eqref{eq:P_ij formula}, take a generic block $B_\omega$ of $\omega$ with its elements ordered as $\{i_1< i_2< \cdots < i_{|B_\omega|}\}$, where $|B_\omega|$ is the number of elements in the block $B_\omega$. The set of random variables $\{X_{i_1}, \cdots, X_{i_{|B_\omega|}}\}$ gives the positions of the Brownian motion associated with this block, measured at times $t_{i_1}, \cdots, t_{i_{|B_\omega|}}$, and has the probability weight $\prod_{k=0}^{|B_\omega|-1}g(X_{i_{k+1}},t_{i_{k+1}}\vert X_{i_k}, t_{i_k})$, with $i_0 = 0$ incorporating the initial condition $X_0=0$ at $t_0=0$. Considering all the independent Brownian motions from each block~\eqref{eq:walker_assign}, we define
\begin{align}\label{eq:partition_weight}
    G_{ij}[\omega](m, \mathbf{X}_\perp) = \left[\prod_{B_\omega\in\omega}\prod_{k=0}^{|B_\omega|-1}g(X_{i_{k+1}},t_{i_{k+1}}\vert X_{i_k}, t_{i_k})\right]_{X_i = X_j=m}
\end{align}
and compute the $P^{(n)}_{ij}[\omega]$ using a formula analogous to~\eqref{eq:P_ij formula}.

\subsubsection*{An explicit example:} For two-time correlation $\cumul{ Q(t_1)Q(t_2) }$, the partitions of the index set $\{1,2\}$ are $\{\{1,2\}\}$ and $ \{\{1\}\{2\}\}$, with weights $\lambda_\omega$ equal to $1$ and $-1$, respectively. This gives the coefficients~\eqref{eq:quenchedcoef}: $ D^{(2)}_{12}= -P^{(2)}_{12}[\{\{1,2\}\}]$ and $S^{(2)}_{12} = P^{(2)}_{12}[\{\{1\}\{2\}\}]$, while
\begin{align}
    D^{(2)}_{01}= P^{(2)}_{01}[\{\{1,2\}\}]-P^{(2)}_{01}[\{\{1\}\{2\}\}]\\
    D^{(2)}_{02}=P^{(2)}_{02}[\{\{1,2\}\}]-P^{(2)}_{02}[\{\{1\}\{2\}\}]
\end{align}
The amplitudes $P^{(2)}_{ij}[\omega]$ are computed from~\eqref{eq:P_ij formula} using the path weights \eqref{eq:partition_weight}. For example, $P^{(2)}_{12}[\{\{1,2\}\}]$ requires evaluating
\begin{align}
    G_{12}[\{\{1,2\}\}](m)=g(m, t_1\vert 0, 0)g(m, t_2\vert m,t_1)
\end{align}
and then using \eqref{eq:P_ij formula} we get $P^{(2)}_{12}[\{\{1,2\}\}]=\tfrac{1}{2}$. Similar calculations follow for the remaining amplitudes, leading to $D^{(2)}_{01}=D^{(2)}_{02}=0$ and $D^{(2)}_{12}=-\tfrac{1}{2}$, $S^{(2)}_{12}=\tfrac{1}{2}$. Using these amplitudes in~\eqref{eq:explicit_quenched}, we recover the well-known result~\cite{Krapivsky_2015, Sadhu_2016}
\begin{align}
	\cumul{ Q(t_1)Q(t_2) }_{\mathcal{Q}} = \frac{(\rho_a + \rho_b)}{2\sqrt{\pi}} \left(\sqrt{t_1 + t_2} - \sqrt{t_2 - t_1}\right)
    \label{eq:non_quenched_two_time_correlation}
\end{align}
\textit{Remark}: The characteristic difference between the annealed and quenched two-time correlations can be seen at large $t_2$ for fixed $t_1$. In this limit, the quenched correlation asymptotically vanishes, while the annealed correlation converges to the half value of the variance. A similar trend has been noted  in~\cite{Benichou_multi_2025} for the tracer correlation in the high-density limit of the SSEP for the uniform initial condition $\rho_a = \rho_b$. 

For the three-time correlation $\cumul{ Q(t_1)Q(t_2) Q(t_3)}$, there are five partitions of the index set $\{1,2,3\}$:
\begin{equation}
\Omega_3\equiv \{\{\{1,2,3\}\},\quad
\{\{1\},\{2,3\}\},\quad
\{\{2\},\{1,3\}\},\quad
\{\{3\},\{1,2\}\},\quad
\{\{1\},\{2\},\{3\}\}\}\nonumber
\end{equation}
The weights $\lambda_\omega$ for the partitions are $+1, -1, -1, -1, +2$, respectively. The amplitude for each partition $\omega$ has a compact expression
\begin{equation}\label{eq:P3ij}
P^{(3)}_{ij}[\omega]=\frac{1}{4}+\frac{\theta_{ij}[\omega]}{4},
\end{equation}
with the results for $\theta_{ij}[\omega]$ given in Table~\ref{tab:thetaij}. Using these amplitudes in \eqref{eq:quenchedcoef} and \eqref{eq:explicit_quenched} we get an explicit expression for the three-time correlation:
\begin{align}
\cumul{Q(t_1)Q(t_2)&Q(t_3)}_{\Qq}\simeq
\frac{\rho_a-\rho_b}{4\sqrt{\pi}}\Bigg\{
\Big[\As{\tfrac{u_{21}-1}{u_{31}-1}}-\As{u_{23}}\Big]\sqrt{t_1}
-\As{u_{13}}\sqrt{t_2}
\cr
& +\Big[\As{\tfrac{u_{32}-1}{u_{31}-1}}-\As{u_{12}}\Big]\sqrt{t_3}
-\As{\tfrac{1}{1+u_{31}}}\sqrt{t_2-t_1}
-\As{\tfrac{1}{1+u_{21}}}\sqrt{t_3-t_1}
\nonumber\\
&-\Big[\As{\tfrac{1}{u_{12}+1}}-\As{1 - u_{12}}\Big]\sqrt{t_3-t_2}
+2\As{\tfrac{1}{1+u_{31}(u_{12}+1)}}\,\sqrt{t_1+t_2}
\nonumber\\
&+\Big[2\As{\tfrac{1}{1+u_{21}(u_{13}+1)}}
-\As{\tfrac{1 - u_{23}}{1+u_{21}}}\Big]\sqrt{t_1+t_3}
\nonumber\\
&+\Big[2\As{\tfrac{1}{1+u_{12}(u_{23}+1)}}
-\As{\tfrac{1-u_{13}}{u_{12}+1}}
-\As{\tfrac{1-u_{12}}{u_{13}+1}}\Big]\sqrt{t_2+t_3}
\Bigg\}.
\label{eq:non_quenched_three_time_correlation}
\end{align}

\begin{table}
\centering
\small
\setlength{\tabcolsep}{6pt}
\renewcommand{\arraystretch}{1.6}
\resizebox{\linewidth}{!}{\begin{tabular}{c ccccc}
\toprule
\diagcorner{$(i,j)$}{$\omega$}
  & $\{123\}$ & $\{1\}\{23\}$ & $\{2\}\{13\}$ & $\{3\}\{12\}$ & $\{1\}\{2\}\{3\}$\\
\midrule
$(0,1)$ & $\As{\tfrac{u_{21}-1}{u_{31}-1}}$ & $\As{u_{23}}$ & $0$ & $0$ & $0$\\
$(0,2)$ & $0$ & $0$ & $\As{u_{13}}$ & $0$ & $0$\\
$(0,3)$ & $\As{\tfrac{u_{32}-1}{u_{31}-1}}$ & $0$ & $0$ & $\As{u_{12}}$ & $0$\\
$(1,2)$ & $0$ & $0$ & $0$ & $-\As{\tfrac{1}{1+u_{31}}}$ & $-\As{\tfrac{1}{1+u_{31}+u_{32}}}$\\
$(1,3)$ & $0$ & $-\As{\tfrac{1-u_{23}}{1+u_{21}}}$ & $-\As{\tfrac{1}{1+u_{21}}}$ & $0$ & $-\As{\tfrac{1}{1+u_{21}+u_{23}}}$\\
$(2,3)$ & $-\As{1-u_{12}}$ & $-\As{\tfrac{1}{1+u_{12}}}$ & $-\As{\tfrac{1-u_{13}}{1+u_{12}}}$ & $-\As{\tfrac{1-u_{12}}{1+u_{13}}}$ & $-\As{\tfrac{1}{1+u_{12}+u_{13}}}$\\
\bottomrule
\end{tabular}}
\caption{Expression for $\theta_{ij}[\omega]$ in the amplitude $P^{(3)}_{ij}[\omega]$ of \eqref{eq:P3ij}, where we denote $u_{ij}=\tfrac{t_i}{t_j}$.}
\label{tab:thetaij}
\end{table}

\begin{figure}
    \centering
    \includegraphics[width=0.8\linewidth]{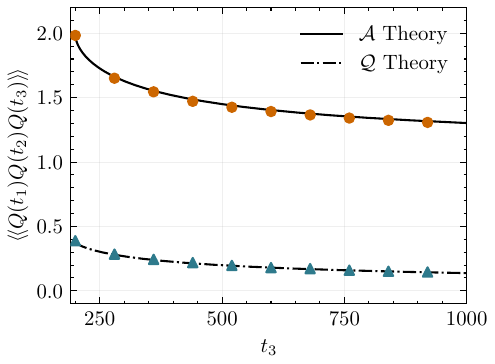}
    \vspace{0.5em}
    \caption{\textbf{Non- interacting three-time correlations:} The solid line represents the theoretical result \eqref{eq:non_annealed_three_time_correlator} for the annealed ensemble, while the dashed line corresponds to the quenched result \eqref{eq:non_quenched_three_time_correlation} for $\rho_a = 1$ and $\rho_b =0.5$ as a function of $t_3$, with $t_1 = 100$ and $t_2 = 200$ fixed. The markers (circles and triangles) denote Monte Carlo simulation results for non-interacting continuous-time random walkers on an infinite lattice, averaged over $10^8$ realizations.}    
    \label{fig:NCTRW three time}
\end{figure}

We have verified these explicit expressions for the three-point correlations~\eqref{eq:non_annealed_three_time_correlator} and~\eqref{eq:non_quenched_three_time_correlation} in the two ensembles using Monte Carlo simulations, and the comparison is shown in~\fref{fig:NCTRW three time}. A similar explicit expression for the four-time correlation in the quenched case is presented in~\ref{appendix:derivation_quenched_four_time_correlator}. Numerical verification of the four-time correlation requires extensive sampling, which we do not pursue in this article.

Similar to the annealed setting, the time dependence of expressions (\ref{eq:non_quenched_two_time_correlation}, \ref{eq:non_quenched_three_time_correlation})
matches, up to numerical prefactors, the time dependence~\cite{Benichou_multi_2025} of the multi-time correlations of tracer position in the high-density limit of the SSEP for the uniform initial condition $\rho_a=\rho_b$, emphasizing the relevance of these  explicit computations.

\section{Interacting gas\label{sec:mft_infinite_interacting}}

For generic diffusive systems characterized by $D(\rho)$ and $\sigma(\rho)$, solving the Euler-Lagrange equations~\eqref{eq:optimal} exactly is generally challenging. Nevertheless, the correlations can be computed systematically, order by order, through a perturbative expansion in the fugacity $\Lambda$, analogous to the approach used for single-time statistics~\cite{Krapivsky2012PRE,Saha_2023,Berlioz_2024,Sadhu2023,Benichou_2024,KMS_interface,Dandekar2023}. Here, building on an earlier work on tracer statistics for a uniform initial state~\cite{Krapivsky_2015}, we extend this perturbative framework to multi-time correlations. We restrict the discussion to systems with constant diffusivity $D=1$, although the method can be generalized to arbitrary $D(\rho)$.

The primary step involves a series solution of the Euler-Lagrange equations~\eqref{eq:optimal} in powers of $\Lambda$. For $\Lambda=0$, the evolution is unbiased, and naturally the solution is $p(x,\tau)=0$, while $q(x,\tau)=q_0(x,\tau)$ describes the average spontaneous evolution of the density profile. This vanishing response field $p(x,\tau)$ at zeroth order in $\Lambda$ decouples the Euler-Lagrange equation for the $p$ field at a given order from the corresponding $q$ field at the same order, making a systematic hierarchical solution possible.

For correlations, we use the representation
\begin{equation}
    \chi(\Lambda(\tau))=\sum_{n\ge 1}\int_0^{1}\mathrm{d}\tau_1\int_{\tau_1}^{1}\mathrm{d}\tau_2\cdots \int_{\tau_{n-1}}^{1}\mathrm{d}\tau_n\,C(\tau_1,\cdots,\tau_n)\Lambda(\tau_1)\cdots \Lambda(\tau_n)
    \label{eq:expansion scgf}
\end{equation}
with the correlations defined in (\ref{eq: anneal scaled correlations},\ref{eq: quench scaled correlations}). This gives a series expansion of the scgf  $\chi=\chi^{(1)}+\chi^{(2)}+\cdots$ in powers of $\Lambda$, such that
\begin{equation}\label{eq:chi expansion}
    \chi^{(1)}(\Lambda)=\int_0^{1}\mathrm{d}\tau\, C(\tau)\Lambda(\tau)\quad \textrm{and }\chi^{(2)}(\Lambda)=\int_0^{1}\mathrm{d}\tau_1\int_{\tau_1}^{1}\mathrm{d}\tau_2\,C(\tau_1,\tau_2)\Lambda(\tau_1) \Lambda(\tau_2)
\end{equation}

For a systematic treatment, we define $p=p_1+p_2+\cdots$ and $q=q_0+q_1+q_2+\cdots$, where the subscript denotes the order in $\Lambda$. The goal is to use a series solution of the Euler-Lagrange equations~\eqref{eq:optimal} in the reduced expressions for the scgf in (\ref{eq:chi annealed},\ref{eq:chi quenched}) and write the scgf order by order in the form \eqref{eq:chi expansion}, from which the correlations can be extracted.

\subsection{Mean current}
At zeroth order, from the Euler-Lagrange equations~\eqref{eq:optimal}, we get, for both ensembles, 
\begin{align}
	\partial_{\tau} q_0(x, \tau) =  \partial_{x}^2 q_0(x, \tau)\qquad \textrm{with }q_0(x,0)=r(x)
	\label{eq:diffusion_equation}
\end{align}
with $r(x)$ given in \eqref{eq: initial density profile}. For the annealed case, the initial condition comes from a series solution of \eqref{eq:boundary anneal}. The initial condition also shows that at linear order, the free energy $\mathcal{F}$ in~\eqref{eq:F} vanishes.

Collecting these leading order results in (\ref{eq:chi annealed},\ref{eq:chi quenched}), for both ensembles, we obtain 
\begin{equation}\label{eq:chi1 sol}
\chi^{(1)}(\Lambda)
=
\int_0^1\mathrm{d}\tau\,\Lambda(\tau)
\int_0^\infty\mathrm{d}x\,
\big(q_0(x,\tau)-q_0(x,0)\big)
\end{equation}
with an explicit solution
\begin{align}
	q_0(x, \tau) = \bar{\rho} - \frac{\Delta \rho}{2}\erf{\frac{x}{2\sqrt{\tau}}}  \label{eq: density profile}
\end{align}
describing the average evolution of the density profile. Here, we denote $\Delta \rho = \rho_a - \rho_b$ and $\bar{\rho} = \tfrac{\rho_a + \rho_b}{2}$.  Integrating over $x$ in \eqref{eq:chi1 sol}, we get $\chi^{(1)}(\Lambda)$ in the desired form \eqref{eq:chi expansion}, resulting in the average current (\ref{eq: anneal scaled correlations},\ref{eq: quench scaled correlations})
\begin{equation}
C(\tau) = \Delta \rho\sqrt{\frac{\tau}{\pi}}\quad \textrm{and equivalently }\langle Q(t) \rangle=\Delta \rho\sqrt{\frac{t}{\pi}}
\end{equation}
for both ensembles. The mean current is proportional to the density difference $\Delta\rho$, but it is subextensive in time.

\subsection{Two-time correlation}

At second order for the scgfs (\ref{eq:chi annealed}, \ref{eq:chi quenched}) we get
\begin{align}
	\chi^{(2)}\left(\Lambda\right) \simeq  \int_{0}^{1} \mathrm{d}\tau \int_{0}^{\infty} \mathrm{d}x &\, \Lambda(\tau)\left( q_1(x, \tau) - q_1(x, 0) \right) - \int_{0}^{1}\mathrm{d}\tau\int_{-\infty}^{\infty}\mathrm{d}x \, \frac{\sigma(q_0(x, \tau))}{2} (\partial_x p_1(x, \tau))^2 \nonumber \\ &  - \mathcal{F}_2(q_0,q_1) 
	\label{eq:mu2}
\end{align}
where $\mathcal{F}_2 = 0$ in the quenched setting, while for the annealed setting
\begin{align}
	\mathcal{F}_2(q_0,q_1) =  \int_{-\infty}^{\infty} \mathrm{d}x \, \frac{(q_1(x,0))^2}{\sigma(q_0(x,0))} \label{eq:F_2_annealed}
\end{align}
denotes the second-order term in the expansion of the free  energy~\eqref{eq:F}, obtained using the initial condition in \eqref{eq:diffusion_equation}.

Evaluating \eqref{eq:mu2} requires knowledge of $q_0(x,\tau)$, which is given in \eqref{eq: density profile}, and the linear-order solutions $q_1(x,\tau)$ and $p_1(x,\tau)$ of the Euler--Lagrange equations~\eqref{eq:optimal}, which read
\begin{subequations}    
	\begin{align}
		\partial_{\tau} q_1(x, \tau) - \partial_{x}^2q_1(x, \tau) & = -\partial_{x}(\sigma(q_0(x, \tau))\partial_{x}p_1(x, \tau)) \label{eq:optimal_q1} \\
		\partial_{\tau}p_1(x, \tau) + \partial_{x}^2p_1(x, \tau) & = -\Lambda(\tau) \Theta(x)
		\label{eq:optimal_p1}
	\end{align}
\end{subequations}
Note that the equation for $p_1(x,\tau)$ is decoupled from $q_1(x,\tau)$ and can be solved independently. It is worth pointing out that the source term $\Lambda(\tau)$ makes the analysis different from the corresponding analysis of single-time statistics~\cite{Krapivsky2012PRE,Krapivsky_2015,KMS_interface,Dandekar2023}.

\subsubsection{Quenched Ensemble: \label{sec:quench perturbation}}
The corresponding boundary conditions~\eqref{eq:q0 pT quench} at linear order give
\begin{align}\label{eq:boundary_conditions_quenched_p1_q1}
    p_1(x, 1) = 0 \quad \textrm{and } q_1(x, 0) = 0
\end{align}
The field $p_1(x,\tau)$ satisfies the anti-diffusion equation~\eqref{eq:optimal_p1} with the boundary condition \eqref{eq:boundary_conditions_quenched_p1_q1}. The corresponding solution 
\begin{align}
    p_1(x,\tau) = \int_{\tau}^1 \mathrm{d}\tau^{\prime} \int_{-\infty}^{\infty} \mathrm{d}z \, \Lambda(\tau^{\prime}) \, \Theta(z) \, g(z,\tau^{\prime} \mid x,\tau)
    \label{eq:sol_p1 1}
\end{align}
is expressed in terms of the Green's function \eqref{eq:brownian propagator}.
Similarly, for the density field, we write the solution $q_1(x, \tau) = -\partial_x \psi(x, \tau)$ with
\begin{align}
    \psi(x, \tau) = \int_{0}^{\tau}\mathrm{d}\tau^{\prime}\int_{-\infty}^{\infty} \mathrm{d}z \, g(x, \tau \mid z, \tau^{\prime})\,\sigma(q_0(z, \tau^{\prime}))\partial_z p_1(z, \tau^{\prime})
    \label{eq:sol psi}
\end{align}

With this solution, the first term in \eqref{eq:mu2} simplifies
\begin{align}
    \mathcal{I} & =\int_{0}^{1} \mathrm{d}\tau\, \Lambda(\tau)\int_{0}^{\infty}\mathrm{d}x \,  q_1(x, \tau) = \int_{0}^{1} \mathrm{d}\tau \, \Lambda(\tau) \psi(0, \tau) \cr
    & = \int_{0}^{1}\mathrm{d}\tau \, \Lambda(\tau) \int_{0}^{\tau}\mathrm{d}\tau_1\int_{\tau_1}^{1}\mathrm{d}\tau_2 \, \Lambda(\tau_2)\int_{-\infty}^{\infty} \mathrm{d}z \, \sigma(q_0(z, \tau_1))g(0, \tau_2\mid z, \tau_1) g(0, \tau\mid z, \tau_1)\label{eq:first integration 2}
\end{align}
while the second term yields an almost identical expression apart from the factor $\tfrac{1}{2}$,
\begin{align}
\int_{0}^{1}\mathrm{d}\tau\int_{-\infty}^{\infty}\mathrm{d}x \, \frac{\sigma(q_0(x, \tau))}{2} (\partial_x p_1(x, \tau))^2 =\frac{1}{2}\mathcal{I}
\end{align}

Collecting the two terms gives $\chi^{(2)}_{\mathcal{Q}}=\tfrac{\mathcal{I}}{2}$, which is in the form \eqref{eq:chi expansion}, allowing us to read off the two-time correlation, for $\tau_2>\tau_1$, 
\begin{align}
    C_{\mathcal{Q}}(\tau_1, \tau_2) \simeq \int_{0}^{\tau_1} \mathrm{d}\tau \int_{-\infty}^{\infty} \mathrm{d}z \,\sigma(q_0(z, \tau)) g(0, \tau_1\mid z, \tau) g(0, \tau_2\mid z, \tau)
    \label{eq:quenched_two_time_intermediate}
\end{align}
with $q_0(x, \tau)$ in~\eqref{eq: density profile}. 

For generic $\sigma(\rho)$, the integration cannot be simplified further. Rather, we use \eqref{eq: density profile} to expand around the uniform initial state, which leads (see~\ref{appendix:quenched_two_time_integrated_current_correlation_MFT}) to the expression~\eqref{eq:quenched_two_time_correlation_general_sigma_constant_D_intro} for the correlation up to second order in the density difference $\Delta\rho=\rho_a-\rho_b$.

\subsubsection{Annealed ensemble: \label{sec:anneal perturbation}}
The corresponding boundary condition \eqref{eq:boundary anneal} at linear order gives
\begin{align}
    p_1(x, 1) = 0 \quad \text{and} \quad 		q_1(x, 0) = \frac{\sigma(r(x))}{2 }\left(p_1(x, 0) - \Theta(x)\int_{0}^{1}\mathrm{d}\tau~\Lambda(\tau)\right)
    \label{eq: boundary condition of optimal q1}
\end{align}
where we used $q_0(x,0)=r(x)$ from \eqref{eq:diffusion_equation}.

The solution for $p_1(x,\tau)$ is identical to that in the quenched case~\eqref{eq:sol_p1 1}. For $q_1(x,\tau)$, since the governing equation~\eqref{eq:optimal_q1} is linear, its solution can be decomposed into homogeneous and inhomogeneous parts.
\begin{align}
    q_1(x, \tau) = q_{ih}(x, \tau) + q_h(x, \tau)
\end{align}
where the inhomogeneous part $q_{ih}(x, \tau)$ satisfies the differential equation 
\begin{align}
    \partial_{\tau}q_{ih}(x,\tau) - \partial_{x}^2q_{ih}(x,\tau) = -\partial_x(\sigma(q_0(x,\tau))\partial_xp_1(x,\tau))
    \label{eq:qi_evolution}
\end{align}
similar to \eqref{eq:optimal_q1} in the quenched case, while the homogeneous part $q_h(x, \tau)$ follows the equation 
\begin{align}
    \partial_{\tau}q_h(x,\tau) - \partial_{x}^2q_h(x,\tau) = 0 
    \label{eq:qh_evolution_equation}
\end{align}    
To maintain the correspondence with the quenched case, we assign the boundary conditions from \eqref{eq: boundary condition of optimal q1} as
\begin{align}
\quad q_{ih}(x,0) = 0, \quad \textrm{and } q_h(x,0) = \frac{\sigma(r(x))}{2}\left(p_1(x,0) - \Theta(x)\int_{0}^{1}\mathrm{d}\tau \, \Lambda(\tau)\right)
    \label{eq:qh_boundary_condition}
\end{align}
This assignment makes the solution for $q_{ih}(x, \tau)$ identical to the solution for $q_1(x, \tau)$ in the quenched case, giving $q_{ih}(x, \tau) = -\partial_x \psi(x, \tau)$ with \eqref{eq:sol psi}. 

This correspondence between parts of the solutions for the two ensembles reveals a simple relation between the corresponding scgfs in~\eqref{eq:mu2}
\begin{align}
\chi^{(2)}_{\mathcal{A}}\left(\Lambda\right) = \chi^{(2)}_{\mathcal{Q}}\left(\Lambda\right)  + \int_{0}^{1}&\mathrm{d}\tau\,\Lambda(\tau)\int_{0}^{\infty}\mathrm{d}x \, \big(q_h(x, \tau) - q_h(x, 0)\big) - \mathcal{F}_2(r(x),q_h(x,0))
        \label{eq:identity interacting particles MFT annealed}
\end{align}
with $\mathcal{F}_2$ in \eqref{eq:F_2_annealed} and $q_0(x,0)=r(x)$ from \eqref{eq:diffusion_equation}. The homogeneous solution has the remarkable property (see \ref{appendix: derivation identity interacting particles MFT annealed}) that the second term in \eqref{eq:identity interacting particles MFT annealed} equals $2\mathcal{F}_2(r,q_h)$, simplifying
\begin{align}
\chi^{(2)}_{\mathcal{A}} = \chi^{(2)}_{\mathcal{Q}}  + \mathcal{F}_2(r(x),q_h(x,0))
        \label{eq:identity interacting particles MFT annealed 2}
\end{align}
Interestingly, the difference between the two scgfs involves only the initial profile $q_h(x,0)$. Writing the terms explicitly using (\ref{eq:F_2_annealed},\ref{eq:qh_boundary_condition}) gives the difference
\begin{align}
\chi^{(2)}_{\mathcal{A}}\left(\Lambda\right) - \chi^{(2)}_{\mathcal{Q}}\left(\Lambda\right) = \frac{1}{4}\int_{-\infty}^{\infty}\mathrm{d}x \, \sigma(r(x))\left(\Theta(x)\int_{0}^{1}\mathrm{d}\tau \Lambda(\tau) - p_1(x, 0) \right)^2 \label{eq:mu2_diff_simplified}
\end{align}
where $p_1(x,0)$ is in~\eqref{eq:sol_p1 1}. 

For the step initial profile $r(x)$ in \eqref{eq: initial density profile}, the integration can be carried out explicitly using the identity \eqref{eq:id two erfc}, which gives
\begin{equation}\label{eq:chi2 diff final}
\chi^{(2)}_{\mathcal{A}}\left(\Lambda\right) - \chi^{(2)}_{\mathcal{Q}}\left(\Lambda\right) 
	=\! \int_{0}^{1} \!\!\mathrm{d}\tau_1 \int_{\tau_1}^{1} \!\! \mathrm{d}\tau_2  \Lambda(\tau_1) \Lambda(\tau_2) \left(\frac{\sigma(\rho_b) + \sigma(\rho_a)}{4\sqrt{\pi}}\left( \sqrt{\tau_1} + \sqrt{\tau_2} - \sqrt{\tau_1 + \tau_2} \right)\right)
\end{equation}
Considering the definition \eqref{eq:chi expansion}, we see that the term inside the brackets in \eqref{eq:chi2 diff final} gives the difference in correlations between the two ensembles, resulting in
\begin{equation}\label{eq:annealed_quenched_difference_MFT}
	C_{\mathcal{A}}(\tau_1, \tau_2) = C_{\mathcal{Q}}(\tau_1, \tau_2) + \frac{\sigma(\rho_b) + \sigma(\rho_a)}{4\sqrt{\pi}}\left( \sqrt{\tau_1} + \sqrt{\tau_2} - \sqrt{\tau_1 + \tau_2} \right)
\end{equation} 
which also gives the corresponding correlations $\cumul{Q(t_1)Q(t_2)}_{\mathcal{A}}$ in \eqref{eq: anneal scaled correlations}. The perturbation expansion~\eqref{eq:annealed_two_time_correlation_general_sigma_constant_D_intro} is obtained by combining \eqref{eq:annealed_quenched_difference_MFT} and \eqref{eq:quenched_two_time_correlation_general_sigma_constant_D_intro}.

The series solution can be systematically used to compute higher-order time correlations. However, the computation is tedious and we defer it to future work. Before concluding this MFT section, we emphasize two aspects. (a) For the uniform initial density $\Delta\rho=0$, the two-point correlations (\ref{eq:annealed_two_time_correlation_general_sigma_constant_D_intro},\ref{eq:quenched_two_time_correlation_general_sigma_constant_D_intro}) have a simple dependence on time. In particular, for the annealed ensemble, the correlation \eqref{eq:annealed_quenched_difference_MFT}
corresponds to the covariance of fractional Brownian motion~\cite{1968_Mandelbrot_Fractional,Sadhu2018ArcsineFBM,Sadhu2021FunctionalsFBM} with Hurst exponent $H=\tfrac{1}{4}$ and has been reported in several earlier works~\cite{2013_Leibovich_Everlasting,Dandekar2023,Krapivsky_2015,Sadhu_2016}. This correspondence breaks down for the transient state with $\rho_a\ne \rho_b$. (b) For the uniform initial state, the correlation for the annealed case can be obtained (see the discussion in~\cite{Krapivsky_2015}) from the quenched correlation by a shift of time.

\section{Exact solution for independent random walkers\label{sec:micro_non_interacting}}

Here, we present an independent derivation of the results in (\ref{eq:mu_annealed_non_interacting_MFT},\ref{eq:mu_quenched_non_interacting_MFT}) for the non-interacting system by studying its microscopic dynamics. Specifically, we consider a lattice gas of independent continuous-time random walkers on an infinite one-dimensional chain, with sites indexed by integers $i$, where particles jump to nearest-neighbor sites with unit rate.
We shall follow an analysis previously used in~\cite{Sadhu_2015} for a gas of independent Brownian particles.

The integrated current $Q(t)$ is the particle flux at the origin up to time $t$, which we write as follows.
\begin{align}
	Q(t) = L(t) - R(t)
	\label{eq:current_definition}
\end{align}
where $L(t)$ is the number of particles that started to the left of the origin ($i\le 0$) and have moved to the right ($i>0$), while $R(t)$ is the number of particles that started to the right ($i>0$) and have moved to the left $(i\le 0)$ by time $t$.

Because of the independence of $L(t)$ and $R(t)$, the generating functional splits as
\begin{equation}\label{eq: micro_gf}
	\left\langle e^{ \int_0^T \mathrm{d}t \, \lambda(t) Q(t) } \right\rangle = \left\langle e^{ \int_0^T \mathrm{d}t \, \lambda(t) L(t) } \right\rangle \left\langle e^{ -\int_0^T \mathrm{d}t \, \lambda(t)R(t) } \right\rangle
\end{equation}

Further simplifications follow from the absence of interactions among the walkers. The generating function of $L(t)$ becomes 
\begin{equation}
    \left\langle e^{ \int_0^T \mathrm{d}t \, \lambda(t) L(t) } \right\rangle = \prod_{i}\left\langle e^{ \int_0^T \mathrm{d}t \, \lambda(t) \Theta_{X_i(t)}} \right\rangle_{k_i} := \prod_{i}^{}F_L(\lambda,k_i)
\end{equation}
where the product over $i$ runs over walkers initially located at $k_i\le 0$, and their positions at subsequent time $t$ are denoted by $X_i(t)$. The indicator function $\Theta_{n}=1$ for integer $n > 0$ and vanishes otherwise.

Similarly, for $R(t)$,
\begin{equation}\label{eq:def FR}
    \left\langle e^{  -\int_0^T \mathrm{d}t \, \lambda(t){R}(t) } \right\rangle = \prod_{j}\left\langle e^{ - \int_0^T \mathrm{d}t \, \lambda(t) \Theta_{-X_j(t)+1} } \right\rangle_{k_j} := \prod_{j}^{}F_R(\lambda,k_j)
\end{equation}
where the product over $j$ runs over all walkers initially located at site $k_j> 0$, and their positions at subsequent time $t$ are denoted by $X_j(t)$. 

In the initial state, the sites are populated by particles following a Poisson distribution with average density $\rho_a$ for sites $i\le 0$ and density $\rho_b$ for sites $i>0$. Averaging over the initial particle distribution, we get
\begin{equation}\label{eq:anneal log average left half}
   \overline{ \left\langle e^{  \int_0^T \mathrm{d}t \, \lambda(t) L(t) } \right\rangle} = \prod_{k\le 0}\sum_{n_k\ge 0}\frac{\rho_a^{n_k}e^{-\rho_a}}{n_k!} \left(F_L(\lambda,k)\right)^{n_k} = e^{\rho_a\sum_{k\le 0}\left(F_L(\lambda,k)-1\right)}
\end{equation}
where $n_k$ denotes the number of particles at site $k$. A similar average follows for \eqref{eq:def FR}. Combining the contributions from both generating functionals in~\eqref{eq: micro_gf}, we arrive at the annealed cgf \eqref{eq:annealed generating functional unscaled},
\begin{align}\label{eq:cgf anneal exact micro}
    \mu_{\mathcal{A}}(\lambda(t))= \rho_a\sum_{k\le 0}\left(F_L(\lambda,k)-1\right) + \rho_b\sum_{k> 0}\left(F_R(\lambda,k)-1\right)
\end{align}

For the quenched cgf \eqref{eq:quenched generating functional unscaled}, we use
\begin{equation}\label{eq:quenched log average left half}
   \overline{ \log \left\langle e^{  \int_0^T \mathrm{d}t \, \lambda(t) L(t) } \right\rangle} = \sum_{k\le 0} \sum_{n_k\ge 0} \left(\log F_L^{n_k}(\lambda,k)\right) \frac{\rho_a^{n_k}e^{-\rho_a}}{n_k!} = \rho_a\sum_{k\le 0}\log F_L(\lambda,k)
\end{equation}
Similarly, averaging over sites on the right half of the origin, we write the quenched cgf
\begin{align}\label{eq:cgf quench exact micro}
    \mu_{\mathcal{Q}}(\lambda(t))= \rho_a\sum_{k\le 0}\log F_L(\lambda,k) + \rho_b\sum_{k> 0}\log F_R(\lambda,k)
\end{align}
Note how the initial average \eqref{eq:quenched log average left half} is dominated by the mean occupation, while in \eqref{eq:anneal log average left half} the initial fluctuations contribute significantly. This distinct contribution of typical and atypical initial fluctuations characterizes the difference between the quenched and annealed ensembles.

Note that $F_L$ and $F_R$ are properties of a single random walker $X(t)$, defined below,
\begin{equation}\label{eq:FLFR definition}
    F_L(\lambda,k)=\left\langle e^{ \int_0^T \mathrm{d}t \, \lambda(t) \Theta_{X(t)} } \right\rangle_{X(0)=k\le 0} \quad \textrm{and } F_R(\lambda,k)=\left\langle e^{ -\int_0^T \mathrm{d}t \, \lambda(t) \Theta_{-X(t)+1} } \right\rangle_{X(0)=k> 0}
\end{equation}
They can be written in a series expansion
\begin{align}
	F_L(\lambda, k) = 1 + \sum_{n\ge 1} \int_0^T \mathrm{d}t_1\int_{t_1}^T \mathrm{d}t_2 \dots \int_{t_{n-1}}^T \mathrm{d}t_n \, \lambda(t_1) \cdots \lambda(t_n) \mathcal{P}_{+}(t_1, \dots, t_n\vert k\leq 0)
	\label{eq:F_t_L_lambda_k_0}
\end{align}
where $\mathcal{P}_{+}(t_1, \dots, t_n\vert k\leq 0)$ is the probability that a random walker starting at a site $k\le 0$ is found on the right half of the origin ($i>0$) at times $(t_1,\cdots,t_n)$. Similarly,
\begin{align}
	F_R(\lambda, k) = 1 + \sum_{n=1}^\infty (-1)^n \int_0^T \mathrm{d}t_1\int_{t_1}^T \mathrm{d}t_2 \dots \int_{t_{n-1}}^T \mathrm{d}t_n \, \lambda(t_1) \cdots \lambda(t_n) \mathcal{P}_{-}(t_1, \dots, t_n\vert k> 0)
	\label{eq:F_t_R_lambda_k_0}
\end{align}
where $\mathcal{P}_{-}(t_1, \dots, t_n\vert k> 0)$ is the probability that a random walker starting at a site $k> 0$ is found on the left half of the origin ($i\le 0$) at times $(t_1,\cdots,t_n)$. These persistence probabilities $\mathcal{P}_{\pm}$ can be expressed in terms of the single random walker propagator 
\begin{align}
	g_{i,j}(t) = I_{i-j}(2t)e^{-2t} 
	\label{eq:ssep_micro_green}
\end{align}
where $I_n(t)$ is the modified Bessel function of the first kind. With these results, (\ref{eq:cgf anneal exact micro},\ref{eq:cgf quench exact micro}) gives an exact cgf valid at arbitrary time scales.  

\subsection*{Long-time limit.} For large $T$, by comparison with (\ref{eq:anneal scaling scgf},\ref{eq:quench scaling scgf}), we define the rescaled variables $\lambda(t)=\tfrac{\Lambda(\tau)}{T}$, $\tau=\tfrac{t}{T}$,  and $y=\tfrac{k}{\sqrt{T}}$. In this limit, the random-walk propagator \eqref{eq:ssep_micro_green} reduces to the Gaussian propagator \eqref{eq:brownian propagator}, $g_{k, k'} ( t - t') \simeq \frac{1}{\sqrt{T}}g(y, \tau \vert y', \tau')$. This leads to the continuum limit,
\begin{equation}
    F_L(\lambda(t),k) \simeq  f(\Lambda (\tau),y)\quad \textrm{and } F_R(\lambda(t),k) \simeq f(-\Lambda (\tau),y)
\end{equation}
expressed in terms of a single function
\begin{align}
	f(\Lambda(\tau), y) = 1 + \sum_{n = 1}^{\infty} \int_{0}^{1}\mathrm{d}\tau_1\int_{\tau_1}^{1}\mathrm{d}\tau_2 \cdots \int_{\tau_{n-1}}^{1}\mathrm{d}\tau_n\Lambda(\tau_1) \dots \Lambda(\tau_n) K_n(y, 0, \tau_1, \tau_2, \dots, \tau_n)\label{eq:f_lambda_y0} 
\end{align}
where $K_n(y, 0, \tau_1, \tau_2, \dots, \tau_n)$ is given in~\eqref{eq:K_n}. This makes it evident that the cgfs (\ref{eq:cgf anneal exact micro},\ref{eq:cgf quench exact micro}) obey the scaling (\ref{eq:anneal scaling scgf},\ref{eq:quench scaling scgf}) with the corresponding scgfs (\ref{eq:mu_annealed_non_interacting_MFT},\ref{eq:mu_quenched_non_interacting_MFT}), confirming the result obtained from MFT.

\section{Exact solution for SSEP\label{sec:micro_SSEP}}
In this section, we verify the expressions~(\ref{eq:annealed_two_time_correlation_general_sigma_constant_D_intro},\ref{eq:quenched_two_time_correlation_general_sigma_constant_D_intro}) for the two-time correlation of the current using the microscopic dynamics of the SSEP on a one-dimensional infinite lattice, which corresponds to $D=1$ and $\sigma(\rho)=2\rho(1-\rho)$. We shall follow an analysis previously used in~\cite{Sadhu_2016} for the SSEP on a finite lattice. 

The SSEP is a generalization of the system of continuous-time random walkers in~\sref{sec:micro_non_interacting}, with on-site exclusion interactions that forbid any two particles from occupying the same site at the same time. Each particle independently attempts to jump to an adjacent site at unit rate, provided the target site is empty. A configuration of the system at time $t$ is specified by the set of occupation variables $\{n_i(t)\}$, where $n_i(t)\in\{0,1\}$ denotes an empty or occupied site, respectively. The corresponding dynamics is
\begin{align}
    n_i(t + \mathrm{d}t) =
    \begin{cases}
        n_i(t) + 1 & \textrm{with probability } (n_{i-1}(t) + n_{i+1}(t))(1 - n_i(t))\mathrm{d}t \\
        n_i(t) - 1 & \textrm{with probability } n_i(t)(2 - n_{i-1}(t) - n_{i + 1}(t))\mathrm{d}t \\
        n_i(t) & \textrm{with probability } 1 - \big(n_{i+1}(t) + n_{i-1}(t) + 2n_i(t) \\ & \quad \quad \quad \quad \quad \quad \quad - 2n_i(t)n_{i+1}(t) - 2n_i(t)n_{i-1}(t)\big)\mathrm{d}t
    \end{cases}
\label{eq:occupation_time_evolution}
\end{align}

Initially, the particles are distributed according to a Bernoulli distribution with domain-wall average density $\rho_a$ for sites $(i \leq 0)$ and $\rho_b$ for sites $(i > 0)$. The time-integrated current $Q_i(t)$ is measured as the total flux of particles from site $i$ to site $i + 1$ during the time interval $[0, t]$. The corresponding evolution is 
\begin{align}
    Q_i(t + \mathrm{d}t) =
    \begin{cases}
        Q_i(t) + 1 & \textrm{with probability } n_i(t)\left(1 - n_{i + 1}(t)\right)\mathrm{d}t \\
        Q_i(t) - 1 & \textrm{with probability } \left(1 - n_i(t)\right)n_{i + 1}(t)\mathrm{d}t \\
        Q_i(t) & \textrm{with probability } 1 - \left(n_i(t) + n_{i+1}(t) - 2n_i(t)n_{i+1}(t)\right)\mathrm{d}t
    \end{cases}
    \label{eq:current_time_evolution}
\end{align}

\subsection{Average}

Following the dynamics (\ref{eq:occupation_time_evolution},\ref{eq:current_time_evolution}), the time evolutions of the average occupation and current satisfy
\begin{align}
    \frac{\mathrm{d}}{\mathrm{d}t}\langle n_j(t) \rangle_{\text{evo}} = \sum_{k}M_{jk} \langle n_k(t)\rangle_{\text{evo}} \quad \text{and} \quad \frac{\mathrm{d}}{\mathrm{d}t}\langle Q_i(t) \rangle_{\text{evo}} = \langle n_i(t) \rangle_{\text{evo}} - \langle n_{i+1}(t) \rangle_{\text{evo}} 
\end{align}
where $M_{i,j}=\delta_{i,j+1}-2\delta_{i,j}+\delta_{i,j-1}$. To distinction between the average over the evolution and the average over the initial distribution, we use the subscript `evo' to indicate the former.

The solution to these linear equations is given~\cite{Gerschenfeld2009Bethe} in terms of the propagator $g_{i,j}(t)$ in~\eqref{eq:ssep_micro_green}, which satisfies
\begin{equation}
\frac{\mathrm{d}}{\mathrm{d}t}{g}_{ij}(t)=\sum_k M_{ik}g_{kj}(t) \quad \textrm{with } g_{ij}(0)=\delta_{ij}
\end{equation}
We write a formal solution for these averages
\begin{align}
    \langle n_j(t) \rangle_{\text{evo}} = \sum_{k} g_{j,k}(t) n_k(0) \qquad \langle Q_i(t) \rangle_{\text{evo}} = \sum_{k}\int_{0}^{t}\mathrm{d}s\,(g_{i,k}(s) - g_{i+1,k}(s)) n_k(0)\label{eq:avg_curr_SSEP}
\end{align}

\subsection{Two-time correlation of current}

We shall see that the two-time correlation of the current $\cumul{ Q_i(t_2)Q_j(t_1)}_{\text{evo}}$ can be expressed in terms of the equal-time correlation of occupation variables $\cumul{n_i(t)n_j(t)}_{\text{evo}}$. An expression for the latter was derived in~\cite{Gerschenfeld2009Bethe}, which we use to determine the current correlation. 

From~\eqref{eq:current_time_evolution} we write
\begin{align}
	\frac{\mathrm{d}}{\mathrm{d}t} \cumul{ Q_i(t)Q_j(t_1)}_{\text{evo}}& =\cumul{ n_i(t)Q_j(t_1)}_{\text{evo}} -\cumul{ n_{i+1}(t)Q_j(t_1) }_{\text{evo}} \cr 
    & := K_{i,j}(t,t_1)\quad \text{for} \quad t\geq t_1. \label{eq:K_i_j_1}
\end{align}
Integrating~\eqref{eq:K_i_j_1} over time $t$ in the interval $[t_1, t_2]$, yields
\begin{align}
	\cumul{ Q_i(t_2) Q_j(t_1) }_{\text{evo}} = \cumul {Q_i(t_1) Q_j(t_1) }_{\text{evo}} + \int_{t_1}^{t_2} \mathrm{d}t K_{i,j}(t, t_1) \label{eq:QQ_unequal_time_2}
\end{align}
The first term involves the current correlation at different lattice sites at equal time $t_1$, which can be obtained again by utilizing \eqref{eq:current_time_evolution} and expressed in terms of $K_{i,j}$.  
\begin{align}
    \frac{\mathrm{d}}{\mathrm{d}t}\cumul{ Q_i(t) Q_j(t) }_{\text{evo}} = K_{i,j}(t,t) + K_{j,i}(t,t) + \delta_{i, j}\Gamma_j(t)\label{eq:rate equation QQ equal time}
\end{align} 
where $K_{i,j}$ is defined in~\eqref{eq:K_i_j_1}, and we define
\begin{equation}
\Gamma_i(t)= \Bigl\langle \bigl[n_i(t)-n_{i+1}(t)\bigr]^2 \Bigr\rangle_{\text{evo}} \label{eq:Gamma_i_t}
\end{equation}

After integrating \eqref{eq:rate equation QQ equal time} over the time interval $[0, t_1]$ and combining it with the result~\eqref{eq:QQ_unequal_time_2}, we arrive at
\begin{align}
	\cumul{ Q_i(t_2) Q_j(t_1) }_{\text{evo}} = \int_{0}^{t_1}\mathrm{d}t\left\{\,\delta_{i,j}\Gamma_{i}(t) + K_{i,j}(t,t) + K_{j,i}(t,t)\right\} + \int_{t_1}^{t_2} \mathrm{d}t K_{i,j}(t, t_1) \label{eq:QQ_unequal_time_3}
\end{align}
This expression forms the basis for computing the current correlations for different choices of initial ensembles. Before considering these cases, we derive the explicit form of $K_{i,j}$.

\subsection*{Computation of $K_{i,j}(t_2, t_1)$}

As defined in~\eqref{eq:K_i_j_1}, $K_{i,j}$ depends on the occupation-current correlation ($\cumul{ n_i(t_2) Q_j(t_1) }$), which, using~(\ref{eq:occupation_time_evolution},\ref{eq:current_time_evolution}), is found to evolve as
\begin{align}
    \frac{\mathrm{d}}{\mathrm{d}t_2}\cumul{ n_i(t_2) Q_j(t_1) }_{\text{evo}} = \sum_{k} M_{i,k}\cumul{ n_k(t_2) Q_j(t_1) }_{\text{evo}}
\end{align}
Its solution in terms of Green's function~\eqref{eq:ssep_micro_green} is
\begin{align}
	\cumul{ n_i(t_2) Q_j(t_1) }_{\text{evo}} = \sum_{k} g_{i,k}(t_2 - t_1) \cumul{ n_k(t_1) Q_j(t_1) }_{\text{evo}} \quad \text{for} \quad t_2 \geq t_1. \label{eq:tau_Q_relation}
\end{align}
This involves an equal-time occupation-current correlation. To compute this correlation, we use (\ref{eq:occupation_time_evolution},\ref{eq:current_time_evolution}) and write
\begin{align}
	\frac{\mathrm{d}}{\mathrm{d}t}\cumul{ n_i(t)Q_j(t)}_{\text{evo}}=\sum_{k} M_{i,k}\cumul{ n_k(t)Q_j(t)}_{\text{evo}}+A_{i,j}(t) 
	\label{eq:rate_equation_tauQ_equal_time}
\end{align}
where we defined 
\begin{equation}
    A_{ij}(t)=
\cumul{n_i(t)n_j(t)}_{\text{evo}}
-
\cumul{n_i(t)n_{j+1}(t)}_{\text{evo}}
+
\Gamma_j(t)
(\delta_{i,j+1}-\delta_{ij})
\label{eq:A_i_j_t_1}
\end{equation}
Note how the quantity depends solely on the equal-time correlation of occupation variables.

The solution of this linear equation~\eqref{eq:rate_equation_tauQ_equal_time} can be written in terms of the Green's function~\eqref{eq:ssep_micro_green},
\begin{align}
	\cumul{ n_i(t) Q_j(t) }_{\text{evo}} = \sum_{k} \int_0^t \mathrm{d}s \, g_{i,k}(t - s) A_{k,j}(s) \label{eq:tauQ_equal_time}
\end{align}
and using this result in~\eqref{eq:tau_Q_relation}, together with the identity $\sum_{k}~g_{i,k}(t - t^{\prime}) g_{k,j}(t^{\prime} - t^{\prime\prime}) = g_{i,j}(t - t^{\prime\prime})$
we obtain
\begin{align}
	\cumul{ n_i(t_2) Q_j(t_1) }_{\text{evo}} = \sum_{k} \int_0^{t_1} \mathrm{d}s \, g_{i,k}(t_2 - s) A_{k,j}(s)\label{eq:tauQ_unequal_time}
\end{align}
Using this solution in the definition~\eqref{eq:K_i_j_1} for $K_{i,j}(t_2, t_1)$, we arrive at a formula
\begin{align}
	K_{i,j}(t_2, t_1) = \int_0^{t_1} \mathrm{d}s \, F_{i,j}(t_2, s), \label{eq:K_i_j_F_i_j}
\end{align}
which depends only on the equal-time occupation correlation through the function 
\begin{align}
F_{ij}(t,s)=\sum_{l} \Bigl( g_{il}(t-s) - g_{i+1,l}(t-s) \Bigr) A_{lj}(s) \qquad \textrm{for } t\ge s.
\label{eq:F_i_j_1}
\end{align}

\subsection*{Current Correlation}
Substituting~\eqref{eq:K_i_j_F_i_j} into~\eqref{eq:QQ_unequal_time_3} expresses the evolution average of two-time correlation of current in terms of equal-time occupation correlations.
\begin{align}
\cumul{Q_i(t_2)Q_j(t_1)}_{\text{evo}}
={}&
\delta_{ij}
\int_0^{t_1}\!\mathrm{d}t\,\Gamma_i(t)
+
\int_0^{t_1}\!\mathrm{d}t^{\prime}
\int_0^{t'}\!dt^{\prime\prime}
\Bigl(
F_{ij}(t^{\prime},t^{\prime\prime})
+
F_{ji}(t^{\prime},t^{\prime\prime})
\Bigr)
\nonumber\\
&+
\int_{t_1}^{t_2}\!\mathrm{d}t^{\prime}
\int_0^{t_1}\!\mathrm{d}t^{\prime \prime}
F_{ij}(t^{\prime},t^{\prime\prime})
\label{eq:QQ_unequal_time_1}
\end{align}
with $\Gamma_i$ and $F_{ij}$ in \eqref{eq:Gamma_i_t} and \eqref{eq:F_i_j_1}, respectively. The expression for $\cumul{n_i(t)n_j(t)}_{\text{evo}}$ is given in~\cite{Gerschenfeld2009Bethe}, which we use next. 

Note that the expression~\eqref{eq:QQ_unequal_time_1} is formally similar to the expression for the corresponding correlation in the finite SSEP~\cite{Sadhu_2016}, with the primary difference being the corresponding propagator.

\subsection{Average over initial state.\label{subsec:SSEP_microscopic_annealed}}

\subsubsection*{Annealed setting:}
For the annealed ensemble the two-time correlation function  
\begin{align}
    \cumul{ Q_i (t_1) Q_j(t_2) }_{\mathcal{A}} & =  \overline{   \cumul{Q_i(t_1) Q_j(t_2)}}_{\text{evo}}  \label{eq:annealed_defined}
\end{align}
where $\overline{\cdots}$ represents the average over the initial configurations. Evaluating this quantity through~\eqref{eq:QQ_unequal_time_1} requires the corresponding annealed average of the mean occupation and the two-time correlation. Both quantities have been evaluated in~\cite{Gerschenfeld2009Bethe}, and at long times they have the following scaling forms
\begin{equation}
	\overline{\langle n_i(t) \rangle}_{\text{evo}}\simeq   \bar{\rho} - \frac{\Delta \rho}{2}\Erf{\frac{i}{2\sqrt{t}}} \label{eq:average occupation SSEP}
\end{equation}
and for $j>i$,
\begin{align}
	\overline{\cumul{ n_j(t)n_i(t) }}_{\text{evo}} \simeq  -(\Delta \rho)^2\,\frac{e^{-\frac{(i+j)^2}{8t}}}{4\sqrt{2\pi t}}\left(1 + \Erf{\frac{i-j}{\sqrt{8t}}}\right)
	\label{eq:tau_tau_equal_time_SSEP}
\end{align}
(In~\cite{Gerschenfeld2009Bethe}, the results are for $(\rho_a,\rho_b) = (1,0)$ and are easily generalized to arbitrary domain-wall densities.)

Using these scaling results (\ref{eq:average occupation SSEP},\ref{eq:tau_tau_equal_time_SSEP}) in~\eqref{eq:QQ_unequal_time_1} determines the long-time result for $\cumul{ Q_i (t_1) Q_j(t_2) }_{\mathcal{A}}$. In this article, we report only the case $i=j=0$, for which we recover the expression~\eqref{eq:annealed_two_time_correlation_general_sigma_constant_D_intro} for $\sigma(\rho)=2\rho(1-\rho)$ obtained from MFT. Details of the algebra are presented in \ref{appendix: annealed two time integrated current correlation micro}.

\subsubsection*{Quenched setting:}
According to the definition~\eqref{eq:generating functional unscaled}, the two-time correlations in the two ensembles differ by
\begin{align}
	\cumul{ Q_i(t_1)Q_j(t_2) }_{\mathcal{A}} - \cumul{ Q_i(t_1)Q_j(t_2) }_{\mathcal{Q}}  = \overline{\langle Q_i(t_1)\rangle_{\text{evo}} \langle Q_j(t_2) \rangle_{\text{evo}}}-\overline{\langle Q_i(t_1)\rangle }_{\text{evo}}  \,\, \overline{\langle Q_j(t_2)\rangle }_{\text{evo}} \label{eq:annealed_quenched_diff}
\end{align}
The right hand side is easily computed using~\eqref{eq:avg_curr_SSEP} and the Bernoulli distribution of the initial occupation variables $n_i(0)$. In the long-time limit, for $i=j=0$, we get (see~\ref{appendix: quenched two time integrated current correlation micro}) 
\begin{align}
	\cumul{ Q_0(t_1)Q_0(t_2) }_{\mathcal{A}} -\cumul{ Q_0(t_1)Q_0(t_2) }_{\mathcal{Q}} \simeq \frac{1}{\sqrt{\pi}}\left(\bar{\rho}(1 - \bar{\rho}) - \frac{(\Delta \rho)^2}{4}\right)\left(\sqrt{t_1} +\sqrt{t_2}-\sqrt{t_1 + t_2}\right)
	\label{eq:annealed_quenched_difference_SSEP}
\end{align}
which recovers the quenched two-time correlation~\eqref{eq:quenched_two_time_correlation_general_sigma_constant_D_intro} for the SSEP upon setting $\sigma(\rho) = 2\rho(1 - \rho)$.

\subsection{Density-Current Correlation\label{sec:density_current_micro_SSEP}}

An intermediate step of our analysis of current-current correlation involves an expression~\eqref{eq:tauQ_equal_time} for the correlation between occupation and current, and which can be evaluated similarly using the results~(\ref{eq:average occupation SSEP},\ref{eq:tau_tau_equal_time_SSEP}).

We present here the result for the simplest case, $\cumul{ n_j(t)Q_0(t) }$, which, in the long-time limit in the annealed setting (see~\ref{appendix:density_current_correlation} for a derivation), has the following explicit expression for $j>0$,
\begin{align}
	\cumul{ n_j(t)Q_0(t) }_{\mathcal{A}}\simeq&
    \frac{\bar{\rho} (1 - \bar{\rho})}{2}\Erfc{ \frac{|j|}{2 \sqrt{t}} }-\frac{(\Delta \rho)^2}{8}\!\!\left\{\erfc{\frac{j}{2\sqrt{2t}}}^2\!\!-\erfc{\frac{j}{2\sqrt{t}}} \right\}\label{eq:a_density_current}
\end{align}

In the quenched setting, similar to~\eqref{eq:annealed_quenched_diff}, we write 
\begin{align}
    \cumul{ n_j(t)Q_0(t)}_{\mathcal{A}} - \cumul{ n_j(t)Q_0(t)}_{\mathcal{Q}}=\overline{\langle n_j(t) \rangle_{\text{evo}}\langle Q_0(t) \rangle_{\text{evo}}} - \overline{\langle n_j(t) \rangle}_{\text{evo}}\,\,\overline{\langle Q_0(t) \rangle}_{\text{evo}}
\end{align}
Computing the initial averages using~\eqref{eq:avg_curr_SSEP} and the Bernoulli distribution of the initial occupation variables $n_i(0)$, we get, in the long-time limit (see~\ref{appendix: density current AA_QQ diff} for a derivation),
\begin{align}
	\cumul{ n_j(t)Q_0(t)}_{\mathcal{A}} - \cumul{ n_j(t)Q_0(t)}_{\mathcal{Q}} \simeq & \frac{\rho_a(1 - \rho_a)}{4}\phi\left(\frac{j}{2\sqrt{t}}\right)  - \frac{\rho_b(1 - \rho_b)}{4}\phi\left(-\frac{j}{2\sqrt{t}}\right)
	\label{eq:q_density_current}
\end{align}
where $\phi(x) = \Erfc{x} - \frac{1}{2}\Erfc{\frac{x}{\sqrt{2}}}^2$.
These results~(\ref{eq:a_density_current},\ref{eq:q_density_current}) are consistent with the expressions~(\ref{eq:tau_Q_annealed_explicit_SSEP},\ref{eq:annealed_quenched_difference_tauQ}) reported in the introduction. 

These exact results for the equal-time density-current correlations characterize how particles rearrange themselves to yield a specific value of the integrated current $Q(t)\equiv Q_0(t)$ across the origin, as reflected in Fig.~\ref{fig:density_current_annealed}. 
A generating function~\cite{Benichou2021PRL, Grabsch2022} for these correlations is 
\begin{equation}\label{eq:w}
w_j(\lambda,t) = \frac{\langle n_j(t) e^{\lambda Q(t)} \rangle}{\langle e^{\lambda Q(t)} \rangle} = \langle n_j(t) \rangle + \lambda  \cumul{n_j(t)Q(t)}+\cdots
\end{equation}
and, by ensemble equivalence~\cite{DerridaSadhu2019I,DerridaSadhu2019II}, it is related to the density profile at long times in the ensemble conditioned on the integrated current $Q(t)$. 

The hydrodynamic limit  $w_{j}(\lambda,t)\simeq h_\lambda\left(\frac{j}{\sqrt{t}}\right)$ for large $t$ corresponds to the optimal density profile at the final time within the MFT for the generating function $\langle e^{\lambda Q(t)}\rangle$ of single-time statistics at time $t$. The corresponding MFT is obtained from the discussion in \sref{sec: mft infinite} for $\Lambda(\tau)=\lambda \,\delta(1-\tau)$, and within this correspondence, $h_\lambda(x)$ is the optimal density $q(x,1)$ in \eqref{eq:optimal}. From \eqref{eq:w}, this implies that the hydrodynamic limit of the density-current correlation in~(\ref{eq:tau_Q_annealed_explicit_SSEP},\ref{eq:annealed_quenched_difference_tauQ}) is proportional (with prefactor $\tfrac{1}{\lambda}$) to the optimal density $q_1(x,1)$ at linear order in \sref{sec:mft_infinite_interacting} for the corresponding ensemble. 

For SSEP in the quenched ensemble, from \sref{sec:quench perturbation} we obtain
\begin{subequations}\label{eq:q1 for sep 1time}\begin{equation}
    q^{\mathcal{Q}}_1(x, 1) = - \int_{0}^{1}\mathrm{d}\tau^{\prime}\int_{-\infty}^{\infty} \mathrm{d}z \, \partial_x g(x, 1 \vert z, \tau^{\prime})\,2 q_0(z, \tau^{\prime}) (1- q_0(z, \tau^{\prime}))\partial_z p_1(z, \tau^{\prime})
\end{equation}
with $q_0(x,\tau)$ in \eqref{eq: density profile} and 
\begin{align}
    p_1(x,\tau) = \lambda \int_{0}^{\infty} \mathrm{d}z \, g(z,1 \vert x,\tau)
\end{align}\end{subequations}
obtained from \eqref{eq:sol_p1 1} using $\Lambda(\tau)=\lambda \delta(\tau-1)$. 

Similarly, for the annealed ensemble, from \sref{sec:anneal perturbation}, we obtain for the SSEP,
\begin{equation}\label{eq:q1 for sep 1time anneal}
    q^{\mathcal{A}}_1(x, 1) = q^{\mathcal{Q}}_1(x, 1) + \int_{-\infty}^{\infty} \mathrm{d}z \, g(x,1 \vert z,0)\, r(z)\big(1 - r(z)\big)\big(p_1(z,0) - \lambda\,\Theta(z)\big)
\end{equation}

\begin{figure}
     \centering
         \includegraphics[width=0.8\linewidth]{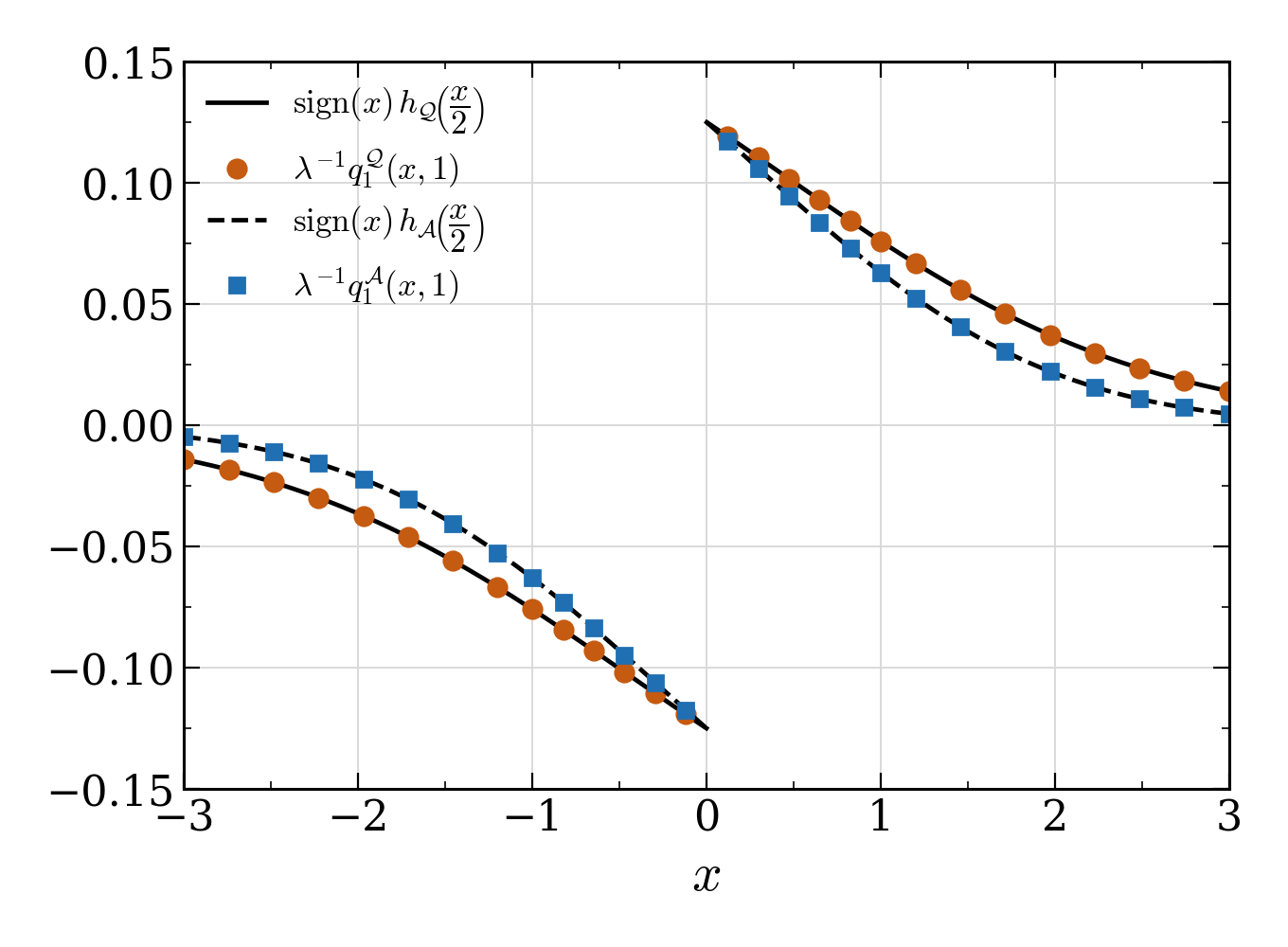}
     \caption{\textbf{Density-current correlation and optimal profile:} A numerical comparison between the density-current correlation~(\ref{eq:tau_Q_annealed_explicit_SSEP},\ref{eq:annealed_quenched_difference_tauQ}) and the optimal density profile~(\ref{eq:q1 for sep 1time},\ref{eq:q1 for sep 1time anneal}) at linear order in $\lambda$ for the SSEP. The results shown are for domain-wall densities $\rho_a = 0.75$ and $\rho_b = 0.25$.}    
     \label{fig:q1hQ match}
\end{figure}

We have confirmed (see \fref{fig:q1hQ match}) that $q_1^{\mathcal{Q}}(x,1)=\lambda \, {\rm sgn}(x)h_{\mathcal{Q}}(\tfrac{x}{2})$ and $q_1^{\mathcal{A}}(x,1)=\lambda \, {\rm sgn}(x)h_{\mathcal{A}}(\tfrac{x}{2})$ with $h_{\mathcal{Q/A}}(x)$ in~(\ref{eq:tau_Q_annealed_explicit_SSEP},\ref{eq:annealed_quenched_difference_tauQ}). This agreement provides an independent microscopic verification of the optimal profile predicted within the MFT formalism for SSEP. 

\section{Summary and Outlook}

In this article, we discussed the MFT formalism for studying multi-time correlations of the time-integrated current in generic one-dimensional diffusive systems on the infinite line, in the transient regime following a domain-wall initial condition. For the simplest case of constant diffusivity
and linear mobility, we showed that the theory is exactly solvable, yielding the closed-form expression~\eqref{eq:mu_nonint_combined} for the SCGF in both
the annealed and the quenched ensembles. The resulting multi-time correlations
take the forms~(\ref{eq:explicit_annealed},\ref{eq:explicit_quenched}), in
which the extensive time dependence is remarkably simple. These results are
corroborated by an independent solution obtained directly from the microscopic
dynamics of a lattice gas of independent random walkers.

For general interacting systems with constant diffusivity and arbitrary mobility, we obtained explicit expressions~(\ref{eq:quenched_two_time_correlation_general_sigma_constant_D_intro},\ref{eq:annealed_two_time_correlation_general_sigma_constant_D_intro}) for the two-time correlation of the integrated current. In particular, we showed that, for a non-stationary initial condition, the two-time correlation in the annealed ensemble is no longer that of fractional Brownian motion. We independently verified these results for the multi-time correlations using microscopic solution of the SSEP.

For the SSEP, we also computed the density-current
correlation~(\ref{eq:tau_Q_annealed_explicit_SSEP},\ref{eq:annealed_quenched_difference_tauQ}),
which describes how the system rearranges its particles in order to sustain an
atypical value of the current. To the best of our knowledge, this result was
not known previously, either for the domain-wall initial condition or for the
quenched ensemble~\cite{Grabsch2022}. These exact results provide an
independent verification (\fref{fig:q1hQ match}) of the optimal density profile
predicted by the MFT formalism.

Our results for the multi-time correlations reveal features that are not captured by the single-time statistics. For example, the single-time statistics show no qualitative difference between the stationary case ($\rho_a=\rho_b$) and the non-stationary case ($\rho_a\ne\rho_b$), apart from the vanishing of
the odd cumulants. Moreover, their essential physics is already captured in the
simple non-interacting limit: the $\sqrt{t}$ scaling of the cumulants, the
tails of the large deviations, and the lasting memory of the initial condition. However, for the multi-time statistics with a domain-wall initial
condition interactions introduce a qualitatively different time
dependence~(\ref{eq:quenched_two_time_correlation_general_sigma_constant_D_intro},%
\ref{eq:annealed_two_time_correlation_general_sigma_constant_D_intro}).

Several interesting extensions follow naturally from the present work. Foremost
among them is the computation of the full multi-time current statistics for
generic interacting diffusive systems, and in particular for the SSEP. A
promising route in this direction is to extend the integrability structures
uncovered within MFT for quadratic
mobility~\cite{Mallick2022PRL, KMP_exact_2022, KPZ_exact_2021} to the
multi-time setting. Another natural candidate is the diffusive hard-rod gas,
whose MFT for the single-time statistics was solved
recently~\cite{Jangid2026SingleFileMFT}.

Beyond this, the fluctuating-hydrodynamics framework developed here can be
extended naturally to investigate temporal correlations in comb-like
structures~\cite{Benichou2015PRL, iomin2018fractional, Venturelli2025JPAMT},
higher-dimensional lattices~\cite{Spohn1991, Venturelli2025, Berlioz_2024,BodineauDerrida2026},
multi-species systems~\cite{Cantini_2026}, systems with several conserved
quantities~\cite{BodineauDerrida2011}, and diffusive systems subject to external
fields~\cite{Berlioz2025PRL}. Developing an analogue of our analysis for
ballistic systems within the ballistic MFT~\cite{2023_Doyon_Ballistic,
2026_Kethepalli_Ballistic} would be equally timely.

\ack
Several parts of this work draw heavily on tools and concepts developed in
earlier joint work with Bernard Derrida, Paul Krapivsky, and Kirone Mallick;
TS gratefully acknowledges these past collaborations. This work was primarily supported
by the Department of Atomic Energy, Government of India, under Project Identification Number RTI-4012 and in part by grant NSF PHY-2309135 to the Kavli Institute for Theoretical Physics (KITP). The computations were carried out on the
computing clusters of the Department of Theoretical Physics, TIFR, Mumbai, and
we thank Ajay Salve and Kapil Ghadiali for their computational support. The
authors acknowledge the use of AI models for assistance with some of the analytical calculations and proofreading.

\appendix

\section{Derivation of the annealed correlations}
\label{sec:derivation}

Here we give a derivation of the annealed pair sum
\eqref{eq:explicit_annealed} starting from the form~\eqref{eq:QA multi corr}.

\subsection{The setup}

In expression~\eqref{eq:QA multi corr}, the function $\psi_{z_0}(t_1,\dots,t_n)$ denotes the probability that a Brownian particle starting at position $-z_0$ to the left of the origin at time $t_0=0$  is found at positions $z(t_i)>0$ (on the right side) at all subsequent times
$t_1<\dots <t_n$. A coordinates shift $X(t)=z(t)+z_0$ transforms $\psi_{z_0}(t_1,\dots,t_n)$ to
the joint probability $\pr{X_1, \cdots, X_n}$ of a standard Brownian motion $X(t)$ with $X(0)=0$ such that $X_i \equiv X(t_i)>z_0$ for all times $t_i$. It simplifies the integral over the initial position $z_0$ in~\eqref{eq:QA multi corr} as follows:
\begin{equation}\label{eq:append_psi_prob}
\int_0^\infty\!dz_0\,\psi_{z_0}
=\int_0^\infty\!dz_0\int\! dX_1\,\dots\int\! d X_n\, \pr{X_1,\dots,X_n}\prod_i\Theta(X_i-z_0)
\end{equation}

The Heaviside step function $\Theta(x)$ restricts the non-zero contribution to the interval $0<z_0<\min_i X_i$, allowing us to write
\begin{equation}
\int_0^\infty\!dz_0\,\psi_{z_0}=\left<{\min_j X_j\;\Theta\big(\min_j X_j\big)}\right>=\sum_{j=1}^{n}\left<{ X_j\,\Theta(X_j)\prod_{k\ne j}\Theta(X_k-X_j 
)}\right>
\label{eq:presteinder}
\end{equation}

\subsection{Integration by parts}
The joint distribution $\pr{X_1, \cdots, X_n}$ is a multivariate Gaussian with zero mean and covariance $C_{ij}\equiv\mathrm{Cov}(X_i,X_j)=2\min(t_i,t_j)$. Using integration by parts (Stein's lemma), we write
\begin{equation}\label{eq:steins}
\left<X_j \,h(X_1,\dots,X_n)\right>=\sum_{i=1}^{n}\mathrm{Cov}(X_i,X_j)\, 
\left<{\partial_{X_i} h}\right>, 
\quad \text{where}\, h=\Theta(X_j)\prod_{k\ne j}\Theta(X_k-X_j)
\end{equation}
Using \eqref{eq:steins} in \eqref{eq:presteinder} and carrying out the derivatives, we get
\begin{align}
\int_0^\infty\!dz_0\,\psi_{z_0}
=&\sum_j C_{jj}\,\left<{\delta(X_j)\!\prod_{k\ne j}\!\Theta(X_k-X_j)}\right>
\cr & +\sum_j\sum_{i\ne j}\big(C_{ij}-C_{jj}\big)\,
\left<{\delta(X_i-X_j)\,\Theta(X_j)\!\prod_{l\ne i,j}\!\Theta(X_l-X_j)}\right>
\label{eq:twoterm}
\end{align}
Further simplification follows from using the expression for $C_{ij}$ and noting that $X_0=0$ at $t_0=0$,
\begin{equation}\label{eq:A5}
\int_0^\infty\!dz_0\,\psi_{z_0}
=2\sum_{j=1}^n\sum_{i=0}^{j-1}(t_j-t_i)\epsilon_{ij}\,
\left<{\delta(X_i-X_j)\,\Theta(X_j)\!\!\prod_{l\ne i,j}\!\!\Theta(X_l-X_j)}\right> 
\end{equation}
where $\epsilon_{0j}=+1$ and $\epsilon_{ij} = -1$ for $i> 0$.

Since $(X_i-X_j)$ is a Brownian increment, the delta function constraint $X_i=X_j$
yields a factor $\tfrac{1}{\sqrt{4\pi(t_j-t_i)}}$, while the remaining average
is conditioned on $X_i=X_j$:
\begin{equation}\label{eq:prefactor_condition_prob}
\left\langle
\delta(X_i-X_j)\,
\Theta(X_j)\!\!\prod_{l\ne i,j}\!\!\Theta(X_l-X_j)
\right\rangle 
=
\frac{P^{(n)}_{ij}}{\sqrt{4\pi(t_j-t_i)}}
\end{equation}
where the term on the right-hand side, $P_{ij}^{(n)}$, is the conditioned probability
\begin{equation}\label{eq:append_condition_prob}
P^{(n)}_{ij}=\left<{\Theta(X_j)\!\!\prod_{l\ne i,j}\!\!\Theta(X_l-X_j)\ \Big|\
X_i=X_j}\right>
=\pr{X_l\ge X_i\ge 0 \forall l\notin\{i,j\}\ \big|\ X_i= X_j}
\end{equation}
Using \eqref{eq:prefactor_condition_prob} in \eqref{eq:A5} we get the pair-sum formula
\begin{equation}
\int_0^\infty\!dz_0\,\psi_{z_0}
=\frac{1}{\sqrt\pi}\sum_{0\leq i<j \leq n} \epsilon_{ij} P^{(n)}_{ij}\sqrt{t_j-t_i}
\label{eq:finalder}
\end{equation}

\section{Derivation of the quenched correlation}
\label{sec:derivation_q}
 
The pair-sum formula for the quenched correlation~\eqref{eq:explicit_quenched} is built from~\eqref{eq:QQ multi corr} following an analysis very similar to that of the annealed case, applied
once per partition. We only highlight what changes relative to ~\ref{sec:derivation}.
 
\subsection{Setup}  
 
The quenched integrand in~\eqref{eq:QQ multi corr}
\begin{equation}\label{eq:append_Phi_Q}
\Phi_{z_0}=\sum_{\omega\in\Omega_n}\lambda_\omega
\prod_{B_\omega\in\omega}\psi_{z_0}\left(t_{b_1},\ldots,t_{b_{|B_{\omega}|}}\right),
\qquad \textrm{with } \lambda_\omega=(-1)^{|\omega|-1}(|\omega|-1)!
\end{equation}
where, similarly to~\eqref{eq:append_psi_prob}, $\psi_{z_0}=\pr{X(t_k)>z_0\ \forall k\in B_\omega}$ is the single-particle
probability restricted here to the times of that block $B_\omega$. Unlike the annealed case, where $\psi_{z_0}$ was written in terms of a single Brownian process, here, due to the partition structure, one Brownian motion $X_{(B_\omega)}$ is assigned to each block $B_\omega$ of the partition $\omega$. For a partition $\omega$,
a product over blocks~\eqref{eq:append_Phi_Q} is then the joint probability of independent standard Brownian motions:
\begin{equation}
\prod_{B_\omega\in\omega}\psi_{z_0}(t_{B_\omega})
=\pr{X_k>z_0\ \ \forall k=1,\dots,n},
\qquad X_k\equiv X_{(B_\omega)}(t_k)
\end{equation}
with $B_\omega$ the block containing $k$. This procedure is carried out for each partition $\omega$ and then summed according to~\eqref{eq:append_Phi_Q}. The integration over $z_0$, exactly
as in~\eqref{eq:append_psi_prob}, gives one positive-minimum average per partition:
\begin{equation}\label{eq:append_F_sum}
\int_0^\infty\!dz_0\,\Phi_{z_0}=\sum_{\omega\in\Omega_n}\lambda_\omega\,F[\omega],
\qquad
F[\omega]\equiv\left<{\min_{k}X_k\ \Theta\left(\min_{k}X_k\right)}\right> 
\end{equation}
The random variables $(X_1,\dots,X_n)$ are distributed according to a multivariate Gaussian with zero mean and covariance $C_{ik}[\omega]=2\min(t_i,t_k)$  when $i\sim k$ (same Brownian motion) and $C_{ik}[\omega]=0$ when $i \not\sim k$ (independent Brownian motions).
 
\subsection{Integration by parts}
Expression \eqref{eq:append_F_sum} simplifies further, similarly to the annealed case~\eqref{eq:twoterm}, but is now weighted by $\lambda_\omega$. Additionally, unlike the annealed case~\eqref{eq:twoterm}, the contributions differ depending on whether $\{i,j\}$ are in the same block or in disjoint blocks. If $\{i,j\}$ are in the same block, the analysis is exactly the same as in the annealed case. The only new contribution arises when the walkers are in disjoint blocks ($i \not\sim j$).
\begin{equation}
\underbrace{(C_{ij}[\omega]-C_{ii}[\omega])}_{-2t_i}
+\underbrace{(C_{ji}[\omega]-C_{jj}[\omega])}_{-2t_j}
=-2(t_i+t_j)
\end{equation}
and for independent walkers the pre-factor in~\eqref{eq:prefactor_condition_prob} is $\tfrac{1}{\sqrt{4\pi(t_i+t_j)}}$. This is the
origin of the new pair-sum class $\sqrt{t_i+t_j}$.
 
\subsection{Assembling the two classes}
Each
 partition contributes a pair sum in which
the contribution (either $\sqrt{t_j-t_i}$ or $\sqrt{t_j+t_i}$) is dictated by whether the pair $\{i,j\}$ is chosen from the same block or two disjoint blocks. Therefore, for each partition $\omega$, the sum separates in the following way:
\begin{equation}
F[\omega]=\frac{1}{\sqrt\pi}\Bigg[\!\!\sum_{\substack{j>i\\ i\sim j}}\!\!\epsilon_{ij}P^{(n)}_{ij}[\omega]\sqrt{t_j-t_i}
-\!\!\sum_{\substack{j>i\\ i\not\sim j}}\!\!P^{(n)}_{ij}[\omega]\sqrt{t_i+t_j}
\Bigg]
\end{equation}
with $P_{ij}^{(n)}[\omega]$ the conditioned probability for the vector $(X_1, \cdots, X_n)$ defined in~\eqref{eq:append_condition_prob}. Summing over partitions with
the weights $\lambda_\omega$ as in~\eqref{eq:append_F_sum} and defining~\eqref{eq:quenchedcoef}
yields~\eqref{eq:explicit_quenched}.

\section{A derivation of the two-time correlation function.}\label{appendix:derivation_annealed_quenched_two_time_correlator}
\subsubsection*{Annealed Setting.}
For $n=2$, the expression~\eqref{eq:non_annealed_n_point} gives
\begin{align}
    \cumul{ Q(t_1)Q(t_2) }_{\mathcal{A}}\simeq \left(\rho_a + \rho_b\right)\int_{-\infty}^{0} \mathrm{d}z_0\,\Psi_{z_0}\left(t_1, t_2\right)\label{eq:2-time_annealed}
\end{align}    
where $\Psi_{z_0}(t_1,t_2)$ in \eqref{eq:Psi n point} can be written explicitly as
\begin{align}\label{eq:psi t1t2}
    \Psi_{z_0}(t_1, t_2) = \frac{1}{4}\left(\Erfc{-\frac{z_0}{2\sqrt{t_1}}} + \Erfc{-\frac{z_0}{2\sqrt{t_2}}}\right) - T_{\text{Owen}}\left(\frac{z_0}{\sqrt{2t_2}}, \sqrt{\frac{t_2}{t_1} - 1}\right)
\end{align}
in terms of Owen's T function~\cite{Owen1980} in \eqref{eq:Owen_T}. Evaluating the integral over $z_0$ in \eqref{eq:2-time_annealed} using the identity \eqref{eq:id TW} gives the explicit result~\eqref{eq:non_annealed_two_time_correlator}, where we have expressed $\arctan$ in terms of $\arcsin$.

\subsubsection*{Quenched Setting.}
For $n=2$, the expression~\eqref{eq:non_quenched_n_point} reduces to
\begin{align}
    \cumul{ Q(t_1)Q(t_2) }_{\mathcal{Q}} \simeq \left(\rho_a + \rho_b\right)  \int_{-\infty}^{0}\mathrm{d}z_0 \,\left( \Psi_{z_0}(t_1, t_2) - \Psi_{z_0}(t_1)\Psi_{z_0}(t_2)\right)\label{eq:2-time_quenched}
\end{align}
Using \eqref{eq:psi t1t2} and
\begin{align}
    \Psi_{z_0}(t) = \frac{1}{2}\Erfc{-\frac{z_0}{2\sqrt{t}}}\label{eq:Psi_1_time_explicit}
\end{align}
from \eqref{eq:Psi n point} leads to the explicit result~\eqref{eq:non_quenched_two_time_correlation}, where we expressed $\arctan$ in terms of $\arcsin$.

\section{A derivation of the three-time correlation function.}\label{appendix:derivation_three_time_correlator}

The integral expressions for the correlations ~(\ref{eq:non_annealed_n_point_full}, \ref{eq:non_quenched_n_point_full}) are similar to those for the corresponding correlations of the tracer position in the high-density limit of the SSEP obtained in~\cite{Benichou_multi_2025}, where the integrals are evaluated explicitly for three-time and four-time correlations. Here, we provide an independent verification along similar lines.

\subsubsection*{Annealed Setting:}
Writing the expression~\eqref{eq:non_annealed_n_point_full} for $n = 3$ gives
\begin{align}
    \cumul{ Q(t_1)Q(t_2)Q(t_3) }_{\mathcal{A}}\simeq \left(\rho_a-\rho_b\right)\int_{-\infty}^{0} \mathrm{d}z_0\,\Psi_{z_0}\left(t_1, t_2, t_3\right)\label{eq:non_annealed_3_point}
\end{align}    
with $\Psi_{z_0}\left(t_1, t_2, t_3\right)$ from \eqref{eq:Psi n point}. The integrations over $z_0$ and $z_3$ are simpler and give
\begin{align}\label{eq:D2}
\int_{-\infty}^{0} \mathrm{d}z_0\,\Psi_{z_0}\left(t_1, t_2, t_3\right) = \frac{\omega}{4\sqrt{\pi}}\int_{0}^{\infty}\mathrm{d}z_1\mathrm{d}z_2\, \Erfc{-\gamma z_2}e^{-\omega^2(z_2 - z_1)^2}\Erfc{\beta z_1}
\end{align}
where we define the parameters $\gamma = \frac{1}{2\sqrt{t_3 - t_2}}$, $\omega = \frac{1}{2\sqrt{t_2 - t_1}}$ and $\beta = \frac{1}{2\sqrt{t_1}}$ (these definitions are restricted to this section). To evaluate the remaining integrations over the $z$-variables, we use Feynman's trick and define
\begin{align}
    \mathsf{I}_1(\alpha) = \frac{\omega}{4\sqrt{\pi}}\int_{0}^{\infty}\mathrm{d}z_1\mathrm{d}z_2\,\Erfc{-\gamma z_2}e^{-\omega^2(z_2 - z_1)^2}\Erfc{\beta\alpha z_1}\label{eq:Psi_3_time_eval_first_step}
\end{align}
such that 
\begin{align}\label{eq:psi3t I1}
\int_{-\infty}^{0} \mathrm{d}z_0\,\Psi_{z_0}\left(t_1, t_2, t_3\right) =  \mathsf{I}_1(1)
\end{align}

To compute $\mathsf{I}_1(1)$, it is simpler to evaluate $\partial_\alpha\mathsf{I}_1(\alpha)$ using the identity \ref{eq:id erf}, leading to
\begin{align}
    \frac{\mathrm{d} \mathsf{I}_1(\alpha)}{\mathrm{d}\alpha} = & -\frac{1}{4\pi^{3/2}\alpha^2\beta}\Bigg\{\frac{\pi}{2} + \mathrm{arctan}\frac{\gamma}{\omega} + \frac{\pi\omega}{2\sqrt{B_1}} + \frac{ \gamma \omega}{\sqrt{B_2}}\left(\pi - \mathrm{arctan}\frac{\sqrt{B_2}}{\omega^2}\right)\Bigg\}
\end{align}
where $B_1 = \alpha^2 \beta^2 + \omega^2$ and $B_2 = \alpha^2\beta^2\gamma^2 + \omega^2(\gamma^2 + \alpha^2 \beta^2)$. Then, $\mathsf{I}_1(1)$ is evaluated by integrating over $\alpha$ using $\mathsf{I}_1(\infty)=0$. This, together with~\eqref{eq:psi3t I1}, leads to the explicit expression~\eqref{eq:non_annealed_three_time_correlator}, where we have expressed $\arctan$ in terms of $\arcsin$.
\subsubsection*{Quenched Setting:}
Evaluating the three-time correlation from~\eqref{eq:non_quenched_n_point_full} in the quenched case is more involved, as it contains several terms
\begin{align}
    \cumul{ Q(t_1)Q(t_2)Q(t_3) }_{\mathcal{Q}}& \simeq \left(\rho_a - \rho_b\right)  \int_{-\infty}^{0}\mathrm{d}z_0 \,\bigg\{\Psi_{z_0}(t_1, t_2, t_3) - \Psi_{z_0}(t_1, t_2)\Psi_{z_0}(t_3) \cr  - \Psi_{z_0}&(t_1, t_3)\Psi_{z_0}(t_2) - \Psi_{z_0}(t_2, t_3)\Psi_{z_0}(t_1) + 2\Psi_{z_0}(t_1)\Psi_{z_0}(t_2)\Psi_{z_0}(t_3)\bigg\}\label{eq:non_quenched_3_point}
\end{align}

The integral of the first term gives the correlation for the annealed case \eqref{eq:non_annealed_3_point}. To evaluate the second integral, we use~(\ref{eq:psi t1t2}, \ref{eq:Psi_1_time_explicit}) and the identity~\eqref{eq:id owenT erfc}, leading to
\begin{align}
\int_{-\infty}^{0}\mathrm{d}z_0 \Psi_{z_0}(t_1, t_2)&\Psi_{z_0}(t_3) = \frac{1}{4\pi^{3/2}}\Bigg\{\pi\left(\sqrt{t_1}+\sqrt{t_2} + 2\sqrt{t_3} - \sqrt{t_1+t_3}\right) - 2\sqrt{t_2 - t_1} \, \mathrm{arctan}\sqrt{\frac{t_3}{t_1}} \nonumber \\ & -2\sqrt{t_3} \, \mathrm{arctan}\sqrt{\frac{t_2}{t_1} - 1} - 2\sqrt{t_2 + t_3} \, \mathrm{arctan}\sqrt{\frac{t_1(t_2 + t_3)}{t_3(t_2 - t_1)}}\Bigg\}
\label{eq:second_term_3_time}
\end{align}
The third and fourth terms in \eqref{eq:non_quenched_3_point} are obtained by permuting the indices in \eqref{eq:second_term_3_time}.

The last integral in \eqref{eq:non_quenched_3_point} can be evaluated using \eqref{eq:Psi_1_time_explicit} and the identity \eqref{eq:int erf multiple}, resulting in 
\begin{align}
2\int_{-\infty}^{0}\mathrm{d}z_0 \,&\Psi_{z_0}(t_1)\Psi_{z_0}(t_2)\Psi_{z_0}(t_3) \nonumber \\ = &\frac{1}{2\pi^{3/2}}\left\{\pi\sum_{i=1}^{3}\sqrt{t_i} - \sum_{\sigma\in S_3}\sqrt{t_{\sigma(2)}+t_{\sigma(3)}}\;\arctan\!\sqrt{\frac{t_{\sigma(1)}}{t_{\sigma(2)}}+\frac{t_{\sigma(1)}}{t_{\sigma(3)}}}\right]
\label{eq:third_term_3_time}    
\end{align}

Incorporating the three results (\ref{eq:non_annealed_3_point}, \ref{eq:second_term_3_time}, \ref{eq:third_term_3_time}), we obtain the final expression~\eqref{eq:non_quenched_three_time_correlation}, where we expressed $\arctan$ in terms of $\arcsin$.

\section{A derivation of the four-time correlation function.\label{appendix:derivation_four_time_correlator}}
We take an approach similar to that used for three-time correlations in \ref{appendix:derivation_three_time_correlator}; however, the analysis is more tedious.
\subsection{Annealed Setting.}
The integral expression for the correlation~\eqref{eq:non_annealed_n_point_full} for $n=4$ is
\begin{align}\label{eq:annealed_4_point}
    \cumul{ Q(t_1)Q(t_2)Q(t_3)Q(t_4)}_{\mathcal{A}}\simeq \left(\rho_a+\rho_b\right)\int_{-\infty}^{0} \mathrm{d}z_0\,\Psi_{z_0}\left(t_1, t_2, t_3, t_4\right)
\end{align}    
The integration over $z_0$ and $z_4$ are simpler and yield an expression similar to \eqref{eq:D2}, 
\begin{align}
	\int_{-\infty}^{0} \mathrm{d}z_0\,\Psi_{z_0}\left(t_1, t_2, t_3, t_4\right) = \frac{\sqrt{\omega_1\omega_2}}{4\pi}\int_{0}^{\infty}\prod_{i=1}^{3}\mathrm{d}z_i\,\erfc{-\gamma z_3}\erfc{\beta z_1}  e^{-\omega_2(z_3 - z_2)^2 - \omega_1(z_2 - z_1)^2}
\end{align}
where we define $\gamma = \frac{1}{2\sqrt{t_4 - t_3}}$, $\omega_2 = \frac{1}{4(t_3 - t_2)}$, $\omega_1 = \frac{1}{4(t_2 - t_1)}$, and $\beta = \frac{1}{2\sqrt{t_1}}$. (These definitions are restricted to this section and differ from similar parameters in \eqref{eq:D2}.)

For the remaining integrals, we use Feynman's trick and define
\begin{align}
    \mathsf{I}_2(\alpha) = \int_{0}^{\infty}\prod_{i=1}^{3}\mathrm{d}z_i\,\erfc{-\gamma z_3}e^{-\omega_2(z_3 - z_2)^2-\omega_1(z_2 - z_1)^2}\erfc{\beta\alpha z_1}
\end{align}
such that
\begin{align}
    \int_{-\infty}^{0} \mathrm{d}z_0\,\Psi_{z_0}\left(t_1, t_2, t_3, t_4\right) = \mathsf{I}_2(1).
\end{align}
It is simpler to evaluate $\partial_{\alpha}\mathsf{I}_2(\alpha)$ using the identities~(\ref{eq:id two erf one exp},\ref{eq:id erf}, \ref{eq:identity},\ref{eq:id three erf one exp},\ref{eq:id two erf one exp}) and we obtain
\begin{align}\label{eq:derivative I4}
\frac{\mathrm{d}\mathsf{I}_2(\alpha)}{\mathrm{d}\alpha} = &-\frac{1}{2\alpha^2\beta\sqrt{\pi\omega_1\omega_2}}\Bigg\{\frac{\pi}{2} + \mathrm{arctan}\frac{\gamma\sqrt{\omega_1 + \omega_2}}{\sqrt{\omega_1\omega_2}} + \mathrm{arctan}\frac{\gamma\sqrt{\omega_2}}{\sqrt{\omega_1(\omega_2 + \gamma^2)}} + \mathrm{arctan}\sqrt{\frac{\omega_2}{\omega_1}} \Bigg\} \nonumber \\ & - \frac{1}{4\alpha^2\beta\sqrt{\pi}}\Bigg\{\pi\left(\frac{1}{\sqrt{B_2}}+\frac{1}{\sqrt{\omega_2B_1}}+\frac{\gamma}{\sqrt{B_3}}\right) + \frac{2\,}{\sqrt{\omega_2B_1}}\mathrm{arctan}\frac{\gamma}{\sqrt{\omega_2}} \nonumber \\ & + \frac{2\gamma\, }{\sqrt{B_3}}\mathrm{arctan}\frac{\omega_2\sqrt{B_1}}{\sqrt{B_3}}\Bigg\} -\frac{1}{2\sqrt{\pi}\alpha^2\beta}\Bigg\{\frac{1}{\sqrt{B_2}}\mathrm{arctan}\left(\frac{\omega_1}{\sqrt{B_2}}\right)  \nonumber \\ & + \frac{\gamma}{\sqrt{B_3}}\Bigg(\mathrm{arctan}\frac{\omega_1\omega_2^2}{(\omega_1 + \omega_2)\sqrt{B_3(\gamma^2 + \omega_2)}} + \mathrm{arctan}\frac{\omega_1\omega_2}{\sqrt{B_3(\omega_1 + \omega_2)}} \nonumber \\ & + \mathrm{arctan}\frac{\omega_1\sqrt{B_3}}{B_2\sqrt{\gamma^2 + \omega_2}}\Bigg)\Bigg\}
\end{align}
where we define the parameters $B_1 = \alpha^2\beta^2 + \omega_1$, $B_2 = \omega_1\omega_2 + \alpha^2\beta^2(\omega_1 + \omega_2)$, and $B_3 = \gamma^2\omega_1\omega_2 + \alpha^2\beta^2(\omega_1\omega_2 + \gamma^2(\omega_1 + \omega_2))$. To evaluate $\mathsf{I}_2(1)$, we integrate~\eqref{eq:derivative I4} with respect to $\alpha$, subject to the boundary condition $\mathsf{I}_2(\infty)=0$. Using properties of $\arctan$, we then obtain the explicit result for the correlation in~\eqref{eq:non_annealed_four_time_correlator}.

The result~\eqref{eq:non_annealed_four_time_correlator} can be compared with~\eqref{eq:explicit_annealed} for $n=4$ using the intensive coefficients $P_{ij}^{(4)}$ in Table~\ref{tab:pij anneal four}.

\begin{table}
\renewcommand{\arraystretch}{2.0}
\begin{center}
\begin{tabular}{cl}
\toprule
(i,j) & $P^{(4)}_{ij}$\\
\midrule
$(0,1)$ &
$\displaystyle
\frac18+\frac{1}{8}\!\left[
\As{\tfrac{u_{21}-1}{u_{31}-1}}
+\As{\tfrac{u_{21}-1}{u_{41}-1}}
+\As{\tfrac{u_{31}-1}{u_{41}-1}}
\right]$
\\
$(0,2)$, $(1,2)$ &
$\displaystyle
\frac18+\frac{1}{8}
\As{\tfrac{u_{32}-1}{u_{42}-1}}$
\\
$(0,3)$ &
$\displaystyle
\frac18+\frac{1}{8}
\As{\tfrac{u_{32}-1}{u_{31}-1}}$
\\
$(0,4)$ &
$\displaystyle
\frac18+\frac{1}{8}\!\left[
\As{\tfrac{u_{42}-1}{u_{41}-1}}
+\As{\tfrac{u_{43}-1}{u_{41}-1}}
+\As{\tfrac{u_{43}-1}{u_{42}-1}}
\right]$
\\
$(1,3)$ &
$\displaystyle
\frac18$
\\
$(1,4)$ &
$\displaystyle
\frac18+\frac{1}{8}
\As{\tfrac{(u_{21}-1)(u_{34}-1)}
{(u_{31}-1)(u_{24}-1)}}$
\\
$(2,3)$, $(2,4)$ &
$\displaystyle
\frac{1}{8}-\frac{1}{8}
\As{1 - u_{12}}$
\\
$(3,4)$ &
$\displaystyle
\frac18-\frac{1}{8}\!\left[
\As{1 - u_{13}}
+\As{1 - u_{23}}
-\As{\tfrac{u_{23}-1}{u_{13}-1}}
\right]$
\\
\bottomrule
\end{tabular}
\end{center}
\caption{Explicit expressions for $P_{ij}^{(4)}$ computed from \eqref{eq:P_ij formula} with $A(x)=\frac{2}{\pi}\arcsin\sqrt{x}$.}\label{tab:pij anneal four}
\end{table}

\subsection{Quenched Setting.}
\label{appendix:derivation_quenched_four_time_correlator}
In the quenched setting, the corresponding expression for the four-time correlation~\eqref{eq:non_quenched_n_point} for $n=4$ is written explicitly as
\begin{align}
    \cumul{ Q(t_1)Q(t_2)Q(t_3)Q(t_4) }_{\mathcal{Q}} \simeq \left(\rho_a + \rho_b\right)  \int_{-\infty}^{0}\mathrm{d}z_0 \,\Phi_{z_0}\left(t_1, t_2, t_3, t_4\right)\label{eq:non_quenched_4_point}
\end{align}
where $\Phi_{z_0}$ is expressed in terms of $\Psi_{z_0}$ and involves $15$ different combinations. For compact notation, we denote $\overline{S}\equiv\{1,2,3,4\}\setminus S$ and write
\begin{align}
\Phi_{z_0}(t_1,t_2,t_3,t_4)=&\Psi(t_1,t_2,t_3,t_4)-\sum_{i=1}^{4}\Psi_{z_0}(t_i)\,\Psi_{z_0}\left(t_{\overline{\{i\}}}\right)-\sum_{j=2}^{4}\Psi_{z_0}(t_1,t_j)\,\Psi_{z_0}\left(t_{\overline{\{1,j\}}}\right)\cr
&+2\sum_{1\le i<j\le 4}\Psi_{z_0}(t_i,t_j)\,\Psi_{z_0}(t_k)\,\Psi_{z_0}(t_l)\;-\;6\prod_{i=1}^{4}\Psi_{z_0}(t_i)
\end{align}
where in the fourth term $\{k,l\}=\overline{\{i,j\}}$.

The integral of the first term gives the corresponding annealed correlation~\eqref{eq:annealed_4_point}, and the remaining integrals can be evaluated following similar methods. This part of the calculation is tedious, and we only provide the final answer, which is expressed in the predicted form~\eqref{eq:explicit_quenched}, with $D^{(4)}_{0j}=0$. For a closed-form expression, we write
\begin{align}
\cumul{ Q(t_1)\cdots Q(t_4)}_{\mathcal{Q}}\simeq\;
\frac{\left(\rho_a+\rho_b\right)}{8\sqrt{\pi}}
\sum_{1\le i<j\le 4}\left( \widehat{S}_{ij}\,\sqrt{t_i+t_j}-\widehat{D}_{ij}\,\sqrt{t_j-t_i}
\right)
\end{align}
where we define $D^{(4)}_{ij}=-\frac{1}{8}\widehat{D}_{ij}$ and $S^{(4)}_{ij}=\frac{1}{8}\widehat{S}_{ij}$. These  functions have a scaling dependence on time via $A(x)=\frac{2}{\pi}\arcsin\sqrt{x}$. The function $\widehat{S}_{ij}$ has the following expression
\begin{align}
\widehat{S}_{12}&=R_{1234}+\mathcal W_{2324}+\mathcal W_{1314}\\
\widehat{S}_{13}&=R_{1324}+\mathcal W_{1214}-2\,\mathcal M_{1324},\qquad 
\widehat{S}_{23}=R_{2314}-2\,\mathcal M_{3214}-2\,\mathcal M_{2314}\\
\widehat{S}_{14}&=R_{1423}+\mathcal W_{1213}+\As{\tfrac{(t_1+t_2)(t_4-t_3)}{(t_1+t_3)(t_4-t_2)}}
-2\,\mathcal M_{1432}-2\,\mathcal M_{1423}\\
\widehat{S}_{24}&=R_{2413}+\As{\tfrac{(t_2-t_1)(t_4-t_3)}{(t_1+t_4)(t_2+t_3)}}
+\As{\tfrac{(t_1+t_2)(t_4-t_3)}{(t_4-t_1)(t_2+t_3)}}
-2\,\mathcal M_{2431}-2\,\mathcal M_{4213}-2\,\mathcal M_{2413}\\
\widehat{S}_{34}&=R_{3412}
+\As{\tfrac{(t_1+t_4)(t_3-t_2)}{(t_3-t_1)(t_2+t_4)}}
+\As{\tfrac{(t_4-t_1)(t_3-t_2)}{(t_1+t_3)(t_2+t_4)}}
+\As{\tfrac{(t_3-t_1)(t_4-t_2)}{(t_1+t_4)(t_2+t_3)}}
+\As{\tfrac{(t_1+t_3)(t_4-t_2)}{(t_4-t_1)(t_2+t_3)}}\nonumber\\
&\qquad-2\,\mathcal M_{4321}-2\,\mathcal M_{3421}-2\,\mathcal M_{4312}-2\,\mathcal M_{3412}
\end{align}
while $\widehat{D}_{ij}$ has the following expression:
\begin{align}
\widehat{D}_{12}&=E_{134}+\mathcal W_{2324} &
\widehat{D}_{13}&=E_{124}\\
\widehat{D}_{23}&=E_{214}-\mathcal V_{124} &
\widehat{D}_{24}&=E_{213}-\mathcal V_{123}\\
\widehat{D}_{14}&=E_{123}+\As{\tfrac{(t_2-t_1)(t_4-t_3)}{(t_3-t_1)(t_4-t_2)}} &
\widehat{D}_{34}&=E_{312}+\mathcal W_{2313}-\mathcal V_{231}-\mathcal V_{132}
\end{align}
Here we define
\begin{align}
E_{ikl}=2\,\mathcal H_{kl}(t_i)-\mathcal U_{kl}(t_i),\qquad
R_{ijkl}=6\,\mathcal H_{kl}\!\Big(\tfrac{t_it_j}{t_i+t_j}\Big)-2\,\mathcal U_{kl}\!\Big(\tfrac{t_it_j}{t_i+t_j}\Big)
\end{align}
in terms of functions 
\begin{align}
\mathcal H_{kl}(s)&=\As{\tfrac{s^{2}}{(s+t_k)(s+t_l)}} &
\mathcal U_{kl}(s)&=\As{\tfrac{s+t_k}{s+t_l}} && \\
\mathcal V_{abc}&=\As{\tfrac{t_b-t_a}{t_b+t_c}} &
\mathcal W_{abcd}&=\As{\tfrac{t_b-t_a}{t_d-t_c}} && \\
\mathcal M_{pqrl}&=\As{\tfrac{t_p^{2}(t_q-t_r)}{(t_p+t_r)\,e_2(t_p,t_q,t_l)}} &
e_2(a,b,c)&=ab+ac+bc & &\nonumber
\end{align}
We have provided some of the identities in \ref{app:identity} that are useful for this analysis.

We recall that the integral expressions coincide, up to numerical prefactors, with the expression for the four-time correlation of the tracer~\cite{Benichou_multi_2025} in the dense limit of the SSEP. The final explicit result is also reported in~\cite{Benichou_multi_2025}, but in terms of $\arctan$, and agrees with our result.  

\section{Derivation of~\texorpdfstring{\eqref{eq:quenched_two_time_correlation_general_sigma_constant_D_intro}}{Eq. (8)} from~\texorpdfstring{\eqref{eq:quenched_two_time_intermediate}}{Eq. (79)} }\label{appendix:quenched_two_time_integrated_current_correlation_MFT}   

Substituting $q_0(z, \tau)$ from~\eqref{eq: density profile} into~\eqref{eq:quenched_two_time_intermediate} and expanding in small $\Delta \rho$, we write
\begin{equation}
    C_{\mathcal{Q}}(\tau_1, \tau_2) \simeq \mathbf{I}_0+\Delta \rho  \, \mathbf{I}_1+(\Delta \rho )^2 \mathbf{I}_2 + \cdots
    \label{eq:1st_2nd_3rd_parts}
\end{equation}
where
\begin{align}
	\mathbf{I}_n = \frac{\sigma^{(n)}\left(\bar{\rho}\right)}{2^n n!}\int_{0}^{\tau_1} \mathrm{d}\tau \int_{-\infty}^{\infty} \mathrm{d}z \, \erf{\frac{z}{2\sqrt{ \tau}}}^n g(0, \tau_1\mid z, \tau) g(0, \tau_2\mid z, \tau)
\end{align}	
These integrals can be explicitly evaluated explicitly. The simplest one is
\begin{align}
    \mathbf{I}_0 & = \frac{\sigma\left(\bar{\rho}\right)}{2\sqrt{\,\pi}}\left(\sqrt{\tau_2 + \tau_1} - \sqrt{\tau_2 - \tau_1}\right) \qquad \textrm{for $\tau_2>\tau_1$.}\label{eq:first part 2}
\end{align}
This contributes the correlation for the uniform initial state. The next term, \(\mathbf{I}_1\), vanishes because its integrand is an odd function of $z$. For the quadratic-order term \(\mathbf{I}_2\), the  $z$ integration is computed using the identity~\eqref{eq:id two erf one exp}, 
\begin{align}
    \mathbf{I}_2 = \frac{\sigma^{\prime\prime}\left(\bar{\rho}\right)}{8\pi^{3/2}}\int_{0}^{\tau_1} \frac{\mathrm{d}\tau}{2\sqrt{\tau_1 + \tau_2 - 2\tau}} & \left\{\pi - 4\mathrm{arctan}\left(\sqrt{\frac{\tau(\tau_1 + \tau_2 - 2\tau)}{2\tau_1\tau_2 - \tau(\tau_1 + \tau_2)}}\right)\right\}
    \label{eq:third part 2}
\end{align}
which is further reduced to the $(\Delta \rho)^2$ term in~\eqref{eq:quenched_two_time_correlation_general_sigma_constant_D_intro}. 

\section{Simplification of \texorpdfstring{\eqref{eq:identity interacting particles MFT annealed}}{Eq. (85)} to \texorpdfstring{\eqref{eq:identity interacting particles MFT annealed 2}}{Eq. (86)}}\label{appendix: derivation identity interacting particles MFT annealed}
We take the second term from the expression~\eqref{eq:identity interacting particles MFT annealed} 
\begin{align}
    \mathcal{L} = \int_{0}^{1} \mathrm{d}\tau \int_{-\infty}^{\infty} \mathrm{d}x \, \Lambda(\tau) \Theta(x) (q_h(x,\tau) - q_h(x,0))
    \label{eq:L}
\end{align}
and show that it equals to $2\mathcal{F}_2$ in \eqref{eq:F_2_annealed}. Using $q_h(x, 0)$ from the boundary condition~\eqref{eq:qh_boundary_condition}, $\mathcal{L}$ can be rewritten as    
\begin{align}
    \mathcal{L}  = 2\int_{-\infty}^{\infty} \mathrm{d}x \, \frac{(q_h(x, 0))^2}{\sigma(q_0(x,0))} - \mathcal{V} = 2\mathcal{F}_2(r, q_h) - \mathcal{V}\label{eq:L_rewritten} 
\end{align}   
where we have used \eqref{eq:F_2_annealed} and defined
\begin{align}
    \mathcal{V} = - \int_{0}^{1} \mathrm{d}\tau \int_{-\infty}^{\infty} \mathrm{d}x \, H(\tau) \Theta(x) q_h(x, \tau) + \int_{-\infty}^{\infty} \mathrm{d}x \, p_1(x,0) q_h(x,0) \label{eq:V}
\end{align}
We shall show that $\mathcal{V}=0$. For this, we use $H(\tau)\Theta(x)$ from the optimal equation~\eqref{eq:optimal_p1} and write
\begin{align}
    \mathcal{V} = \int_{0}^{1} \mathrm{d}\tau \int_{-\infty}^{\infty} \mathrm{d}x \, \left( \partial_{\tau} p_1(x, \tau) +  \partial_{x}^2 p_1(x, \tau) \right) q_h(x, \tau) + \int_{-\infty}^{\infty} \mathrm{d}x \, p_1(x, 0) q_h(x,0) \label{eq:V_substituted}
\end{align}
Integrating by parts with respect to $\tau$ in the first term and using the boundary condition \( p_1(x, 1) = 0 \), we write
\begin{align}
    \mathcal{V}= & - \int_{0}^{1} \mathrm{d}\tau \int_{-\infty}^{\infty} \mathrm{d}x \, \left(\partial_{\tau} q_h(x, \tau) p_1(x, \tau) -  \partial_{x}^2 p_1(x, \tau) q_h(x, \tau) \right)
\label{eq:V_simplified}
\end{align}
Inserting \(\partial_{\tau} q_h(x,\tau)\) from the optimal equation~\eqref{eq:optimal_q1} and simplifying using \(\partial_{x}^2 q_h(x, \tau) p_1(x, \tau) - \partial_{x}^2 p_1(x, \tau)q_h(x, \tau) = \partial_x \left( \partial_{x} q_h(x, \tau) p_1(x, \tau) - \partial_{x} p_1(x, \tau) q_h(x, \tau) \right)\), the above expression reduces to        
\begin{align}
    \mathcal{V} = - \int_{0}^{1} \mathrm{d}\tau \int_{-\infty}^{\infty} \mathrm{d}x \, \partial_x \left(\partial_{x} q_h(x, \tau) p_1(x, \tau) - \partial_{x} p_1(x, \tau) q_h(x, \tau) \right)
     \label{eq:V_final}
\end{align}
which vanishes since \( p_1(x, \tau) \) and \( q_h(x, \tau) \) decay to zero as \( x \to \pm \infty \). The resulting equation~\eqref{eq:L_rewritten} gives $\mathcal{L}=2\mathcal{F}_2$, which reduces \eqref{eq:identity interacting particles MFT annealed} to \eqref{eq:identity interacting particles MFT annealed 2}.
    
\section{Two-Time Current Correlation for SSEP}\label{appendix: annealed two time integrated current correlation micro}

The two-time correlation $\cumul{ Q_0(t_1)Q_0(t_2) }_{\text{evo}}$ in~\eqref{eq:QQ_unequal_time_1} depends on $F_{0,0}(t,s)$ and $\Gamma_0(t)$. From~\eqref{eq:F_i_j_1} we write
\begin{align}
    F_{0,0}(t,s) = F^{(1)}_{0, 0}(t,s) + F^{(2)}_{0, 0}(t,s),
    \label{eq:F_0_0}
\end{align}
where 
\begin{align}
    F^{(1)}_{0, 0}(t, s) &= \sum_{k = 1}^{\infty}\big(I_{k+1}(2(t-s)) - I_k(2(t-s))\big)e^{-2(t-s)}\big(A_{k+1,0}(s) - A_{-k,0}(s)\big)
    \label{eq:F_1_0_0}\\
    F^{(2)}_{0, 0}(t, s) &= \big(I_0(2(t-s)) - I_1(2(t-s))\big)e^{-2(t-s)}(A_{0,0}(s) - A_{1,0}(s))
    \label{eq:F_2_0_0}
\end{align}

To compare with the results obtained from MFT in the large-$T$ limit, we introduce the scaled variables $x = \tfrac{k}{\sqrt{T}}$, $y = \tfrac{j}{\sqrt{T}}$, $\alpha = \tfrac{s}{T}$, and $\tau = \tfrac{t}{T}$, and then take the large-$T$ limit. The leading-order behavior of $\Gamma_0(t)$ from~\eqref{eq:Gamma_i_t} in scaled coordinates is   
\begin{align}
    \Gamma_0(t) \simeq \Gamma( T\tau) = 2\bar{\rho}(1 - \bar{\rho})
    \label{eq:Gamma_0_asymptotics}
\end{align}
and this gives the first term on the right-hand side of~\eqref{eq:QQ_unequal_time_1},
\begin{align}
    T\int_{0}^{\tau_1} \mathrm{d}\tau \, \Gamma_0(T \tau) \simeq 2 \bar{\rho} (1 - \bar{\rho})T \tau_1.
    \label{eq:third_integral}
\end{align}
Similarly, we obtain the asymptotic form of $ A_{k,0}(s)$ in the large-$T$ limit,
 \begin{align}
     A_{k,0}(s) \simeq \begin{cases}\frac{(\Delta \rho)^2}{4\sqrt{2\pi} \, \alpha T}\left(\frac{\mathrm{sgn}(x)}{\sqrt{2\pi}}e^{-\frac{x^2}{4\alpha}} - \frac{x}{4\sqrt{\alpha}} e^{-\frac{x^2}{8\alpha}}\erfc{\frac{\vert x \vert}{2\sqrt{2\alpha}}}\right) & k \neq 0,1 \\ (- 1)^{k+1}\bar{\rho}(1 - \bar{\rho}) & k = 0, 1
     \end{cases} 
  \label{eq:A_k_0_asymptotics}
 \end{align}
from~\eqref{eq:A_i_j_t_1} using~(\ref{eq:average occupation SSEP},\ref{eq:tau_tau_equal_time_SSEP}) while from~\eqref{eq:ssep_micro_green} we get
\begin{align}
    g_{k,j}(t) \simeq \frac{1}{\sqrt{4\pi\tau T}}\exp\left(-\frac{(x - y)^2}{4\tau}\right).
    \label{eq:gtilde_asymptotics}
\end{align}
Substituting the asymptotic forms of $A_{k,0}(s)$ and $g_{i,j}(t)$ into~\eqref{eq:F_1_0_0}, and completing the summation over $k$ using an integral approximation we get
\begin{align}
    &F^{(1)}_{0, 0}(t, s) \simeq \frac{(\Delta \rho)^2}{4 \pi^{3/2} T^{3/2}} \left\{-\left(\frac{1}{\sqrt{\tau - \alpha} \left(\tau + \alpha\right)}\right) + \frac{\arctan\left(\sqrt{\frac{\tau + \alpha}{\tau - \alpha}}\right)}{\left(\tau + \alpha\right)^{3/2}} \right\}
\label{eq:F_1_0_0_asymptotics}
\end{align}
The second and the third terms of~\eqref{eq:QQ_unequal_time_1} can be computed using the asymptotic form of $F^{(1)}_{0,0}(t,s)$ as follows
\begin{align}
	2 T^2 & \int_{0}^{\tau_1} \mathrm{d}\tau \int_{0}^{\tau} \mathrm{d}\alpha \, F^{(1)}_{0,0}(\tau T, \alpha T) + T^2 \int_{\tau_1}^{\tau_2} \mathrm{d}\tau \int_{0}^{\tau_1} \mathrm{d}\alpha \, F^{(1)}_{0,0}(\tau T, \alpha T) \simeq \frac{(\Delta \rho)^2 \sqrt{T} }{4 \pi^{3/2}} \nonumber \\ &\Bigg\{ \pi(\sqrt{\tau_1} + \sqrt{\tau_2}) - 4 \sqrt{\tau_1} \arctan{\left( \sqrt{\frac{\tau_2 - \tau_1}{\tau_1}} \right)} - 4 \sqrt{\tau_1 + \tau_2} \arctan{\left( \sqrt{\frac{\tau_2 + \tau_1}{\tau_2 - \tau_1}} \right)} \Bigg\}
	\label{eq:first_integral}
\end{align}
Similarly, for the sums involving $F^{(2)}_{0,0}(t,s)$, we get
\begin{align}
2 T^2 \int_{0}^{\tau_1} \mathrm{d}\tau &\int_{0}^{\tau} \mathrm{d}\alpha \, F^{(2)}_{0,0}(\tau T, \alpha T) + T^2 \int_{\tau_1}^{\tau_2} \mathrm{d}\tau \int_{0}^{\tau_1} \mathrm{d}\alpha \, F^{(2)}_{0,0}(\tau T, \alpha T) \nonumber \\ & \simeq -2 \bar{\rho} (1 - \bar{\rho}) \tau_1 T + \frac{\bar{\rho} (1 - \bar{\rho})\sqrt{T}}{\sqrt{\pi}}\left( \sqrt{\tau_1} + \sqrt{\tau_2} - \sqrt{\tau_2 - \tau_1} \right)
\label{eq:second_integral}
\end{align}

Adding the results~\eqref{eq:first_integral}, \eqref{eq:second_integral}, and~\eqref{eq:third_integral}, we recover the expression~\eqref{eq:annealed_two_time_correlation_general_sigma_constant_D_intro} for the SSEP, where \( \tau_1 T=t_1 \) and \( \tau_2 T=t_2 \).

\section{Derivation of~\texorpdfstring{\eqref{eq:annealed_quenched_difference_SSEP}}{Eq. (127)}}\label{appendix: quenched two time integrated current correlation micro}
Using~\eqref{eq:avg_curr_SSEP} in~\eqref{eq:annealed_quenched_diff}, we get
\begin{align}
\overline{\langle Q_0(t_1)\rangle_{\text{evo}} \langle Q_0(t_2) \rangle_{\text{evo}}} & - \overline{\langle Q_0(t_1)\rangle_{\text{evo}} }  \, \, \, \overline{\langle Q_0(t_2)\rangle_{\text{evo}} } \nonumber \\ =
\int_0^{t_1} \mathrm{d}t \int_0^{t_2}& \mathrm{d}s \sum_{k,m = -\infty}^{\infty} \left[ g_{0,k}(t) - g_{1,k}(t) \right] \left[ g_{0,m}(s) - g_{1,m}(s) \right]~\overline{\overline{n_m(0) n_k(0)}}, \label{eq:additional_term}
\end{align}
where the \emph{connected} correlation of the occupation variables at the initial time is
\begin{align}
    \overline{\overline{n_m(0) n_k(0)}}=\rho_m(1-\rho_m)\delta_{m,k}, \quad \text{with} \quad \rho_m = \begin{cases} \rho_a, & m \leq 0, \\ \rho_b, & m > 0. \end{cases} \label{eq:initial_correlation}
\end{align}
Substituting the above expression and $g_{j,k}(t)$ from~\eqref{eq:ssep_micro_green} into~\eqref{eq:additional_term} and applying the Bessel-function identity \(\sum_{k} I_{k + \alpha}(x) I_{k}(y) = I_{\alpha}(x + y)\), we get
\begin{align}
     &\overline{\langle Q_0(t_1) \rangle \langle Q_0(t_2) \rangle} - \overline{\langle Q_0(t_1) \rangle }  \, \overline{\langle Q_0(t_2) \rangle }\nonumber \\ & = (\rho_a(1 - \rho_a) + \rho_b(1 - \rho_b))\int_{0}^{t_1} \mathrm{d}t \int_{0}^{t_2} \mathrm{d}s \, \mathrm{e}^{-2(t + s)} \left( I_0(2(t + s)) - I_1(2(t + s)) \right)\label{eq:covariance_bessel}
\end{align}
Integrating over $t$ and $s$, we obtain
\begin{align}
    \overline{\langle Q_0(t_1) \rangle \langle Q_0(t_2) \rangle} & - \overline{\langle Q_0(t_1) \rangle }  \, \overline{\langle Q_0(t_2) \rangle }\nonumber \\ & = \frac{1}{4}\left(\rho_a(1 - \rho_a) + \rho_b(1 - \rho_b)\right)\left( \zeta(2 t_1) + \zeta(2 t_2)  - \zeta(2 (t_1 + t_2)) \right)\label{eq:covariance_f}    
\end{align}
Writing in terms of scaled coordinates $t=\tau\,T$, and using $\rho_a = \bar{\rho} + \tfrac{\Delta \rho}{ 2}$, $\rho_b = \bar{\rho} - \tfrac{\Delta \rho}{ 2}$, gives the expression~\eqref{eq:annealed_quenched_difference_SSEP}.

\section{Derivation of the Density-Current Correlation~\texorpdfstring{\eqref{eq:a_density_current}}{Eq. (128)}}\label{appendix:density_current_correlation}

Similarly to the computation of the current correlation, we separate the terms in the summation in \eqref{eq:tauQ_equal_time} for $k = 0$ and $1$, and write $\cumul{ n_j(t) Q_0(t) }_{\text{evo}} = \mathbb{I}_1(j, t) + \mathbb{I}_2(j, t)$, where
\begin{align}
    \mathbb{I}_1(j, t) &= \int_{0}^{t} \mathrm{d}s \, \big( g_{j,0}(t - s) A_{0,0}(s) + g_{j,1}(t - s) A_{1,0}(s) \big) \label{eq:c_j_1}\\
 \text{and} \quad  \mathbb{I}_2(j, t) &=
\sum_{k = 1}^{\infty} \int_{0}^{t} \mathrm{d}s \, \big( g_{j,-k}(t - s) A_{-k,0}(s) + g_{j,k+1}(t - s) A_{k+1,0}(s) \big) \label{eq:c_j_2}
\end{align}
To calculate $\mathbb{I}_1(j, t)$ in the large-$T$ limit,  
we substitute the asymptotic form $A_{0, 0}(s)$ and $A_{1, 0}(s)$ from~\eqref{eq:A_k_0_asymptotics} and $g_{j, k}(t - s)$ from~\eqref{eq:gtilde_asymptotics} into~\eqref{eq:c_j_1}, resulting in
\begin{align}
    \mathbb{I}_1(j, t) &\simeq \frac{\bar{\rho}(1 - \bar{\rho})y}{4\sqrt{\pi}}\int_{0}^{\tau}\mathrm{d}\alpha\, \frac{1}{(\tau - \alpha)^{3/2}}\exp\left(-\frac{y^2}{4(\tau - \alpha)}\right)\cr 
    &\simeq \frac{\bar{\rho} (1 - \bar{\rho})}{2}\operatorname{sgn}(y)\Erfc{ \frac{|y|}{2 \sqrt{\tau}} }\label{eq:J3}
\end{align}
with $j=y\sqrt{T}$ and $t=\tau T$. Similarly,
\begin{align}
    \mathbb{I}_2(j, t) &\simeq - \frac{(\Delta \rho)^2}{16\pi}\Bigg\{\frac{1}{\sqrt{\pi}} \int_{0}^{\tau} \mathrm{d}\alpha\, \frac{1}{\alpha \sqrt{\tau - \alpha}} \int_{0}^{\infty} \mathrm{d}x \, e^{ - \frac{x^2}{4 \alpha}} \left(e^{-\frac{(x + y)^2}{4 (\tau - \alpha)}} - e^{-\frac{(x - y)^2}{4 (\tau - \alpha)} } \right) \nonumber \\  + & \frac{1}{2 \sqrt{2}} \int_{0}^{\tau} \mathrm{d}\alpha\,\frac{1}{\alpha^{3/2}\sqrt{\tau - \alpha}} \int_{0}^{\infty} \mathrm{d}x \, x \Erfc{\frac{x}{2\sqrt{2 \alpha}}} e^{-\frac{x^2}{8\alpha}}\left(e^{-\frac{(x-y)^2}{4(\tau - \alpha)}} - e^{-\frac{(x+y)^2}{4(\tau - \alpha)}}\right)\Bigg\}
\label{eq:I_2}
\end{align}
In the first term in~\eqref{eq:I_2} the $x$ integral is Gaussian and is easy to compute. The remaining $\alpha$ integral can be performed using the identity
\begin{equation}
    \int_0^\infty \mathrm{d}t \frac{\erf{zt}}{(1+t^2)^{3/2}}=1-e^{z^2}\erfc{z}\
\end{equation}
For the second term, the integral over $x$ is evaluated by integration by parts and using the identities (\ref{eq:identity},\ref{eq:Owen_identity}). Combining them, we get
\begin{align}
\mathbb{I}_2(j, t) \simeq &-\frac{(\Delta \rho)^2}{4 \pi} \Bigg\{ \operatorname{sgn}(y)\left(1 - e^{- \frac{y^2}{4 \tau}}\right) - \, \Erf{ \frac{y}{2\sqrt{\tau}}} \nonumber \\ & + \frac{1}{2} \int_{0}^{\tau} \mathrm{d}\alpha \, \Bigg(\frac{ (\alpha - \tau) \, }{  \sqrt{ \alpha \tau} (\tau + \alpha)}e^{-\frac{y^2}{4 \tau}} \, \Erf{\frac{y \sqrt{\alpha}}{2 \sqrt{\tau ( \tau - \alpha)}}}\nonumber \\ & + \frac{y\sqrt{\pi} e^{ - \frac{y^2}{4 (\tau + \alpha)}} }{(\tau + \alpha)^{3/2}} 4 T_{\text{Owen}}\left( \frac{y \sqrt{\alpha}}{\sqrt{2 \tau (\tau + \alpha)}}, \sqrt{\frac{\tau + \alpha}{\tau - \alpha}} \right)  \Bigg)\Bigg\}
 \label{eq:I_2_final_mid}
\end{align}
with Owen's T function~\cite{Owen1980}
\begin{align}
	T_{\text{Owen}}(h, a) = \frac{1}{2 \pi}\int_{0}^{a} \mathrm{d}x \, \frac{e^{-\frac{1}{2} h^2 (1 + x^2)}}{1 + x^2}
	\label{eq:Owen_T}
\end{align}

This cumbersome expression~\eqref{eq:I_2_final_mid} reduces to a surprisingly simple form
\begin{align}
\mathbb{I}_2(j, t) \simeq &-\frac{(\Delta \rho)^2}{8}\operatorname{sgn}(y)\left\{\erfc{\frac{\vert y \vert}{2\sqrt{2\tau}}}^2-\erfc{\frac{\vert y \vert }{2\sqrt{\tau}}} \right\}
 \label{eq:I_2_final}
\end{align}
using the property~\eqref{eq:Owen_identity} of Owen's T function.

Adding the expressions \eqref{eq:J3} and \eqref{eq:I_2_final} and writing them in terms of the microscopic coordinates using $y = \tfrac{j}{\sqrt{T}}$ and $\tau = \tfrac{t}{T}$, we get the result~\eqref{eq:a_density_current}.

\section{Derivation of~\texorpdfstring{\eqref{eq:q_density_current}}{Eq. (Q)}}\label{appendix: density current AA_QQ diff}
Using the averages in~\eqref{eq:avg_curr_SSEP} we write
\begin{align}
    \overline{\langle n_j(t) \rangle_{\text{evo}}\langle Q_0(t) \rangle_{\text{evo}}} - \overline{\langle n_j(t) \rangle_{\text{evo}}}\,\overline{\langle Q_0(t) \rangle_{\text{evo}}} = \!\!\sum_{k,l}\!\! \int_{0}^{t} \!\!\!\!\mathrm{d}s \, &g_{j,k}(t) \big(g_{0,l}(s) - g_{1,l}(s) \big)\overline{ \overline{n_{k}(0) n_{l}(0)}}
\label{eq:covariance_tauQ}
\end{align}
Using~\eqref{eq:initial_correlation} and the asymptotic form of $g_{j, k}(t)$ from~\eqref{eq:gtilde_asymptotics} we write, in terms of the continuous variables $y = \tfrac{j}{\sqrt{T}}$, $\alpha = \tfrac{s}{T}$, and $\tau = \tfrac{t}{T}$, for large $T$,
\begin{align}
    \overline{\langle n_j(t) \rangle \langle Q_0(t) \rangle} - \overline{\langle n_j(t) \rangle} \, \overline{\langle Q_0(t) \rangle} \simeq & -\frac{1}{8 \pi \sqrt{\tau}}\Bigg(\rho_{a}(1 - \rho_{a})\int_{-\infty}^{0} \mathrm{d}x\, x e^{-\frac{(y-x)^2}{4\tau}} \int_{0}^{\tau}\mathrm{d}\alpha \, \frac{e^{-\frac{x^2}{4\alpha}}}{\alpha^{3/2}} \nonumber \\ & + \rho_{b} (1 - \rho_{b}) \int_{0}^{\infty} \mathrm{d}x \, x e^{-\frac{(y-x)^2}{4\tau}} \int_{0}^{\tau}\mathrm{d}\alpha \, \frac{e^{-\frac{x^2}{4\alpha}}}{\alpha^{3/2}}\Bigg)
    \label{eq:covariance_split_rescaled}
\end{align}
Performing the integral over $\alpha$ and using the identity~\eqref{eq:id one exp one erf} for the $x$ integral, we obtain~\eqref{eq:q_density_current}.

\section{A list of useful identities}\label{app:identity}
Some of the following identities can be found in Refs.~\cite{Owen1980,Brychkov2016, korotkov2020integrals,Prudnikov1986}. For these identities, we assume $\alpha, \beta, \gamma, a, a_1, a_2, a_3$, and $b>0$. 
\begin{align}\label{eq:int erf multiple}
\int_{0}^{\infty}\mathrm{d}z\,\prod_{i=1}^{3}\Erfc{\alpha_i z} = \frac{1}{\sqrt{\pi}}\sum_{i=1}^{3}\frac{1}{\alpha_i} - \frac{1}{\pi^{3/2}}\sum_{\sigma \in S_3}\frac{\sqrt{\alpha_{\sigma_1}^2 + \alpha_{\sigma_2}^2}}{\alpha_{\sigma_1}\alpha_{\sigma_2}}\mathrm{arctan}\frac{\sqrt{\alpha_{\sigma_1}^2 + \alpha_{\sigma_2}^2}}{\alpha_{\sigma_3}}
\end{align}
\begin{subequations}
\begin{align}
\int_{0}^{\infty}\mathrm{d}z\,\prod_{i=1}^{4}\Erfc{\alpha_i z} = \frac{1}{\sqrt{\pi}}\sum_{i=1}^{4}\frac{1}{\alpha_i} + \sum_{\sigma \in S_4}\!\left(\frac{1}{2}f_2(\alpha_{\sigma_1}, \alpha_{\sigma_2}, \alpha_{\sigma_3}, \alpha_{\sigma_4}) - f_1(\alpha_{\sigma_1}, \alpha_{\sigma_2}, \alpha_{\sigma_3})\!\right)
\end{align}
\begin{align}
   \text{where} \quad f_1(\alpha, \beta, \gamma) &= \frac{\sqrt{\alpha^2 + \beta^2}}{\pi^{3/2}\alpha\beta}\mathrm{arctan}\frac{\sqrt{\alpha^2 + \beta^2}}{\gamma}\\
   \text{and} \quad f_2(\alpha, \beta, \gamma, \delta) & = \frac{\sqrt{\alpha^2 + \beta^2}}{\pi^{3/2}\alpha\beta}\mathrm{arctan}\frac{\sqrt{\alpha^2 + \beta^2}\sqrt{\alpha^2 + \beta^2 + \gamma^2 + \delta^2}}{\gamma\delta}
\end{align}    
\end{subequations}
\begin{align}\label{eq:id erf}
    \int_{0}^{\infty} \mathrm{d}x\, x \exp{(-a x^2)} \prod_{i=1}^{2}&\Erf{a_ix}  = \frac{1}{\pi a }\sum_{\sigma \in S_2}\frac{a_{\sigma_1}}{\sqrt{a + a_{\sigma_1}^2}}\arctan\frac{a_{\sigma_2}}{\sqrt{a + a_{\sigma_1}^2}} 
\end{align}	
\begin{align}
\int_{0}^{\infty}\mathrm{d}x\, \exp\left(-ax^2\right)\prod_{i=1}^{2}\erf{a_ix} = \frac{1}{\sqrt{\pi a}}\mathrm{arctan}\frac{a_1a_2}{\sqrt{a^2 + aa_1^2 + aa_2^2}}
	\label{eq:id two erf one exp}
\end{align}
\begin{align}\label{eq:id two erfc}
\int_{0}^{\infty} \mathrm{d}x \, \Erfc{a x}\Erfc{b x} = \frac{a + b - \sqrt{a^2 + b^2}}{\sqrt{\pi}\, a b} 
\end{align}
\begin{align}\label{eq:id three erf one exp}
\int_{0}^{\infty}\mathrm{d}x\,x&\exp\left(-bx^2\right)\prod_{i=1}^3\erf{a_ix} = \frac{1}{2\pi b}\sum_{\sigma \in S_3}^{}\frac{a_{\sigma_1}}{\sqrt{b + a_{\sigma_1}^2}}\mathrm{arctan}\frac{a_{\sigma_2}a_{\sigma_3}}{\sqrt{(b+a_{\sigma_1}^2)(b + a_{\sigma_1}^2 + a_{\sigma_2}^2 + a_{\sigma_3}^2)}}
\end{align}
\begin{align}
	\int_{0}^{\infty} \mathrm{d}x \, e^{- ax - b^2x^2}\Erf{b x} = \frac{\sqrt{\pi}}{4b}\exp\left(\frac{a^2}{4b^2}\right)\left(\Erfc{\frac{a}{2\sqrt{2}b}}\right)^2
	\label{eq:id one exp one erf} 
\end{align}
 \begin{align}
 	\int_{0}^{\infty} \mathrm{d}x\, \Erf{ax + b} e^{-x^2} = \frac{\sqrt{\pi}}{2}\Bigg\{1+&\erf{\frac{b}{\sqrt{1+a^2}}}\erfc{\frac{ab}{\sqrt{1+a^2}}} - 4T_{\text{Owen}}\left(\frac{\sqrt{2}ab}{\sqrt{1+a^2}},\frac{1}{a}\right) \Bigg\}
 	\label{eq:identity}
\end{align}
\begin{align}\label{eq:id TW}
\int_{0}^{\infty}\mathrm{d}z\, T_{Owen}\left(\sqrt{2}\alpha z, a\right) = \frac{a}{4\alpha\sqrt{\pi(1+a^2)}}
\end{align}
\begin{align}\label{eq:id owenT erfc}
\int_{0}^{\infty}\mathrm{d}z\,T_{\text{Owen}}\left(\sqrt{2}\alpha z, a\right)\Erfc{\beta z} = &\frac{\mathrm{arctan}\left(a\right)}{2\pi^{3/2}\beta} + \frac{a}{2\pi^{3/2}\alpha\sqrt{(1 + a^2)}}\mathrm{arctan}\frac{\alpha\sqrt{1 + a^2}}{\beta} \nonumber \\ & - \frac{\sqrt{\alpha^2 + \beta^2}}{2\pi^{3/2}\alpha\beta}\mathrm{arctan}\frac{a\alpha}{\sqrt{\alpha^2 + \beta^2}} 
\end{align}
\begin{subequations}
\begin{align}    \int_{0}^{\infty}\mathrm{d}z\,T_{\text{Owen}}\left(\sqrt{2}\alpha z, a\right)\Erfc{\beta z}\Erfc{\gamma z} = & \frac{1}{2\pi^{3/2}}\Bigg\{g_0(\alpha, \beta, \gamma, a) + g_0(\alpha, \gamma, \beta, a) \nonumber \\ & - \frac{\sqrt{\beta^2 + \gamma^2}}{\beta\gamma}\mathrm{arctan}\frac{a\sqrt{\beta^2 + \gamma^2}}{\sqrt{B_4}}\Bigg\}
\end{align}

\begin{align}
    \text{where} \quad g_0(\alpha, \beta, \gamma, a) = &\frac{a}{\alpha\sqrt{1+a^2}}\left\{\mathrm{arctan}\frac{\alpha\sqrt{1+a^2}}{\beta} - \mathrm{arctan}\frac{\alpha\gamma\sqrt{1+a^2}}{\beta\sqrt{B_4}}\right\} \nonumber \\ & + \frac{\sqrt{\alpha^2 + \beta^2}}{\alpha \beta}\Bigg\{\mathrm{arctan}\frac{a\gamma\alpha}{\sqrt{B_4(\alpha^2 + \beta^2)}} - \mathrm{arctan}\frac{a\alpha}{\sqrt{\alpha^2 + \beta^2}}\Bigg\} \nonumber \\ & + \frac{1}{\beta}\mathrm{arctan}\left(a\right)\\
    \text{and} \quad B_4 =& (1 + a^2)\alpha^2 + \beta^2 + \gamma^2
\end{align}
\end{subequations}
\begin{subequations}
\begin{align}
\int_{0}^{\infty}\mathrm{d}z\,&T_{\text{Owen}}\left(\sqrt{2}\alpha z, a\right)T_{\text{Owen}}\left(\sqrt{2}\beta z, b\right) = \frac{1}{8\pi^{3/2}}\Bigg\{\frac{a}{\alpha\sqrt{1+a^2}}\mathrm{arctan}\frac{b\alpha\sqrt{1+a^2}}{\sqrt{B_5}} \nonumber \\ & + \frac{b}{\beta\sqrt{1+b^2}}\mathrm{arctan}\frac{a\beta\sqrt{1+b^2}}{\sqrt{B_5}} + \frac{\sqrt{\alpha^2 + \beta^2}}{\alpha\beta}\mathrm{arctan}\frac{ab\alpha\beta}{\sqrt{B_5(\alpha^2 + \beta^2)}}\Bigg\} 
\end{align}
\begin{align}
    \text{where} \quad B_5 = (1+a^2)\alpha^2 + (1+b^2)\beta^2
\end{align}
\end{subequations}
\begin{align}\label{eq:Owen_identity}
    T_{\text{Owen}}\left(h, a\right) + T_{\text{Owen}}\left(a h, \frac{1}{a}\right) = \frac{1}{4}\left\{1 - \erf{\frac{h}{\sqrt{2}}}\erf{\frac{ah}{\sqrt{2}}}\right\}\qquad \textrm{for }h>0,~ a> 0
\end{align}

\section*{References}
\bibliographystyle{iopart-num}
\bibliography{references}

\end{document}